\documentclass{aastex62}

\usepackage{amsmath}
\usepackage{amssymb}
\usepackage{subfig}
\usepackage{placeins}

\submitjournal{A$\&$A}

\shortauthors{Lehmann et al.}

\begin{document}

\title{Density Waves and the Viscous Overstability in Saturn's Rings}

\correspondingauthor{Marius Lehmann}
\email{marius.lehmann@oulu.fi}

\author[0000-0002-0496-3539]{Marius Lehmann}
\affil{Astronomy Research Unit, University of Oulu, Finland}

\author{J\"urgen Schmidt}
\affil{Astronomy Research Unit, University of Oulu, Finland}

\author[0000-0002-4400-042X]{Heikki Salo}
\affil{Astronomy Research Unit, University of Oulu, Finland}

%
%
%
%



\begin{abstract}This paper addresses resonantly forced spiral density waves in a dense planetary ring which is close to the threshold for viscous 
overstability. 
We solve numerically the hydrodynamical equations for a dense thin disk in the vicinity of an inner Lindblad resonance with a perturbing 
satellite. Our numerical scheme is one-dimensional so that the spiral shape of a density wave is taken into account through a suitable approximation of the 
advective terms arising from the 
fluid orbital motion. This paper is a first attempt to model the co-existence of resonantly 
forced density waves and short-scale free overstable wavetrains as observed in Saturn's rings, by conducting large-scale hydrodynamical integrations. 
These integrations reveal that the two wave types undergo complex interactions, not taken into account in existing models for the damping of density waves. In 
particular it is found that, depending on the relative magnitude of both wave types, the presence of viscous overstability can lead to a damping of an unstable 
density wave and vice versa.
The damping of the short-scale viscous overstability by a density wave is investigated further by employing a simplified model of an axisymmetric ring 
perturbed by a nearby 
Lindblad resonance. A linear hydrodynamic stability analysis as well as local N-body simulations of this model system are performed and support the results of 
our large-scale hydrodynamical integrations.

 \end{abstract}


\section{Introduction}

The Cassini mission to Saturn has revealed a vast abundance of structures in the planet's ring system, spanning a wide range of length scales.
The finest of these structures have been detected by several Cassini instruments (\citet{collwell2007,thomson2007,hedman2014a}) and are periodic 
and quasi-axisymmetric\footnote{Upper limits for the cant-angle determined for these structures are within 1-3 degrees.} with 
wavelengths of some $100\,\text{m}$. It is generally accepted that this periodic micro structure originates from the viscous overstability 
mechanism which has been studied so far only in terms of axisymmetric models
(\citet{schmit1995,schmit1999,spahn2000a,salo2001,schmidt2003,latter2008,latter2009,latter2010,latter2013,lehmann2017}).
On much greater scales, typically 10's to 100's of kilometers, numerous spiral density waves propagate through the rings, as these are excited at radii where 
the orbiting ring particles are in resonance with the gravitational perturbation of one of the moons orbiting the ring system.

The process of excitation and damping of resonantly forced density waves has been thoroughly studied, mostly in terms of hydrodynamic models 
\citep{goldreich1978b,goldreich1978c,goldreich1979c,shu1984,shu1985a,shu1985b,borderies1985,bgt1986,lehmann2016}.
Throughout the literature one typically distinguishes between \emph{linear} and \emph{nonlinear} density waves.
The former are the ring's response to a relatively small, resonantly perturbing force in the sense that the excited surface mass density perturbation is small 
compared to the equilibrium value. In this case the governing hydrodynamic equations can be linearized and as a consequence the density wave appears sinusoidal 
in shape.

The studies by \citet{shu1985b}, \citet{bgt1986} (BGT86 henceforth) and \citet{lehmann2016} (LSS2016 
henceforth) considered the damping behavior of nonlinear 
density waves in a dense planetary ring, 
such as Saturn's B ring. For a nonlinear density wave the surface density perturbation is of the same order of magnitude as the equilibrium 
value.
Within a fluid description of the ring dynamics, the damping of a density wave is governed by different components of the pressure tensor.
The model by \citet{shu1985b} computes the pressure tensor from the kinetic second order moment equations, using a Krook-collision term.
The model predicts reasonable damping lengths of a density wave for assumed ground state optical 
depths (or surface mass densities) that do not exceed a certain critical value (which depends on the details of the collision term).
For optical depths larger than this critical value, the wave damping becomes very weak so that the resulting wavetrains propagate with ever increasing 
amplitude and nonlinearity. That said, the model fails to describe the damping of nonlinear waves in dense ring regions with high mutual collision frequencies 
of the ring particles, such as the wave excited at the 2:1 inner Lindblad resonance (ILR) with the moon Janus, propagating in Saturn's B ring. The main 
reason for this behavior of the model at large collision frequencies is most likely the neglect of nonlocal contributions to the (angular) 
momentum transport (\citet{shukhman1984,araki1986}) in their kinetic model.
On the other hand, BGT86 compute the pressure tensor from a fluid model (\citet{borderies1985}), as well as by using empirical 
formulae, which yield the correct qualitative 
behavior of the pressure tensor in a dense ring with a large volume filling factor. The computed damping lengths for optical depths relevant to Saturn's dense 
rings are fairly long and the authors suspect this to be a consequence of the fluid approximation. 

\citet{borderies1985} have shown that density waves are unstable in a sufficiently dense ring (such as Saturn's B ring), whereas they are stable in 
dilute rings of small optical depth.
\citet{schmidt2016} pointed out that the instability condition of spiral density waves is identical to the criterion for spontaneous viscous overstability 
(\citet{schmit1995}) in the limit of long wavelengths.
In LSS2016 we derived the damping of nonlinear density waves from a different view point compared to the approaches by 
BGT86 and 
\citet{shu1985b}, which are based on the streamline formalism (see \citet{longaretti1991}). We considered the density wave as a pattern that forms in response 
to this instability, using techniques that are widely applied in the 
studies of pattern formation in systems outside of equilibrium (\citet{cross1993}). Consequently, the wave damping is described in terms of a nonlinear 
amplitude 
equation. The resulting 
damping behavior is very similar to what is predicted by the BGT86 model.

While the models by BGT86 and LSS2016 can predict steady state profiles of density waves alone in an 
overstable 
ring region (see also \citet{stewart2016}), they do not take into account the 
possible presence of additional wave structures that can spontaneously arise in response to the viscous overstability, 
independent of a perturbing satellite. A first attempt to study the presence of multiple modes in a narrow ring
within the streamline formalism was due to \citet{longaretti1989}, but further improvements are required to model the (nonlinear) interaction of different 
modes.
The possibility of co-existence of resonant spiral density waves and short-scale near-axisymmetric periodic micro 
structure was discovered by analyzing stellar occultations of Saturn's A ring, recorded with the Cassini Visual and Infrared Mapping 
Spectrometer (\citet{hedman2014a}). 
This paper is concerned with a modeling of this co-existence and a qualitative understanding of interactions between a resonantly forced density wave and the 
short-scale waves 
generated by the viscous overstability.  
In our one-dimensional hydrodynamical scheme we need to assume that both the density wave and the short-scale waves are non-axisymmetric 
with the same 
azimuthal periodicity. However, since the short-scale waves resulting from spontaneous viscous overstability have wavelengths of some $100\,\text{m}$ (implying 
very small cant-angles of $10^{-3}-10^{-4}$ degrees), their dynamical evolution is expected to be very similar to that of the extensively studied axisymmetric 
modes (see the aforementioned papers). Hydrodynamical integrations presented in this paper confirm this expectation. 

In Section \ref{sec:theo} we outline the basic hydrodynamic model equations. Section \ref{sec:numerics} explains the numerical scheme applied to perform 
large-scale integrations of the hydrodynamical equations. Sections \ref{sec:forcing}, \ref{sec:azideriv} and \ref{sec:sgnum} discuss specific terms appearing 
in these equations that arise from the 
forcing by the satellite, the advection due to orbital motion of the ring fluid, as well as the collective self-gravity forces, respectively.
Results of large-scale hydrodynamical integrations are presented in Section \ref{sec:results}. Here we first describe the excitation process of a density wave 
as 
it follows from our integrations. Subsequently we test our scheme against the nonlinear models by BGT86 and 
LSS2016 in a marginally stable ring. In addition, we present some illustrative examples of density waves which propagate through a 
ring region which contains sharp radial gradients 
in the background surface mass 
density. 
We then consider waves that propagate in an overstable ring.
In order to facilitate an interpretation of the results from our large-scale integrations, we introduce a simplified axisymmetric
model to describe the perturbation of a ring due to a nearby ILR. We perform a linear hydrodynamic stability analysis of this model to compute linear growth 
rates of axisymmetric overstable waves in the perturbed ring. By employing the same model we then perform local N-body simulations of viscous overstability in 
a perturbed ring. Finally, Section \ref{sec:disc} provides a discussion of the main results.

\vspace{0.5cm}

%

\section{Hydrodynamic Model}\label{sec:theo}
From the vertically integrated isothermal balance equations for a dense planetary ring  we derive the model equations
(\citet{stewart1984,schmidt2009}
 \begin{align}\label{eq:nleq}
\begin{split}
\partial_{t} \tau & = - \left[\Omega-\Omega_{L} \right]  \, \partial_{\theta}\tau  -u \partial_{r} \tau - \tau \partial_{r} u, \\[0.1cm]
  \partial_{t} u  &  = - \left[\Omega-\Omega_{L} \right] \,  \partial_{\theta}u -u\partial_{r} u  + 2\Omega  v  -\partial_{r}\left[\phi^{d} + 
\phi^{s} \right] - 
\frac{1}{\sigma} \partial_{r} \hat{P}_{rr} ,
\\[0.1cm] 
\partial_{t} v & = - \left[\Omega-\Omega_{L} \right] \, \partial_{\theta}v - u \partial_{r}v -\frac{1}{2}\Omega  u -\frac{1}{r}\partial_{\theta} 
\phi^{s} - 
\frac{1}{\sigma} \partial_{r} \hat{P}_{r\theta},
\end{split}
\end{align}
in a cylindrical frame $(r,\theta,z=0)$ with origin at $r=r_{L}$, rotating rigidly with angular frequency $\Omega_{L} = \Omega(r_{L})$ where $r_{L}$ denotes 
the 
radial 
location of a specific inner Lindblad resonance (ILR) with a perturbing satellite and 
\begin{equation}
\Omega = \left[\frac{G M_{P}}{r^3}\right]^{1/2}          
\end{equation}
with Saturn's mass $M_{P}=5.96 \cdot 10^{26}\, \text{kg}$ and the gravitational constant $G=6.67\cdot 10^{-11}$. 
In what follows we will also make use of the radial distance
\begin{equation}\label{eq:xx}
 x=r-r_{L}
\end{equation}
as well as its scaled version $\tilde{x}=x/r_{L}$.

The quantity $\sigma$ is the rings' surface mass density and $\tau=\sigma/\sigma_{0}$ with the ground state surface mass density $\sigma_{0}$. The 
symbols $u$, $v$ stand for the radial and azimuthal components of the 
velocity on top of the orbital velocity $\left[\Omega \, -\Omega_{L}\right] r$ in the rigidly rotating frame. Furthermore, $\hat{P}$ is the pressure tensor 
(see below).
The central planet is assumed spherical so that $\Omega=\kappa$, the latter denoting the epicyclic frequency of ring particles. 
The rings' ground state which describes the balance of central gravity and centrifugal force is subtracted from above equations and we neglect the large-scale 
viscous evolution of the rings which occurs on time scales much longer than those considered in this study.

We neglect curvature terms containing factors $1/r$ since these scale as $\lambda /r \sim 
10^{-4}$ compared to radial derivatives. Here $\lambda$ denotes the typical radial wavelength of a spiral density wave near its related Lindblad 
resonance where $\tilde{x} \ll 1$.
From all terms containing derivatives with respect to $\theta$ we retain only the advective terms arising from the Keplerian motion, i.e.\ 
the first terms on the right hand sides of Equations (\ref{eq:nleq}). All other $\theta$-derivatives scale as $(m \lambda) /r$ compared to radial derivatives 
($m$ denoting the number of spiral arms of the density wave), i.e.\ the same as curvature terms.

Poisson's equation for a thin disk
\begin{equation}\label{eq:poissoneq}
 (\partial_{r}^2 +\partial_{z}^2) \phi^{d}=4 \pi G \sigma \delta(z), 
\end{equation}
establishes a relation between the self-gravity potential $\phi^{d}$ and the surface density $\sigma$.

The viscous stress is assumed to be of Newtonian form such that in the cylindrical frame we can write
\begin{equation}\label{eq:pten}
\begin{split}
\begin{array}{@{}*{22}{l@{}}}
\hat{P} &=  \begin{pmatrix}  \hat{P}_{rr}  \hspace{0.2 cm} & \hat{P}_{r\theta} \\[0.18cm] 
\hat{P}_{\theta r}  \hspace{0.2 cm}  & \hat{P}_{\theta\theta}  \end{pmatrix}\\[0.5cm]
\quad & =\begin{pmatrix}   p -\eta\left(\frac{4}{3}+\hat{\gamma}\right)\partial_{r}u   \hspace{0.2 cm} & -\eta\left(-\frac{3}{2}\Omega + 
\partial_{r}v \right) 
\\[0.15 cm] 
-\eta\left(-\frac{3}{2}\Omega +\partial_{r}v \right)  \hspace{0.2 cm}  & p + \eta \left(\frac{2}{3}-\hat{\gamma}\right)\partial_{r} u  \end{pmatrix} .
\end{array}
\end{split}
\end{equation}
It is thus completely described by radial gradients of the velocities $u$, $v$, the dynamic shear viscosity $\eta$ as well as the isotropic pressure $p$ 
(see below).
The ratio of the bulk and shear viscosity is denoted by $\hat{\gamma}$, which is assumed to be constant (\citet{schmit1995}).
The isotropic pressure and the dynamic shear viscosity take the simple forms
\begin{equation}
 p = p_{0} \tau^{p_{\sigma}} \label{eq:pres},
 \end{equation}
 \vspace{-0.5cm}
 \begin{equation}
 \eta =\nu_{0} \sigma_{0} \tau^{\beta+1} \label{eq:shearvis}.
\end{equation}
In this study we assume $p_{\sigma}=1$, i.e.\ the equation of state for an ideal gas. The ground state pressure can be defined in terms of an effective ground 
state velocity dispersion $c_{0}$ such that (\citet{schmidt2001b})
\begin{equation}
 p_{0} = \sigma_{0} c_{0}^2.
\end{equation}
The ground state is characterized by $\sigma_{0}=\text{const.}$, $u_{0}=0$, 
$v_{0}=0$, together with the 
parameters in (\ref{eq:pres}) and (\ref{eq:shearvis}).

We neglect azimuthal contributions due to collective self-gravity forces. This neglect is adequate as long as the exerted satellite torque is much smaller than
the unperturbed viscous angular momentum luminosity of the ring. That is, the (self-gravitational) angular momentum luminosity carried by the wave is 
negligible compared to the viscous luminosity.
The linear inviscid satellite torque deposited at the resonance site reads (\citet{goldreich1979c})
\begin{equation}\label{eq:lintor}
T^{s}= -m \pi^2 \frac{ \sigma_{0}}{\mathcal{D} \, \Omega_{L}^2} \left(\epsilon \, r_{L} \Omega_{L}\right)^4 \left[ \partial_{r}\phi^{s}  -2 m \,\phi^{s} 
\right]_{r_{L}}^2
\end{equation} 
where
\begin{equation}\label{eq:eps}
 \epsilon = \frac{2 \pi G \sigma_{0}}{r_{L} \mathcal{D}}
\end{equation}
and  (\citet{cuzzi1984})
\begin{equation}\label{eq:det}
 \mathcal{D}=3\left(m-1\right)\Omega_{L}^2.
\end{equation}
The viscous angular momentum luminosity in the unperturbed disk is given by (\citet{lynden1974})
\begin{equation*}
L^{visc} = 3 \pi \nu_{0} \sigma_{0}\Omega r^2.
\end{equation*}
In addition it should be mentioned that we are not concerned with the long-term redistribution of ring surface mass density which occurs in response to the
presence of very strong 
density waves (BGT86) so that we assume $\sigma_{0}=const$ as mentioned before. 

For the sake of definiteness we will restrict to parameters corresponding to the Prometheus 7:6 ILR, located at $r\sim 126,000\,\text{km}$ in Saturn's A ring.
We take values of the rings' ground state shear viscosity $\nu_{0}$ and surface mass density $\sigma_{0}$ (see Table \ref{tab:hydropar}) that can be estimated 
from corresponding values obtained by \citet{tiscareno2007b} for this ring region. 
The nominal values of $\beta$ and $\gamma$ correspond to values found in N-body simulations with an optical depth $\tau^{dyn}=1$ [see 
LSS2016 (Section 3)].
Besides the nominal values we will use a range of values for $\beta$ [Equation (\ref{eq:shearvis})] and also $T^{s}$ [Equation 
(\ref{eq:lintor})], in order to explore a variety of qualitatively different scenarios for the damping of 
density waves.
The adopted value for the ground state velocity dispersion $c_{0}$ is larger then what results from local non-gravitating N-body simulations for optical depths 
relevant to this study (e.g.\ \citet{salo1991b}) but corresponds roughly to expected values for Saturn's A ring from self-gravitating N-body simulations 
exhibiting gravitational wakes and assuming meter-sized particles (\cite{daisaka2001,salo2018}). Furthermore, the value is still small enough to ignore 
pressure effects on the density waves' dispersion relation (Section \ref{sec:excite}).

Our hydrodynamic model exhibits spontaneous viscous overstability on finite wavelengths if the viscous parameter $\beta$ exceeds a critical value.
To see this, let us ignore the satellite forcing $\phi^{s}$ for the time being.
We restrict to short radial length scales so that $\Omega=\Omega_{L}$ can be considered constant except that we use
\begin{equation*}
 \left[\Omega -\Omega_{L} \right]\partial_{\theta} = -\frac{3}{2} \Omega_{L} x \partial_{y}
\end{equation*}
in Equations (\ref{eq:nleq}).
Our 1D numerical method to solve Equations (\ref{eq:nleq}) assumes that any mode which forms has $m$-fold azimuthal periodicity (see Section 
\ref{sec:azideriv}).
Hence let us introduce non-axisymmetric oscillatory perturbations such that
\begin{equation}\label{eq:linperos}
 \begin{pmatrix} \tau   \\ u\\ v   \end{pmatrix} = \begin{pmatrix} 1   \\ 0\\ 0  \end{pmatrix} + \begin{pmatrix} 
\tau^{'} \\ u^{'} \\ v^{'} \end{pmatrix} \exp\left\{\omega t + i \left(k_{x} +\frac{3}{2}\frac{m-1}{m} k_{y} \Omega_{L} t \right)x + i k_{y} y \right\},
\end{equation}
with complex oscillation frequency $\omega=\omega_{R} + i \, \omega_{I}$ and real-valued radial and azimuthal wavenumbers $k_{x}>0$ and 
$k_{y}=\frac{m}{r_{L}}$, respectively.
The time-dependent contribution to the radial wavenumber in (\ref{eq:linperos}) stems from the winding of the perturbations due to Keplerian shear [see 
\citet{mvs1987} and Equation 
(\ref{eq:lamwrap})].
Since we know that the linear growth and the nonlinear saturation of spontaneous viscous overstability occurs on wavelengths of typically hundreds of meters
it turns out that we can neglect the effect of winding in (\ref{eq:linperos}).
That is, for the relevant modes the time it takes for the winding term to become equal to $k_{x}$ is given by 
\begin{equation*}
 t \sim  \frac{2}{3 m} \frac{r_{L}}{\lambda_{x}} \,\text{ORB}.
\end{equation*}
With $\lambda_{x}=2\pi /k_{x}\sim 100\,\text{m}$ this yields some 100,000 orbits, which is much longer than the time scale of the nonlinear evolution 
of the modes (i.e.\ thousands of orbits, see \citet{latter2010,latter2013};~LSS2017).
Furthermore, Poisson's equation (\ref{eq:poissoneq}) yields the relationship
\begin{equation}\label{eq:wkbsgos}
 \phi^{'\,d} = -\frac{2 \pi G \sigma_{0}}{k_{x}} \tau^{'}
\end{equation}
for a single wavelength mode (\citet{binney1987}).

In the remainder of this section we apply dimensional scalings such that time is scaled with $1/\Omega_{L}$ and length is scaled with 
$c_{0}/\Omega_{L}$. 
Inserting (\ref{eq:linperos}) and (\ref{eq:wkbsgos}) into (\ref{eq:nleq}) and linearizing with respect to the perturbations (the primed quantities), results in 
the eigenvalue 
problem 
\begin{equation}\label{eq:evp}
\begin{split}
 0 & =  - \omega^3 +
 \omega^2 \left[- \left(\frac{7}{3} + \gamma\right) k_{x}^2 \nu_{0} + \frac{9}{2} i m \tilde{x}\right]\\
 \quad & +  \omega \left[-1 + 2 g k_x - k_{x}^2 - \left(\frac{4}{3} +  \gamma \right) k_{x}^4 \nu_{0}^2  \right. \\ 
 \quad &   \left.  +i \left(7 + 3 \gamma \right) k_{x}^2 m \tilde{x} \nu_{0}  + \frac{27}{4} m^2 \tilde{x}^2 \right]\\
\quad &  -k_{x}^2 \left(3 + 3 \beta - 2 g k_x + k_{x}^2\right) \nu_{0} \\
\quad & +  \frac{1}{2} i m \tilde{x} \left(3 - 6 g k_x + 3 k_{x}^2 + \left(4 + 3 \gamma \right) k_{x}^4 \nu_{0}^2\right)  \\
\quad & +  \frac{3}{4} \left(7 + 3 \gamma \right) k_{x}^2 m^2 \tilde{x}^2 \nu_{0}  - \frac{27}{8} i m^3 \tilde{x}^3 \\
\end{split}
\end{equation}
for $\omega=\omega_{R} + i \omega_{I}$. The non-dimensional distance $\tilde{x}$ is defined as below Equation (\ref{eq:xx}).
This equation can be used to obtain the growth rate $\omega_{R}(k_{x})$ and oscillation frequency $\omega_{I}(k_{x})$ of a given mode $k_{x}$.
This procedure has been carried out for axisymmetric modes (with $m=0$) in several papers [see \citet{lehmann2017} (~LSS2017 
hereafter) and references therein for 
more details].
It can be shown that the growth rates $\omega_{R}(k_{x})$ following from Equation (\ref{eq:evp}) are independent of $m$ (i.e.\ independent of $k_{y}$)
and agree with those of previous studies.

In the remainder of the paper the symbol $k$ denotes the \emph{radial} wavenumber of a given mode.
The threshold for viscous oscillatory overstability, i.e. a vanishing growth rate $\omega_{R}(k)=0$, can be obtained
by setting $\omega = i \omega_{I}$ and solving the imaginary and real parts of Equation (\ref{eq:evp}) for $\omega_{I}$
and $\beta$, respectively, for a given wavenumber $k$. This yields the critical frequency pair\footnote{The third critical frequency is associated with the 
diffusive viscous 
instability, not considered in this paper.}
\begin{equation}\label{eq:omcrit}
 \omega_{c}(k)= \frac{3}{2} m \tilde{x} \pm \sqrt{1 -2 g k + k^2 +\left(\frac{4}{3}+\gamma \right) \nu_{0}^2 k^4}
\end{equation}
and the critical value of the viscosity 
parameter 
\begin{equation}\label{eq:betc}
\begin{split}
  \beta_{c}(k) & = \frac{1}{3}\left(\gamma - \frac{2}{3}\right) -\frac{2}{3}\left(\frac{4}{3}+\gamma \right) g  k\\
  & \quad + \frac{1}{3}\left(\frac{4}{3}+\gamma\right)  k^2 + \frac{1}{27} \left(28 + 33 \gamma + 9 \gamma^2 \right) \nu_{0}^2 k^4 ,
\end{split}
\end{equation}
which describes the stability boundary for viscous overstability and which is also independent of $m$. 
The frequencies (\ref{eq:omcrit}) are Doppler-shifted by $\frac{3}{2} m \tilde{x}$ as compared to the frequencies of axisymmetric modes.
Note that due to the fact that Equations (\ref{eq:nleq}) are defined in a frame rotating with $\Omega_{L}$ this Doppler-shift is very small as $\tilde{x}\sim 
10^{-3}$ 
for all cases considered in this paper. The Doppler-shift can therefore be neglected.
These results show that linear free non-axisymmetric short-scale modes due to spontaneous viscous overstability in our hydrodynamic model behave 
essentially the same as axisymmetric modes with $m=0$.

The curve $\beta_{c}(k)$ possesses a minimum at 
finite wavelength if $g>0$, i.e.\ for a non-vanishing collective self-gravity force.
This wavelength is roughly two times the Jeans-wavelength $\lambda_{J}=c_{0}^2/(G \sigma_{0})$.
In the above equations we define 
\begin{equation}\label{eq:gg}
 g = \frac{ \pi G \sigma_{0} }{\Omega c_{0}},
\end{equation}
denoting the inverse of the hydrodynamic Toomre-parameter (a full list of symbols is provided in Table \ref{tab:scalings}).

\begin{table}[h!]
\caption{Hydrodynamic Parameters} 
\label{tab:hydropar} 
\centering 
\begin{tabular}{l l} 
\hline\hline 
parameter & Prometheus 7:6 ($Pr76$)  \\ 
\hline 
         $c_{0}$ [$10^{-3} \,\text{m}\text{s}^{-1}$] &    1.5   \\
 $\nu_{0}$ [$10^{-2} \,\text{m}^{2}\text{s}^{-1}$] & 1    \\
        $\gamma$  &    4.37    \\
$\beta$ &     0.85  \\
        $\sigma_{0}$ $[\text{kg}\,\text{m}^{-2}]$ &     350  \\
           $r_{L}$ [$10^8\,\text{m}]$  &  1.26  \\ 
             $T^{s}$ [$10^{10}\,\text{kg}\,\text{m}^2]$  &   4.56  \\
              $L^{visc}$ [$10^{10}\,\text{kg}\,\text{m}^2]$  &   7200 \\
\hline 
\end{tabular}
\end{table}

\section{Numerical Methods}\label{sec:numerics}

For numerical solution of Equations (\ref{eq:nleq}) we apply a finite difference Flux Vector Splitting method employing a Weighted Essentially 
Non-Oscillatory (WENO) reconstruction of the flux vector components. The method is identical to that used in LSS2017, apart from 
the reconstruction of the flux 
vector.

We define the flux-conservative variables
\begin{equation*}
 \mathbf{U} = \begin{pmatrix} \tau  \\ \tau u \\ \tau v   \end{pmatrix}
\end{equation*}
so that Equations (\ref{eq:nleq}) can be written as
\begin{equation}\label{eq:nleqnum}
 \partial_{t} \mathbf{U} = - \partial_{r}\mathbf{F} + \mathbf{S}
\end{equation}
 with the flux vector
\begin{equation*}
 \mathbf{F} = \begin{pmatrix} \tau u   \\ \tau u^2 + \tau c_{0}^2 \\ \tau u v  \end{pmatrix}
\end{equation*}
and the source term \begin{equation}\label{eq:source}
 \mathbf{S} = \begin{pmatrix} -\left[\Omega-\Omega_{L}\right] \partial_{\theta}\tau  \\ -\left[\Omega-\Omega_{L}\right] \partial_{\theta}\left(u \tau\right) 
 + 2\Omega  \tau v - \tau \partial_{r} (\phi^{d} +\phi^{s}) + \frac{1}{\sigma_{0}} \partial_{r} \hat{\Pi}_{rr}  \\-\left[\Omega-\Omega_{L}\right] 
\partial_{\theta}\left(v \tau\right) -\frac{1}{2}\Omega \tau u + \frac{1}{\sigma_{0}} \partial_{r} \hat{\Pi}_{r\theta}    \end{pmatrix}.
\end{equation}
In the last expression
\begin{equation*}
\hat{\Pi}= p\, \hat{U} -\hat{P}  
\end{equation*}
is the viscous stress tensor with $\hat{U}$ denoting the unity tensor.

We solve (\ref{eq:nleqnum}) on a radial domain of size $L_{r}$. The domain is discretized by defining nodes $r_{j}$ ($j=1,2,\ldots,n$) with constant 
inter-spacing $h=r_{j+1}-r_{j}$.
We adopt periodic boundary conditions in all integrations. Since a density wave is not periodic in radial direction this requires the radial domain 
size $L_{r}$ to be large enough so
that the Lindblad resonance is located sufficiently far from the inner domain boundary and that an excited density wave is fully damped 
before reaching the outer domain boundary. 
The discretization of the flux derivative $\partial_{r}\mathbf{F}$ is outlined in Appendix \ref{sec:weno}.
The source term (\ref{eq:source}) contains radial derivatives of the stress tensor which are evaluated with central discretizations of 12th order.
Furthermore, the evaluation of the derivatives with respect to $\theta$ and the self-gravity force $\partial_{r} \phi^{d}$ appearing in (\ref{eq:source}) will 
be discussed in Sections 
\ref{sec:azideriv} and \ref{sec:sgnum},
respectively.

\begin{table*}
\caption{List of Symbols} 
\label{tab:scalings} 
\centering 
\begin{tabular}{l l} 
\hline\hline 
Symbol & Meaning  \\ 
\hline 
   $G$ &  gravitational constant \\ 
     $M_{P}$ &  planet's mass \\ 
          $M^{s}$ & mass of perturbing satellite \\ 
           		$a^{s}$ & semimajor axis of perturbing satellite \\
 		$e^{s}$ & eccentricity of perturbing satellite \\
               $T^{s}$ & linear inviscid satellite torque \\    
           		$\phi^{s}$  & satellite potential \\   
               $L^{visc}$ & viscous angular momentum luminosity \\ 
    $\Omega$ & Kepler frequency\\
    $\Omega_{L}$ & Kepler frequency at $r=r_{L}$\\
    $\kappa$ & epicyclic frequency\\
       $\Omega_{Z}$ & vertical frequency of ring particles\\
    $ x=r-r_{L}$ & radial coordinate \\ 
     $ \tilde{x}=\frac{r-r_{L}}{r_{L}}$ & scaled radial coordinate \\ 
    $t$ & time \\ 
      		$\omega^{s}$ & satellite forcing frequency in the frame rotating with $\Omega_{L}$\\  
      		$\hat{\omega}^{s}$ & satellite forcing frequency in the inertial frame \\  
            	$\omega$ & complex frequency of overstable waves \\  
   		$\sigma$ & surface mass density \\
   		$\tau=\frac{\sigma}{\sigma_{0}}$ & scaled surface mass density \\
   		$\tau^{dyn}$ & dynamical optical depth \\
		$u$, $v$ & planar velocity components \\
		$\phi^{d}$  & self-gravity potential \\
		$\phi^{p}$  & planetary potential \\
		  $c_{0}=\sqrt{\frac{p_{0}}{\sigma_{0}}}$ & effective isothermal velocity dispersion \\
          $\nu_{0}$ & ground state kinematic shear viscosity       \\
         	$p$ & isotropic pressure \\
     		$\eta$ & dynamic shear viscosity\\
     		$\beta$ & viscosity parameter \\
     		$\gamma$ & constant ratio of bulk and shear viscosity \\
     		$\hat{P}$ & pressure tensor \\	
     		$g=\frac{\pi G \sigma_{0}}{\Omega c_{0}}$ & inverse ground state Toomre-parameter \\	
 $\hat{\gamma}$ & phase variable of a fluid streamline \\
 		$q$ & nonlinearity parameter of a fluid streamline \\
 		$\Delta$ & phase angle of a fluid streamline \\
 		$a$ & semimajor axis of a fluid streamline \\
 		$e$ & eccentricity of a fluid streamline \\ 
\hline 
\end{tabular}
\end{table*}
Due to the presence of the satellite forcing terms in (\ref{eq:source}) it turns out that the simple time step criterion arising from a 
one-dimensional 
advection-diffusion problem, which was used in LSS2017, is unnecessarily strict.
This criterion reads
\begin{equation}\label{eq:timestep}
 \Delta t \leq   \text{min}\left(\frac{h^2}{2 \hat{\nu}},\frac{2 \hat{\nu}}{\Lambda^2}\right) ,
\end{equation}
where $\Lambda$ is identified with the maximal eigenvalue of the Jacobian 
\begin{equation}\label{eq:jacobian}
\hat{A}= \frac{\partial \mathbf{F}(\mathbf{U})}{\partial  \mathbf{U}}
\end{equation}
of Equations (\ref{eq:nleqnum}) for the whole grid and $\hat{\nu}$ is to be identified with the maximal value of the coefficient in front of the term 
containing 
the second radial derivative $\partial_{r}^2 u$ in (\ref{eq:nleq}), which is
\begin{equation*}
 \hat{\nu} = \nu_{0} \left(\frac{4}{3}+\gamma\right) \tau^{\beta}.
\end{equation*}
The three eigenvalues of (\ref{eq:jacobian}) read
\begin{align*}
\begin{split}
\Lambda_{1}  & = u, \\
\Lambda_{(2/3)} & = u \pm c_{0}.
\end{split}
\end{align*}
For most integrations presented in this paper the grid spacings $h$ are large enough so that the second term in (\ref{eq:timestep}) is by far 
the smallest
and can take values down to some $10^{-5}\, \text{ORB}$. We find, however, that time steps in the range $\Delta t=1-5\cdot 10^{-4}\, \text{ORB}$ are 
suitable for all presented integrations, indicating that the criterion (\ref{eq:timestep}) cannot be appropriate. We have checked for some integrations 
with 
strong satellite forcing that reducing the time step by a factor of $0.5$ does not lead to any notable changes.
For later use we also define the mean kinetic energy density within the computational domain as
\begin{equation}\label{eq:ekin}
 e_{kin} = \frac{1}{L_{r}} \int\limits_{[L_{r}]} \mathrm{d}r \, \frac{1}{2} \sigma \left( u^2 + v^2 \right). 
\end{equation}

\vspace{0.5cm}

\section{Satellite Forcing Terms}\label{sec:forcing}

For simplicity, we restrict to density waves that correspond to a particular inner Lindblad resonance\footnote{In the current 
approximation a Lindblad
resonance coincides with a mean motion resonance.} of first order, so that the forcing satellite orbits exterior to the considered ring portion. The 
wave is 
excited by a particular Fourier mode of the gravitational potential due to this satellite with mass $M^{s}$ and semi-major axis $a^{s}$ and reads (cf.\ Section 
5 in LSS2016)
\begin{equation*}
 \phi^{s}(t,\hat{\theta}) = -\frac{G M^{s}}{a^{s}} b_{1/2}^{m} \, \exp \left\{ i \left(m \hat{\theta} - \hat{\omega}^{s} t \right) \right\},
\end{equation*}
valid in an inertial frame denoted by $(r,\hat{\theta})$. 
The symbol
\begin{equation*}
b_{1/2}^{m}   = \frac{2}{\pi} \int_{0}^{ \pi} \mathrm{d}\Psi \frac{\cos\left(m \Psi\right)}{\sqrt{1+ \rho^2 - 2 \rho 
\cos\left( \Psi\right)}}
\end{equation*} 
is a Laplace-coefficient with
 \begin{equation*}
\rho = \frac{r}{a^{s}}.
\end{equation*} 
In the current approximation the forcing frequency reads
\begin{equation}\label{eq:omfin}
 \hat{\omega}^{s} = m  \Omega^{s} = \left(m-1 \right) \Omega_{L}
\end{equation}
with the satellite mean motion $\Omega^{s}$.
Upon changing to the frame rotating with frequency $\Omega_{L}$, denoted by $(r,\theta)$, we have
\begin{equation*}
\begin{split}
(m \hat{\theta} -\hat{\omega}^{s} t)  & \rightarrow \left( m \left[\theta+\Omega_{L} t\right] -\hat{\omega}^{s} t\right)\\
\quad & \equiv  m\theta -\omega^{s} t,
\end{split}
\end{equation*}
yielding the forcing frequency in the rotating frame
\begin{equation}\label{eq:omegaf}
\begin{split}
 \omega^{s} & =  \hat{\omega}^{s} - m \Omega_{L}\\[0.4cm]
  \quad & = -\Omega_{L},
  \end{split}
\end{equation}
where we used (\ref{eq:omfin}).
Therefore, the radial forcing component appearing in Equations (\ref{eq:nleq}) reads
\begin{equation}\label{eq:fsatrad}
  - \partial_{r}\phi^{s} = \frac{G M^{s}}{a^{s}}\, \partial_{r}b_{1/2}^m \exp \left\{ i \left(m \theta + \Omega_{L} t \right) \right\} .
\end{equation}
Similarly, the azimuthal component is given by
\begin{equation}\label{eq:fsatazi}
   - \frac{1}{r}\partial_{\theta}  \phi^{s}   = \frac{G M^{s}}{a^{s}} \, b_{1/2}^m m  \exp \left\{ i \left(m \theta + \Omega_{L} t +\pi/ 2\right) 
\right\}.
\end{equation}
These terms are evaluated at $r=r_{L}$.

 \vspace{0.5cm}

\section{Azimuthal Derivatives}\label{sec:azideriv}

The persistent spiral shape of a (long) density wave is generated by the resonant interaction between the ring material and the 
perturbing satellite potential, as well as the collective self-gravity force. Since our integrations are one-dimensional, it is not possible to describe 
azimuthal structures directly. 
Therefore we need to restrict Equations (\ref{eq:nleq}) to a radial cut which we choose to be $\theta=0$ without loss of generality.
The information about the azimuthal structure of the density pattern (the number of spiral arms $m$) is then contained solely in the terms describing azimuthal 
advection due to orbital motion. i.e.\ the first terms on the right hand sides of Equations (\ref{eq:nleq}). In what follows we refer to these terms 
simply as ``azimuthal derivatives``.
Thus, the requirement is to prescribe proper values of the azimuthal derivatives at $\theta=0$ for each time step of the integration.


We again adopt the cylindrical coordinate frame $(r,\theta)$ of Section \ref{sec:theo} which rotates with angular velocity $\Omega_{L}$ relative to an 
inertial frame 
denoted by 
$(r,\hat{\theta})$ so that
\begin{equation*}
\theta= \hat{\theta} - \Omega_{L} t.
\end{equation*}
If we linearize Equations (\ref{eq:nleq}) with respect to the variables $\tau$, $u$, $v$ and $\phi^{d}$, so that we restrict these to describe linear density 
waves, it is possible to solve the equations in the complex plane by splitting the solution vector $\mathbf{\Psi}$ into its real 
and imaginary parts:
\begin{equation}\label{eq:vectorsim}
\begin{split}
 \mathbf{\Psi} &  = \mathbf{\Psi}_{R} + i \, \mathbf{\Psi}_{I}\\[0.4cm]
  & =  \begin{pmatrix} \tau(r,\theta,t) \\ u(r,\theta,t) \\ v(r,\theta,t) \end{pmatrix}  \\[0.4cm]
  & = \begin{pmatrix} \tau_{R}(r,\theta,t) \\ u_{R}(r,\theta,t) 
\\  v_{R}(r,\theta,t) \end{pmatrix} + i \, \begin{pmatrix} \tau_{I}(r,\theta,t) \\ u_{I}(r,\theta,t) \\  v_{I}(r,\theta,t) \end{pmatrix}. 
\end{split}
\end{equation}

An evolving $m$-armed linear density wave can then be described through the complex vector of state 
\begin{equation}\label{eq:vecsim}
\begin{pmatrix} \tau(r,\theta,t) \\ u(r,\theta,t) \\  v(r,\theta,t) \end{pmatrix}  =   \begin{pmatrix} \mathcal{A_{\tau}}\left(r,t\right) 
   \\ \mathcal{A}_{u}\left(r,t\right)  \\ \mathcal{A}_{v}\left(r,t\right)  \end{pmatrix} \exp \left\{ i (m \theta -\omega^{s} t)\right\} 
\end{equation}
with $\mathcal{A_{\tau}}\left(r,t\right)$, $\mathcal{A}_{u}\left(r,t\right)$ and $\mathcal{A}_{v}\left(r,t\right)$ being complex amplitudes in this notation 
which depend on time and the radial coordinate. The time dependence of the amplitudes is generally much slower than the oscillatory terms and vanishes once 
the integration reaches a steady state.
%
When inserted to the linearized Eqs. (\ref{eq:nleq}) we obtain two sets of three equations that are possibly coupled through the azimuthal derivatives 
and self-gravity, depending on the applied implementation (cf.\ Appendices \ref{sec:azimb} and \ref{sec:wkbsg}).
Note that in practice we will exclusively use the full nonlinear equations (\ref{eq:nleq}). For sufficiently small amplitudes 
$\mathcal{A_{\tau}}\left(r,t\right)$, $\mathcal{A}_{u}\left(r,t\right)$ , $\mathcal{A}_{v}\left(r,t\right)$ the nonlinear terms in (\ref{eq:nleq}) are 
negligible and the equations are essentially linear.

In order to describe \emph{nonlinear density waves}, it is necessary to make an approximation for the azimuthal dependency of the wave.
To obtain such an approximation we assume that (\ref{eq:vecsim}) holds also in the nonlinear case. We will discuss the validity of this assumption
a bit more at the end of Section \ref{sec:azima}. 
We have found two such implementations for the azimuthal derivatives (simply referred to as \emph{Methods A} and \emph{B}) that yield 
a 
stationary final 
state for Equations (\ref{eq:nleq}) in the nonlinear regime.
It turns out that the application of \emph{Method A} (Section \ref{sec:azima}) results in nonlinear wave profiles that agree better with existing nonlinear 
models and therefore the results 
presented in subsequent 
sections are based on integrations using this method. We additionally outline \emph{Method B} (Appendix \ref{sec:azimb}) as we have found it to work 
well in the weakly 
nonlinear regime. Note that for sufficiently linear waves, both methods are exact down to the numerical error.

 \vspace{-0.5cm}
\subsection{Method A}\label{sec:azima}

One implementation of the azimuthal derivatives can be derived if one considers the vector of state of the weakly nonlinear model of 
LSS2016 
in 
the first order approximation, where 
contributions from higher wave harmonics are omitted. 
That is, we have 
\begin{equation}\label{eq:azideriv}
\begin{split}
 & \begin{pmatrix} \tilde{\phi}^{d}(r,\theta,t) \\ \tilde{u}(r,\theta,t) \\ \tilde{v}(r,\theta,t) \end{pmatrix} = \\[0.4 cm]
 &   \begin{pmatrix} -\dfrac{i 2 \tilde{\mathcal{D}} \tilde{\Omega}}{\tilde{\mathcal{D}} \tilde{x}  +  \left(\tilde{\omega}^{s} - m 
[\tilde{\Omega}-\tilde{\Omega}_{L}] \right)^2}  \\[0.6 cm] \dfrac{i 2 
\tilde{\Omega} \left( \tilde{\omega}^{s} - m [\tilde{\Omega}-\tilde{\Omega}_{L}] \right)}{ \tilde{\mathcal{D}} \tilde{x}  +  \left(\tilde{\omega}^{s} - m 
[\tilde{\Omega}-\tilde{\Omega}_{L}] \right)^2}   \\[0.6cm]  1   \end{pmatrix}\\ 
\quad & \times \mathcal{A}\left(r\right) \, \exp \left\{i \left( \, \Phi(r) +  m \theta - \omega^{s} t \right) \right\}\hspace{0.2cm}  + c.c. \hspace{0.2cm} ,
 \end{split}
\end{equation}
where $\tilde{\phi}^{d}$, $\tilde{u}$, $\tilde{v}$, as well as $\tilde{\Omega}$, $\tilde{\Omega}_{L}$, $\tilde{\mathcal{D}}$ and $\tilde{\omega}^{s}$ 
are understood to be scaled according to Table 1 in LSS2016 and $\tilde{x}$ is scaled 
according to Table \ref{tab:scalings}. Note that $\tilde{\omega}^{s}$ is the scaled version of (\ref{eq:omegaf}).
Equation (\ref{eq:azideriv}) corresponds to Equations (35) and (45) in LSS2016, except that (\ref{eq:azideriv}) is written in 
the rotating frame and is \emph{not} expanded to the lowest order in $\tilde{x}$, although small corrections due to pressure and viscosity are neglected.

From the solution of the Poisson-Equation we have (\citet{shu1984})  
\begin{equation}\label{eq:poissonazi}
 \tau = i \, \frac{ \textrm{sgn}(k)}{2 \pi G \sigma_{0}}\partial_{r} \phi^{d} 
\end{equation}
where $\phi^{d}$ is assumed to be given by the unscaled first component of (\ref{eq:azideriv}) and $k= \partial_{r} 
\Phi(r)$ denotes the wavenumber of the 
density wave. Note that 
both sides of Equation 
(\ref{eq:poissonazi}) are understood to be real-valued since the $\mathrm{sgn}$-function takes different signs for the two conjugate complex exponential modes 
in (\ref{eq:azideriv}). The
 azimuthal derivative of $\tau$ is then given by
\begin{equation}\label{eq:azitau}
 \partial_{\theta} \tau = -\frac{m }{2 \pi G \sigma_{0}}\partial_{r} \phi^{d} \, 
\end{equation}
if we assume $k>0$.
Equation (\ref{eq:azitau}) shows that the azimuthal derivative of $\tau$ is directly proportional to the radial component of the self-gravity force, the 
computation of which we discuss in Section \ref{sec:sgnum}.
The azimuthal derivatives of $u$ and $v$ can be directly obtained from (\ref{eq:azideriv}) and read
\begin{align}
 \partial_{\theta} u & = -\frac{ 2 m \Omega \left( \omega^{s} - m [\Omega-\Omega_{L}] \right)}{ \mathcal{D} \tilde{x}  +  \left(\omega^{s} - m 
[\Omega-\Omega_{L}] \right)^2} v \approx 2 m v,\label{eq:aziu}\\[0.4cm]
 \partial_{\theta} v & = m \frac{ \mathcal{D} \tilde{x}  +  \left(\omega^{s} - m [\Omega-\Omega_{L}] \right)^2}{ 2  \Omega \left( \omega^{s} - m 
[\Omega-\Omega_{L}] \right)} u \approx -\frac{1}{2} m  u .\label{eq:aziv}
\end{align}
The approximate expressions follow if we neglect $\mathcal{D} \tilde{x}$, which is fully justified since $\tilde{x} \sim 10^{-3}$ for all cases 
considered here.

As mentioned before, Equation (\ref{eq:vecsim}) can only be used as an \emph{approximation} for a nonlinear 
density wave.
Assume that the latter is correctly described by an infinite series
\begin{equation}\label{eq:nlwave}
\begin{pmatrix} \tau(r,\theta,t) \\ u(r,\theta,t) \\  v(r,\theta,t) \end{pmatrix} =  \sum\limits_{l=1}^{\infty} \begin{pmatrix} 
\mathcal{A}_{\tau,l}\left(r,t\right) 
   \\ \mathcal{A}_{u,l}\left(r,t\right)  \\ \mathcal{A}_{v,l}\left(r,t\right)    \end{pmatrix} \exp\left\{ l i\left( m \theta - \, \omega^{s} t \right) 
\right\} +c.c. \, ,
\end{equation}
where the terms with $l>1$ describe the wave's higher harmonics.
It is not straightforward to estimate the error which the approximation (\ref{eq:vecsim}) ultimately places on computed density wave profiles.
We can, however, quantify a bit more the errors of the azimuthal derivatives themselves. 
%
%
Let us for the time being assume that the surface density $\tau$ in a (steady state) density wave can be described through [\citet{Borderies83}, see also 
Section \ref{sec:pertring} and Appendix \ref{sec:gspert}]
\begin{equation}\label{eq:taupro}
\tau(r,\phi) = \frac{1}{1-q(r) \cos \left(m \phi + \Phi(r) \right)}
\end{equation}
in a cylindrical frame $(r,\phi,z=0)$ rotating with the 
satellite's mean motion frequency $\Omega^{s}= \tilde{\omega}^{s}/m$ [see Equation (\ref{eq:omfin})]. Furthermore, $q$ is the nonlinearity parameter fulfilling 
$0\leq q < 1$ and $\Phi(r)$ is the radial phase function of the density wave [cf.\ (\ref{eq:azideriv})]. 
Clearly, for $q$ not much smaller than unity the variation of the surface density $\tau$ deviates significantly from a simple harmonic.
Taking the azimuthal derivative of (\ref{eq:taupro}) yields
\begin{equation}\label{eq:aziexact}
 \partial_{\phi}\tau = -m q(r) \frac{\sin \left(m \phi + \Phi(r)\right)}{\left[1-q(r) \cos \left(m \phi + \Phi(r)\right) \right]^2}.
\end{equation}
By expanding this expression to first order in $q$, which amounts to our linear treatment of the azimuthal derivative in Equation (\ref{eq:azitau}), we obtain 
\begin{equation}\label{eq:aziap}
 \partial_{\phi} \tau = -m q(r) \sin  \left(m \phi + \Phi(r) \right) + \mathcal{O} \left(q^2\right).
\end{equation}
This would imply that for $0.1 <q <0.5$ the (azimuthally averaged) error made when replacing (\ref{eq:aziexact}) by (\ref{eq:aziap}) [and hence also 
(\ref{eq:azitau})] takes quite large values of $\sim 10-60 
\,\%$.

Despite the considerable error that may be induced by the approximation (\ref{eq:azitau}) [and (\ref{eq:aziu}), (\ref{eq:aziv})] we will see 
in Section \ref{sec:modelcomp} that the 
resulting error in the radial density wave profiles is actually small. As for the approximation (\ref{eq:azitau}) the reason is that in our 
integrations we evaluate 
this term by using an 
accurate (nonlinear) expression for the self-gravity force $\partial_{r} \phi^{d}$ (Section \ref{sec:sgnum}) so that the actual error 
resulting from (\ref{eq:azitau}) is much smaller than what would result from the linearized expression (\ref{eq:aziap}). This can be understood by 
realizing that Equation (\ref{eq:poissonazi}) holds also
for the higher harmonics [$l>1$ in (\ref{eq:nlwave})] of $\tau$ and $\phi^{d}$ (see Appendix B of LSS2016 for more details) 
which stems from the fact that Poisson's equation is linear.
Thus, the actual error which is then made with the approximation (\ref{eq:azitau}) is that the 
contributions of the higher harmonics $l>1$ in (\ref{eq:nlwave}) are \emph{underestimated} by factors of $1/l$, but not entirely neglected.
On the other hand, from LSS2016 (Section 4.5) follows that the approximations (\ref{eq:aziu}), (\ref{eq:aziv}) hold also for 
the second harmonics of the velocity fields upto a factor $1/l$ (with $l=2$) and the same is expected to apply to all higher 
harmonics\footnote{The analysis in 
LSS2016 is restricted to second order harmonics.} $l\geq 3$. Hence, the error made with (\ref{eq:aziu}) and (\ref{eq:aziv}) 
is also a suppression of the higher harmonics $l\geq 2$ by factors $1/l$.

Finally, note that our approximations for the azimuthal derivatives (\ref{eq:azitau}), (\ref{eq:aziu}) and (\ref{eq:aziv}) imply that \emph{any}
mode which forms during an integration on top of the equilibrium state will be non-axisymmetric with azimuthal periodicity $m$. 
For this reason the short-scale overstable waves which appear in our integrations (Section \ref{sec:osdwaves}) are non-axisymmetric with the same $m$ as the 
resonantly forced density wave.
As outlined in Section \ref{sec:theo} it is expected that the dynamical evolution of these modes is very similar to that 
of axisymmetric modes. This expectation will be confirmed in Section \ref{sec:osnofor}.

 \vspace{0.5cm}

\section{Self-Gravity}\label{sec:sgnum}

For most integrations we use the same implementation of collective radial self-gravity forces as described in detail in LSS2017. 
The model approximates the ring material as a collection of infinite straight wires (neglect of curvature) and predicts a self-gravity force at grid point $j$:
\begin{equation}\label{eq:sgsum}
f^{d}_{j}=-2 G h \sigma_{0} \sum_{i=1, i\neq j}^{n}  \tau(r_{i})\, \frac{r_{j} - r_{i}}{|r_{j} - r_{i}|^2}.
\end{equation}
where we defined $f^{d} = -\partial_{r}\phi^{d}$.
This relation (\ref{eq:sgsum}) does not include the force 
generated by mass contained in the bin $j$ itself, which can be approximated through
\begin{equation*}
 \Delta f^{d}(0) = 2 G \sigma_{0} \left[\partial_{r}\tau(0) \, h +  \mathcal{O}\left(h^3\right)\right].
\end{equation*}
If $\tau(r_{i})$ is periodic with period $n$ the sum (\ref{eq:sgsum}) can be replaced by the convolution 
\begin{equation}\label{eq:conv}
f^{d}_{j}= \sum_{i=-n}^{n-1}  \tau_{i}\, f^{kern}_{j-i}
\end{equation}
of $\tau_{i}=\tau(r_{i})$ with the force kernel, which reads
\begin{equation*}
f^{kern}_{j-i} = -2 G h \sigma_{0} \frac{r_{j} - r_{i}}{|r_{j} - r_{i}|^2}.
\end{equation*}
Equation (\ref{eq:conv}) can then be solved with a FFT method.
However, since the density pattern $\tau_{i}$ of a resonantly forced density wave is not periodic we need to pad one half of the array 
$\tau_{i}$ with zeros in order to avoid false contributions from grid points outside the actual grid (\citet{binney1987}), e.g.\ gravitational coupling of 
material inside the resonance with material at far positive distances from resonance across the boundaries.

For comparison with existing models for nonlinear density waves in Section \ref{sec:modelcomp} we also perform integrations which 
adopt the WKB-approximation
for the radial self-gravity force. The corresponding implementation is described in Appendix \ref{sec:wkbsg}.

 \vspace{0.5cm}

\section{Results}\label{sec:results}

\subsection{Excitation of Density Waves}\label{sec:excite}

In this section we illustrate the excitation of a resonantly forced density wave as it results from our integrations.
We use integrations employing the $Pr76$-parameters (Table \ref{tab:hydropar}) with different values of the forcing strength to 
elucidate nonlinear effects. All integrations were carried out with $L_{r}=450\,\text{km}$, $\Delta t=5\cdot 
10^{-4}\,\text{ORB}$, and $h=45\,\text{m}$. Furthermore,
\emph{Method A} for the azimuthal derivatives (Section \ref{sec:azima})
and the \emph{Straight Wire} self-gravity model (Section \ref{sec:sgnum}) were employed. 
The initial state of each integration is the Keplerian shear flow with $\tau(t=0)=1$, $u(t=0)=0$, $v(t=0)=0$ (Section 
\ref{sec:theo}) and the satellite forcing is introduced at time $t=0$.

It is expected that during the excitation process the envelope of a density wave evolves in radial direction with the local group velocity 
(\citet{toomre1969,shu1984}). 
For a linear density wave described by the perturbed surface density
 \begin{equation}\label{eq:lindw}
\sigma = \sigma_{0} +  A\left(r\right) \cdot \exp\Bigg\{i \int_{}^{r}k \left( s\right)\, \mathrm{d}s \Bigg\}  \cdot 
\exp\left\{i\left(m\theta-\omega^{s} t\right) \right\}] 
\end{equation} 
with wavenumber $k$, one obtains in the frame rotating with frequency $\Omega_{L}$ the dispersion relation (\citet{goldreich1978b,shu1984})
\begin{equation}\label{eq:lindisp}
 \kappa^2 - \left(\omega^{s}- m [\Omega-\Omega_{L}] \right)^2 + k^2 c_{0}^2 - 2 \pi G \sigma_{0} |k| =0.
 \end{equation}
Taking the derivative with respect to $k$ on both sides and re-arranging terms yields the group velocity (\citet{toomre1969})
\begin{equation}\label{eq:vg}
 v_{g} = \frac{\mathrm{d}\omega^{s}}{\mathrm{d}k} = \mathrm{sgn}(k) \frac{ \pi G \sigma_{0} - |k| c_{0}^2}{ m [\Omega-\Omega_{L}] - \omega^{s}},
\end{equation}
where $\omega^{s}$ is given by (\ref{eq:omegaf}) and $\text{sgn}(k)$ denotes the sign of $k$.
By defining
\begin{equation*}
 D= \kappa^2 -\left(\omega^{s} - m \left[\Omega - \Omega_{L}\right] \right)^2
\end{equation*}
and expanding this expression about the Lindblad resonance $r=r_{L}$ (using the approximation $\kappa=\Omega$) so that $D = \mathcal{D} \tilde{x} + 
\mathcal{O}\left( x^2 \right)$ with $\mathcal{D}$ given by (\ref{eq:det}),
one obtains from (\ref{eq:lindisp}) the wavenumber dispersion for linear density waves
\begin{equation}\label{eq:disprel}
 k=\frac{ \tilde{x}}{\epsilon  r_{L}},
\end{equation}
where $\epsilon$ is given by (\ref{eq:eps}) and where the effects of pressure, expressed through the term quadratic in $k$ in (\ref{eq:lindisp}), are ignored.

An expression for the nonlinear group velocity can be obtained from the nonlinear dispersion relation of spiral density waves [e.g.\ Equation (87) of 
\citet{shu1985a}] in the WKB-approximation
\begin{equation}\label{eq:nldisp}
 \kappa^2 - \left(\omega^{s}- m [\Omega-\Omega_{L}] \right)^2 +  I(q^2) \, k^2 \, c_{0}^2 - 2 \pi G \sigma_{0} |k| H(q^2) =0.
\end{equation}
In this expression the contributions due to pressure and self-gravity are modified and depend on the nonlinearity parameter $q$ with 
$0\leq q <1$ (see Section \ref{sec:pertring} and 
references therein). The integral
\begin{equation*}
 H(q^2) = \frac{1}{\pi} \int\limits_{-\infty}^{\infty} \mathrm{d}u \frac{\sin ^2 u}{u^2} I\left(\frac{q^2 \sin^2 u}{u^2} \right)
\end{equation*}
describes the nonlinear effects of self-gravity (\citet{shu1985a}) and, similarly,
\begin{equation*}
 I(q^2)= \frac{2}{q^2}\left[ \left(1-q^2\right)^{-1/2} -1\right],
\end{equation*}
describes nonlinear pressure effects.
The resulting group velocity reads
\begin{equation}\label{eq:vgnl}
 v_{g}^{nl} = \mathrm{sgn}(k) \frac{ \pi G \sigma_{0} \left[ H + 2 k \, q  q^{'} \tfrac{\partial H}{\partial q^2} \right] - |k| \,  c_{0}^2 \left[ I + k \, q  
q^{'} \tfrac{\partial I}{\partial q^2}  \right]}{ m [\Omega-\Omega_{L}] - \omega^{s}},
 \end{equation}
where a prime stands for the derivative $\partial / \partial k$.
The integral functions $H(q^2)$ and $I(q^2)$ fulfill $H(q^2)\geq 1$ and $I(q^2) \geq 1$ for $q\geq 0$. Furthermore, it can be verified that all other 
quantities enclosed in the 
brackets are real-valued and positive. In the linear limit $q\to 0$ the nonlinear group velocity $v_{g}^{nl}$ is identical to (\ref{eq:vg}), as 
expected. 
The wavenumber of the density wave  $k$ and the nonlinearity parameter $q$ depend on the radial distance from resonance and 
for a tightly wound density wave we have $q\propto k$ (\citet{shu1985a};~BGT86, see also Section \ref{sec:pertring}).
For typical values of the velocity dispersion $c_{0}$ in Saturn's dense rings the self-gravity term in (\ref{eq:vgnl}) will always dominate the pressure term 
so that the nonlinear 
group 
velocity is expected to be larger than the linear limit (\ref{eq:vg}).

\begin{figure}[h!]
\centering
\includegraphics[width = 0.47 \textwidth]{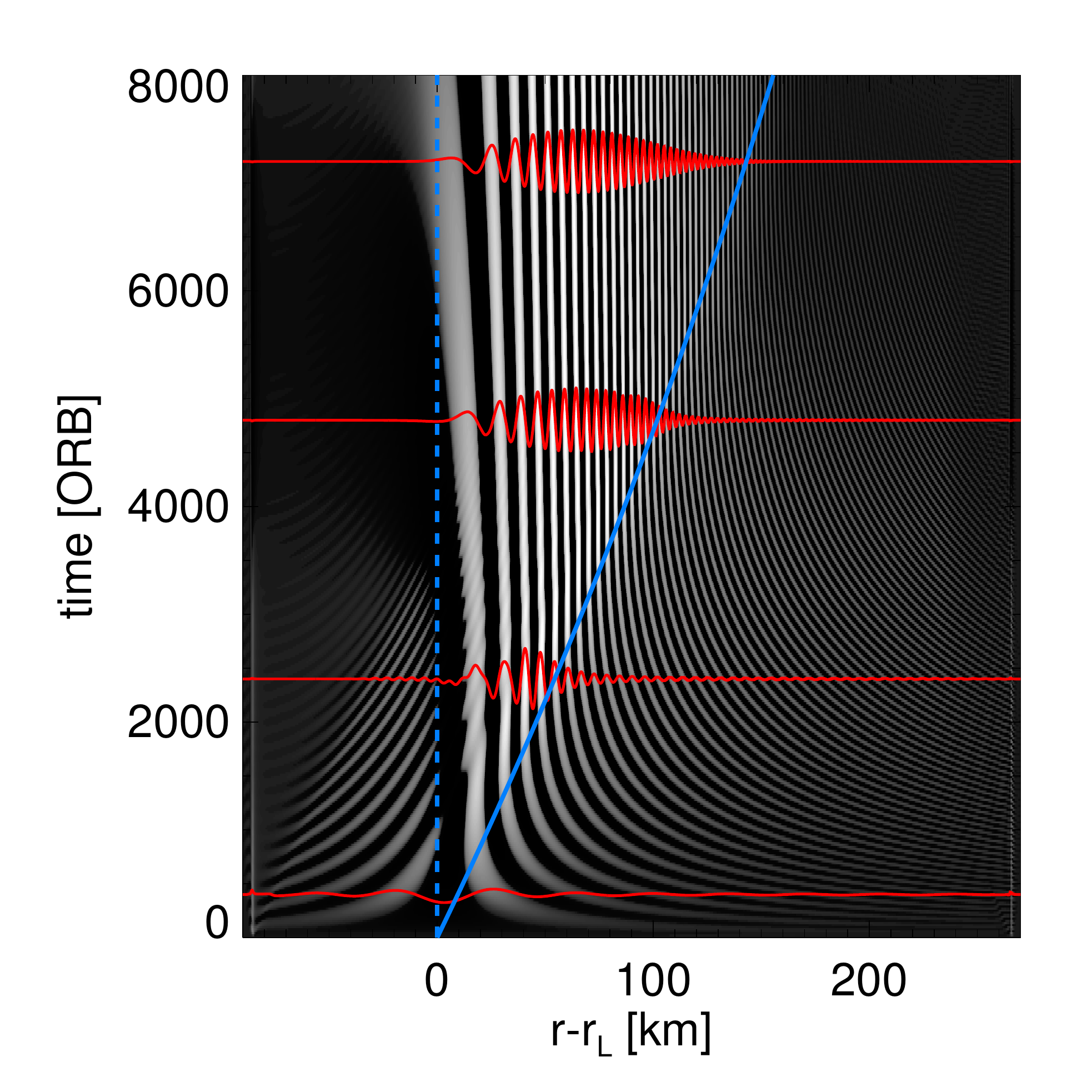}\\
\caption{Stroboscopic space-time diagram showing the evolution of the scaled surface density $\tau$ for an integration with the 
$Pr76$-parameters and an 
associated scaled torque $\tilde{T}^{s}=10^{-4}$. Brighter regions correspond to larger $\tau$-values. The blue solid line marks the path of a signal traveling 
with the linear 
group velocity (\ref{eq:vg}) starting from resonance $r=r_{L}$ at time $t=0$. Also shown are profiles of $\tau$ at different stages of the evolution. Due 
to the plot being stroboscopic, the density wave pattern eventually becomes stationary as the oscillation with frequency $\omega^{s}=-\Omega_{L}$ is 
effectively removed.}
\label{fig:sptdpr76f001}
\end{figure}

\begin{figure}[h!]
\centering
\includegraphics[width = 0.47 \textwidth]{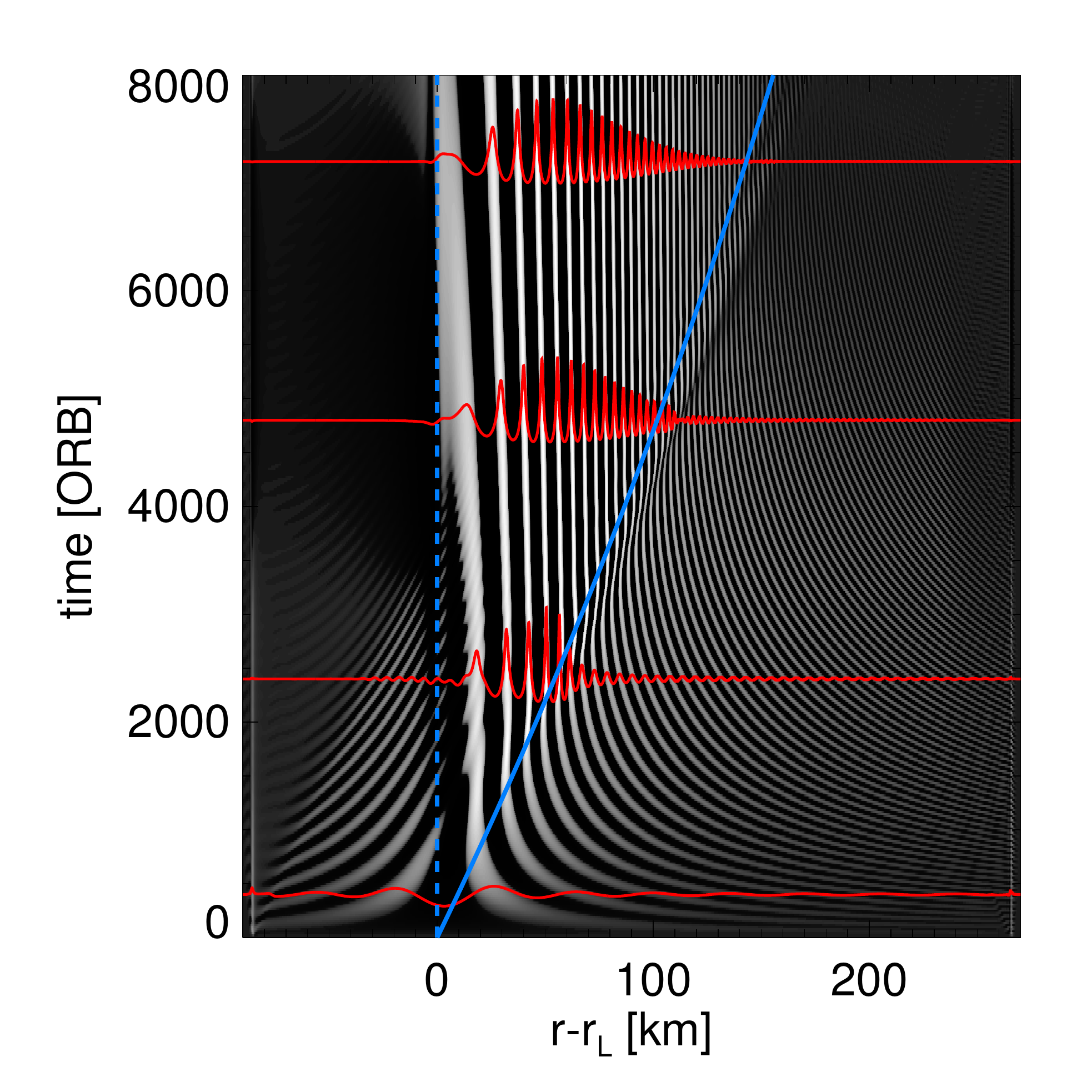}\\
\caption{Same as Figure \ref{fig:sptdpr76f001} except that $\tilde{T}^{s}=1$.}
\label{fig:sptdpr76f10}
\end{figure}

\begin{figure}[h!]
\centering
\includegraphics[width = 0.47 \textwidth]{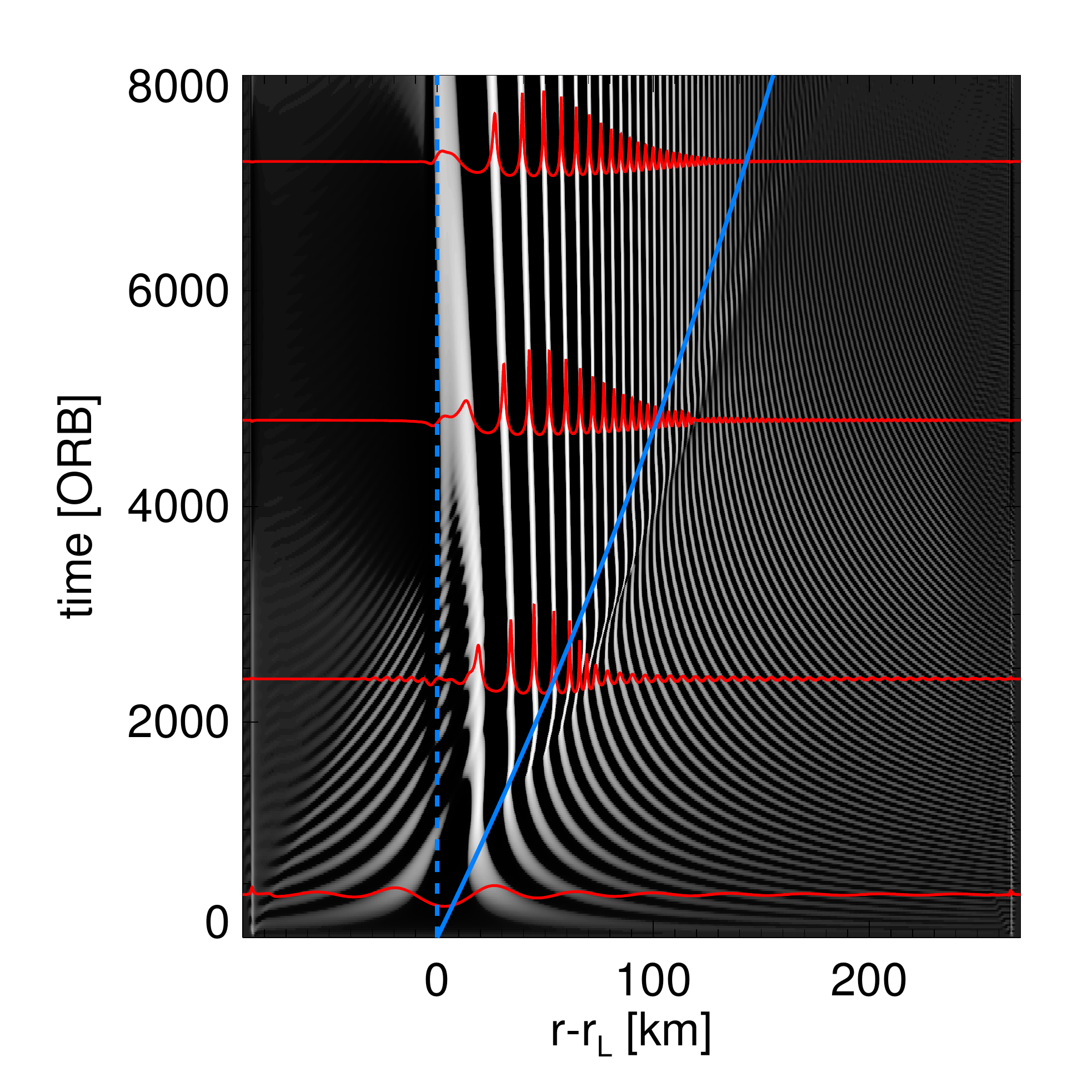}\\
\caption{Same as Figure \ref{fig:sptdpr76f001} except that $\tilde{T}^{s}=4$.}
\label{fig:sptdpr76f20}
\end{figure}

Figures \ref{fig:sptdpr76f001}, \ref{fig:sptdpr76f10} and \ref{fig:sptdpr76f20} show stroboscopic space-time diagrams [time-resolution of $\Omega_{L}/(2\pi)$] 
of integrations 
with scaled linear satellite torques of $\tilde{T}^{s}=10^{-4}$, $\tilde{T}^{s}=1$ and $\tilde{T}^{s}=4$, where $\tilde{T}^{s}=1$ corresponds to the nominal 
forcing strength for the Prometheus 7:6 ILR (Table \ref{tab:hydropar}). 
In these figures the gray shading measures the value of $\tau$ so that brighter regions correspond to larger values of $\tau$.
%
%
Since at $t=0$ the spatially constant satellite forcing is introduced and the disk is homogeneous, initially the 
hydrodynamic quantities $u$, $v$ and $\tau$ oscillate uniformly (with infinite wavelength). Due to Keplerian shear the pattern starts to wrap up 
at a constant rate.  
This transient behavior was derived by \citet{mvs1987} who studied the interaction of a satellite with an initially homogeneous disk in the 
vicinity of a Lindblad resonance and in the linear limit.
They showed that the wavelength of the pattern evolves as
\begin{equation}\label{eq:lamwrap}
 \lambda_{p}(t) = \frac{4 \pi r_{L}}{3\left(m-1 \right)\Omega_{L} t}.
\end{equation}

This result was obtained in the absence of collective forces.
\citet{mvs1987} argued that after sufficiently long time the transient behavior vanishes and the system settles on a stationary solution which is governed by 
collective effects (self-gravity, pressure and viscosity). They proved this for the case of a simple friction law assuming a force $\mathbf{f}=-Q\mathbf{u}$
 with $\mathbf{u}=(u,v)$ in the momentum equation.
In the present situation self-gravity is the dominant collective force and the disk excites a long 
trailing 
density wave propagating outward from the ILR with group velocity approximately given by (\ref{eq:vg}) (\citet{goldreich1978b,shu1984}).

As the wavelength of the pattern decreases with 
time, at a certain radial location and at a certain time the wavelength will fulfill the dispersion relation 
(\ref{eq:nldisp}) [and also (\ref{eq:lindisp}) if $q$ is sufficiently small]. As soon as this is the case, the wavelength is ``locked'' to this value.
In the figures \ref{fig:sptdpr76f001}-\ref{fig:sptdpr76f20}, the region which becomes ``locked`` is enclosed by the 
dashed and solid blue lines.
The former marks the resonance, while the latter is the predicted path of the wave front assuming it propagates with the linear group velocity 
(\ref{eq:vg}).
All wave structures outside this region eventually damp as they become increasingly wound up. An exception are the short-scale waves generated by viscous 
overstability (Sections \ref{sec:theo} and \ref{sec:osdwaves}). 
Also plotted are radial profiles of $\tau$ at four different times during the excitation process.

In Figure \ref{fig:sptdpr76f001} the blue solid line describes well the propagation of the wave front, until a steady state is 
reached (around 8,000 orbital periods) and the wave's amplitude profile remains stationary.
We find a number of differences when comparing the figures. First of all, with increasing torque value the wave profiles attain the typical peaky 
appearance of nonlinear density wavetrains in thin disks (\citet{shu1985a};~BGT86;~\citet{salo2001}).
Furthermore, the group velocity of the waves increasingly departs from the linear prediction (\ref{eq:vg}), albeit mildly. 
One notes that there remains a very slow phase-drift of the wave pattern towards the resonance, indicating an increasing phase velocity 
with decreasing distance from resonance. Theoretically, at resonance the wavenumber of the density wave (\ref{eq:disprel}) vanishes so that the wave's phase 
velocity $ \omega^{s}/k$ diverges. It can therefore in general not be expected from a numerical method to correctly describe the wave pattern at the exact 
resonance location. 

\begin{figure}[h!]
\centering
\includegraphics[width = 0.37 \textwidth]{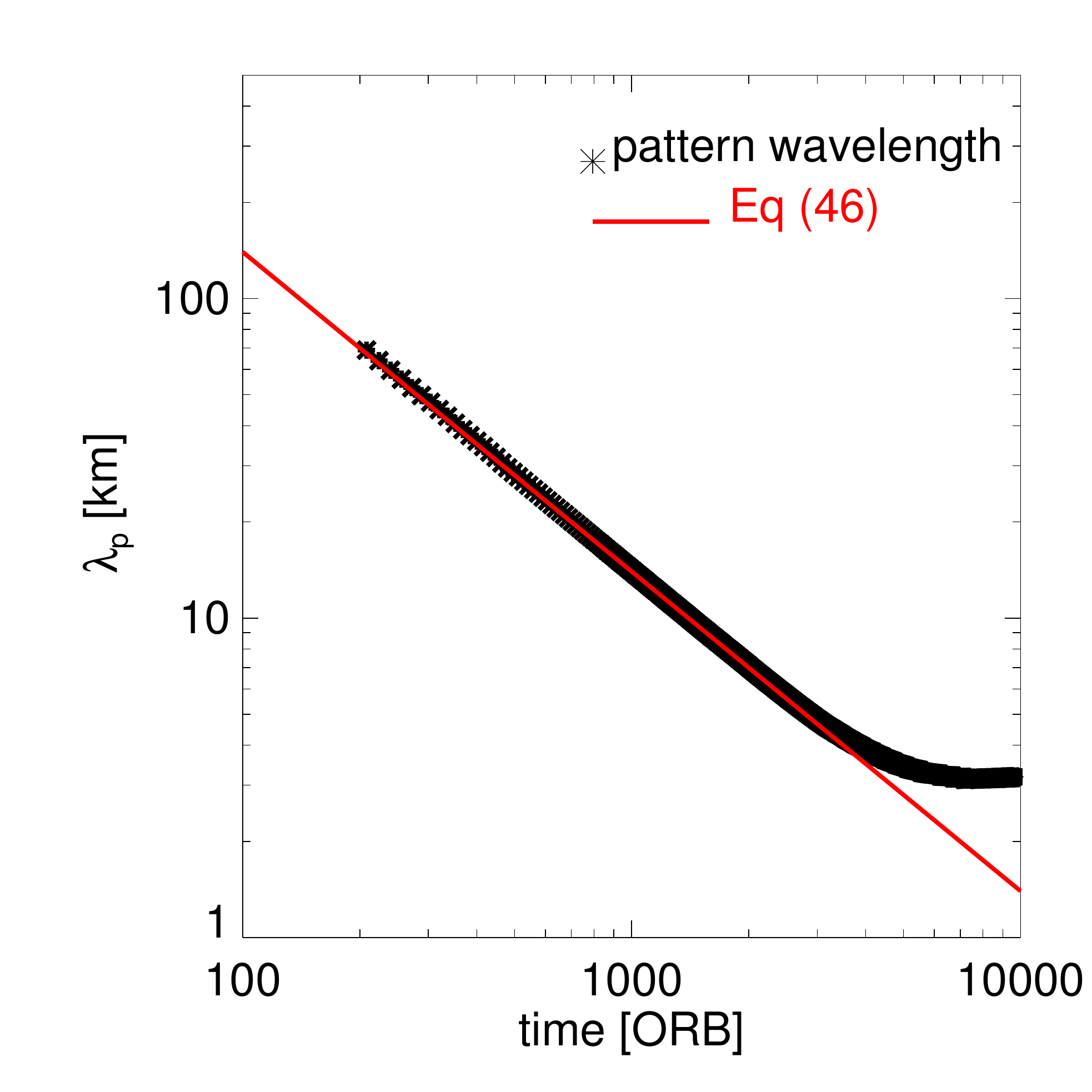}\\
\caption{Average wavelength of the forming pattern in the course of the 
integration shown in Figure \ref{fig:sptdpr76f001}. The wavelength is obtained by counting the number of complete wave cycles of the (sinusoidal) surface 
density $\tau$ within the radial region $50\,\text{km} \leq r-r_{L} \leq 150\,\text{km}$.}
\label{fig:lamwrap}
\end{figure}

Figure \ref{fig:lamwrap} shows for the integration with $\tilde{T}^{s}=10^{-4}$ (Figure \ref{fig:sptdpr76f001}) the average wavelength of 
the forming pattern, sampled within the radial region $50\,\text{km} \leq r-r_{L} \leq 150\, \text{km}$. The agreement with Equation (\ref{eq:lamwrap}) is 
excellent for about 3,000 
orbital periods. After that deviations become notable as a steady state is reached where self-gravity prevents further shortening of the wavelength. 
The closer to the resonance $r=r_{L}$, the earlier a steady state is attained as the resonant density wave pattern emerges at the resonance and propagates 
outward with its local group velocity.

In Saturn's rings an initial transient pattern as seen in our integrations might be observable for density waves driven by the co-orbital satellites 
Janus and Epimetheus.
These satellites interchange orbits every 4 years so that their resonance locations in the rings shift periodically by tens of kilometers. 
Every time a resonance location is changed the wave excited at the preceding location continues to propagate while a new density wave is launched at the
new location (\citet{tiscareno2006b}).

%

\subsection{Comparison with the Nonlinear Models of BGT1986 and LSS2016}\label{sec:modelcomp}

In this section we compare results of our hydrodynamical integrations with the nonlinear models of BGT86 and 
LSS2016, which we refer to as the BGT and the 
WNL (Weakly Nonlinear) model, respectively.
This section is restricted to stable waves in the sense that $\beta < \beta_{c}(\lambda)$ for all wavelengths $\lambda$ [cf.\ Equation (\ref{eq:betc})], i.e.\ 
no overstability occurs in the system.
All hydrodynamical integrations presented in this section were conducted with $L_{r}=450\,\text{km}$, time steps $\Delta t=5\cdot 10^{-4}\,\text{ORB}$, and 
spatial resolution $h=45\,\text{m}$ and used the $Pr76$-parameters (Table \ref{tab:hydropar}). If not stated otherwise, all integrations employed \emph{Method 
A} for the azimuthal derivatives (Section \ref{sec:azima})
and the \emph{Straight Wire} self-gravity model (Section \ref{sec:sgnum}).
Presented BGT model wave profiles were computed using the method of BGT86 (see their Section IVa), using the pressure tensor 
(\ref{eq:pten}) with (\ref{eq:pres}) and (\ref{eq:shearvis}). This model takes into account secular changes in the background surface mass density 
$\sigma_{0}$ that accompany the steady state density wave in order to ensure conservation of angular momentum at all radii in the steady state. These 
modifications are neglected in 
our hydrodynamical integrations as well as the WNL model since the latter neglect the angular momentum luminosity carried by the density wave. To facilitate a 
comparison between the three different methods the profiles of $\tau$ resulting from the BGT model are scaled with the modified background surface mass density 
$\sigma_{0}(r)$ which will not be shown.

Figure \ref{fig:pr76f0103} (Appendix \ref{sec:appendixb}) displays steady state profiles of the hydrodynamic quantities $\tau$, $u$, $v$ as these
result from integrations together with profiles obtained using the WNL model (LSS2016).
The profiles in the left and right columns result from integrations which applied \emph{Method A} and \emph{Method B} for the azimuthal derivatives, 
respectively. The self-gravity is computed with the \emph{Straight Wire} model. As in the previous section, the satellite forcing strengths are indicated by 
the fractional torque $\tilde{T}^{s}$ such that $\tilde{T}^{s}=1$ corresponds to the 
nominal forcing strength at the Prometheus 7:6 ILR and results in a nonlinear density wavetrain. The value $\tilde{T}^{s}=9\cdot 10^{-2}$ corresponds to a 
weakly 
nonlinear wave.
For the latter case both methods A and B produce very similar results in good agreement with the WNL model.
Inspection of the nonlinear case $\tilde{T}^{s}=1$ reveals significant departures at larger distances from resonance between both methods,
and \emph{Method A} produces a clearly better match with the WNL model.
All integrations presented in the following sections were conducted with \emph{Method A}.

In Figure \ref{fig:pr76waveletf20} (Appendix \ref{sec:appendixb}) we present wave profiles along with their Morlet wavelet powers (\citet{torrence1998}) for 
the 
case $\tilde{T}^{s}=4$. Also 
for this strongly nonlinear wave, 
we observe an overall good agreement for both the amplitude profiles and wavenumber dispersions. Note that the WNL model takes 
into account only the first two harmonics of the wave [cf.\ Equation (\ref{eq:nlwave})], which is clearly visible in the wavelet power. This is also the reason
why $\tau$ can take values below 0.5 (see LSS2016 for details).

Finally, Figure \ref{fig:sgcompf10} (Appendix \ref{sec:appendixb}) compares profiles obtained from integrations with the \emph{Straight Wire} self-gravity 
(left 
panels)
and the WKB self-gravity (Appendix \ref{sec:wkbsg}) using Equation (\ref{eq:wkbforce}) (right panels) for the cases $\tilde{T}^{s}=9\cdot 10^{-2}$ (upper 
panels) and $\tilde{T}^{s}=4$ 
(lower panels). Comparison with 
corresponding BGT model wave profiles shows that the WKB-approximation is fully adequate for the weakly nonlinear wave with $\tilde{T}^{s}=9\cdot 10^{-2}$ in 
that
it yields indistinguishable results from the \emph{Straight Wire} self-gravity. For the strongly nonlinear case $\tilde{T}^{s}=4$ the WKB-approximation has a 
notable 
effect. As expected, its application yields overall an even better agreement with the BGT (and WNL) model.
We have verified that 
reducing the time step or the 
grid spacings by factors of $1/2$ does not change the outcome of all integrations presented in this section.
The remaining differences between the integrated wave profiles and the model profiles are most likely due to the approximative implementation of the 
azimuthal derivatives. Nevertheless, the results presented here make us confident that our numerical integrations yield qualitatively correct behavior even of 
strongly nonlinear density waves.
\subsection{Wave Propagation through Density Structures}\label{sec:dstruct}

In this section we present a few illustrative examples of hydrodynamical integrations of density waves propagating through an inhomogeneous ring.
We restrict to the cases of jumps in the equilibrium surface density, but in principle we could also vary
%
%
%
%
%
%
other parameters with radial distance, such as the 
viscosity parameter $\beta$. All integrations adopted the $Pr76$-parameters and employed \emph{Method A} for the azimuthal 
derivatives (Section \ref{sec:azima}) as well as the \emph{Straight Wire} self-gravity model (Section \ref{sec:sgnum}).
Figure \ref{fig:sptdbargap} (Appendix \ref{sec:appendixc}) shows space-time plots of a density wave passing a region of increased surface density 
($\tau_{0}=3$, left panel) 
as well as a region of decreased surface density ($\tau_{0}=0.5$, right panel), in both cases of radial width 
$40\,\text{km}$.
The jumps in the equilibrium surface density, whose locations are revealed in the space-time plots, act like additional sources for 
the 
density wave in the sense 
that the wave profile can change at these locations prior to the expected arrival time of the 
wave front at these locations, the latter being indicated by the solid blue line (cf.\ Figures \ref{fig:sptdpr76f001}-\ref{fig:sptdpr76f20}). 
It is, however, not clear how this is affected by the assumption imposed by 
our azimuthal derivatives that the hydrodynamic 
quantities describe an $m$-armed pattern right from the start of the integration.

Figure \ref{fig:pr76cleanbargap} (Appendix \ref{sec:appendixc}) shows steady state profiles of $\tau$ along with corresponding wavelet-power spectra of density 
waves passing through regions 
of varying equilibrium surface density. 
For reference, the first row shows a density wave in a homogeneous ring.
The second case, with a region of increased $\tau_{0}$, bears some similarities with Figures 4 and 5 in \citet{hedman2016}, showing profiles of the Mimas 5:2 
density wave in Saturn's B ring which passes through a region of radial width $\sim 60\,\text{km}$ where the normal optical depth increases sharply from about 
1.5 to values $3-5$.
In the region of enhanced surface density in Figure \ref{fig:pr76cleanbargap} the wave damping is 
reduced due to its decreased wavenumber. Therefore, after passing the barrier the wave amplitude is enhanced as compared with the wave in the homogeneous 
region. 
The last case represents a situation with a narrow region of mildly decreased 
surface density $\tau_{0}=0.5$. In this region the wavenumber is enhanced, resulting in stronger 
wave damping.

\begin{figure*}
\centering
\includegraphics[width = 0.45 \textwidth]{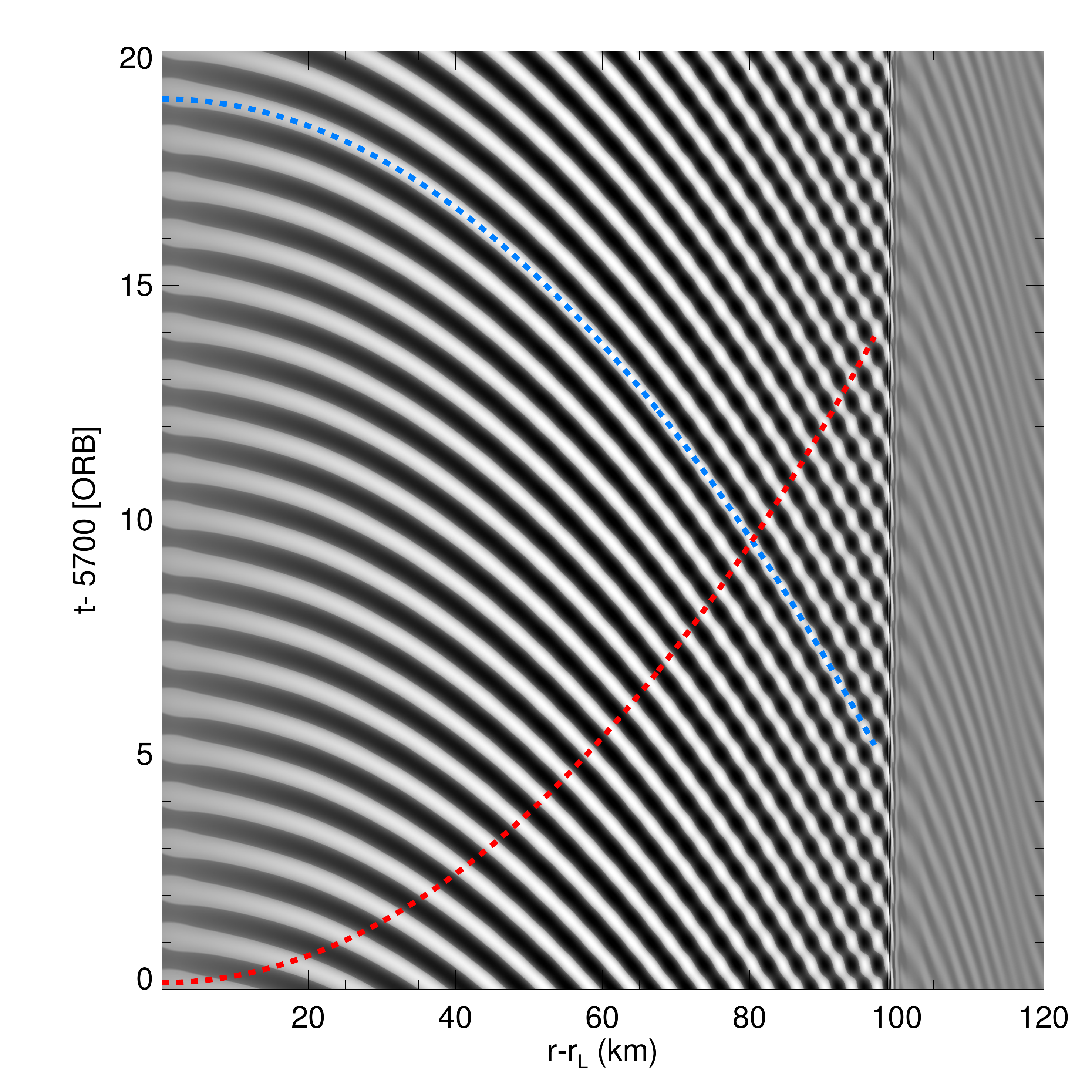}
\includegraphics[width = 0.45 \textwidth]{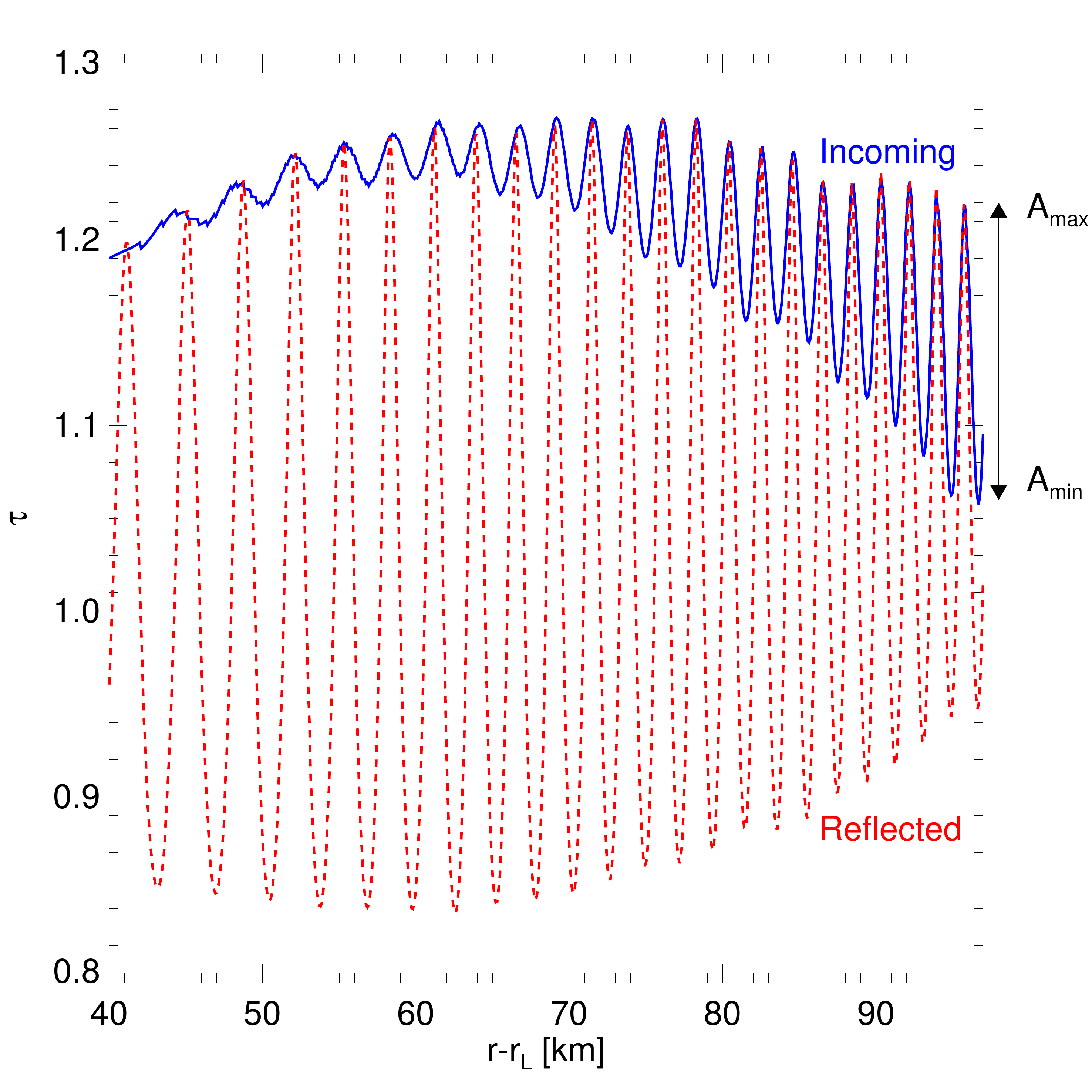}
\caption{Plots illustrating the reflection of a (long) trailing density wave at a sharp boundary at $r-r_{L}\sim 100 \,\text{km}$ where $\tau_{0}$ is reduced 
by 
a factor of $1/5$. In the space time plot (left panel) the blue (red) dashed curve traces a line of equal phase of the incoming (reflected) wave so that it 
follows a density maximum. The plot in the right panel shows $\tau$ evaluated along these curves. As explained in the text, one can estimate the amplitude 
ratio 
of the incoming and the reflected waves from the indicated values $A_{max}$ and $A_{min}$ of $\tau$ [Equation (\ref{eq:ratio})]. Since the considered wave is 
weakly nonlinear it follows $H(q)\gtrsim 1$ in the nonlinear dispersion relation (\ref{eq:nldisp}) so that the linear limit (\ref{eq:disprel}) is not fully 
accurate. To compensate for this we used a slightly increased value of $\sigma_{0}$ (by $0.25 \%$) to compute the wavenumber $k$ from 
(\ref{eq:disprel}), which is used in 
(\ref{eq:phasecurve}), to obtain a better fit to the locations of equal phase in the left panel.}
\label{fig:reflect}
\end{figure*}

If a density wave encounters a sharp discontinuity in the background surface density, such as a sharp ring edge, it is theoretically expected 
that it (partially) reflects at the boundary (see \citet{longaretti2018} and references therein). For the examples presented in Figure 
\ref{fig:pr76cleanbargap} the jumps in
the background surface density are not sufficiently sharp to cause a notable reflection.
However, Figure \ref{fig:reflect} shows a space-time plot (left panel) of a wave with $\tilde{T}^{s}=9 \cdot 10^{-2}$ as it encounters 
a sharp edge near $r-r_{L}=100\,\text{km}$ where $\tau$ changes from $1$ to $0.2$. The plot clearly shows that the long trailing wave is partially reflected as 
a long leading wave, which rapidly damps as it propagates back towards the resonance $r=r_{L}$. The remaining part of the incoming trailing wave is transmitted 
into the rarefied region and quickly attenuates as it propagates with strongly reduced wavelength.
Note that the long trailing wave has a negative phase velocity $-\Omega_{L}/k$ in our coordinate frame while its group velocity [Equation 
\ref{eq:vg}] is positive since $k>0$. For the reflected leading wave which has $k<0$ the situation is exactly the opposite.
%
%
%
Close to the edge at distances $r-r_{L} \lesssim 100\,\text{km}$,  where the reflected wave has a substantial amplitude, the resulting density pattern behaves 
as a left-traveling (negative phase velocity) wave which additionally undergoes a standing-wave motion. The standing-wave motion rapidly 
diminishes with increasing distance from the edge due to the rapid damping of the reflected wave. 
The blue and red dashed curves are curves of constant phase 
of the incoming and reflected wave, respectively, parametrized as [cf.\ Equation (\ref{eq:lindw})]
\begin{equation}\label{eq:phasecurve}
 t\left(x\right) = t_{0} \pm \frac{1}{\omega^{s}} \int\limits_{0}^{x} k \left(s\right) \mathrm{d} s ,
\end{equation}
where $k\left(s\right)$ is the wavenumber of a long density wave [Equation (\ref{eq:disprel})].
The initial value $t_{0}$ is chosen such that the curves follow the path of a density 
maximum of the corresponding wave. 
The plot in the right panel shows the surface density $\tau$ evaluated along these lines of equal phase, represented by the solid and 
dashed curves.
From the solid curve we can estimate the amplitude ratio of the incoming and the reflected waves by measuring the variation of $\tau$ near the edge
as indicated by the arrows. That is, we have
\begin{align*}
A_{max}& = A_{I}+A_{R},\\
A_{min}& = A_{I}-A_{R},
\end{align*}
where the subscripts $I$ and $R$ stand for the incoming and the reflected wave, respectively.
Hence,
\begin{equation}\label{eq:ratio}
 \frac{A_{R}}{A_{I}} = \frac{A_{max}+A_{min}}{A_{max}-A_{min}} \approx 0.6,
\end{equation}
which means that the major fraction of the incoming wave is reflected. We note that it is not clear how the reflection is affected by our approximation of the 
azimuthal derivatives (\ref{eq:azitau}).
Note also that the (viscous) time-scale on which the initially imposed density jumps change in a notable manner, is much longer then the applied 
integration times.
%
%

%
%
%

 \vspace{1cm}

\subsection{Density Waves and Viscous Overstability}\label{sec:osdwaves}

\begin{figure}[h!]
\centering
\vspace{-0.2cm}
\includegraphics[width = 0.38 \textwidth]{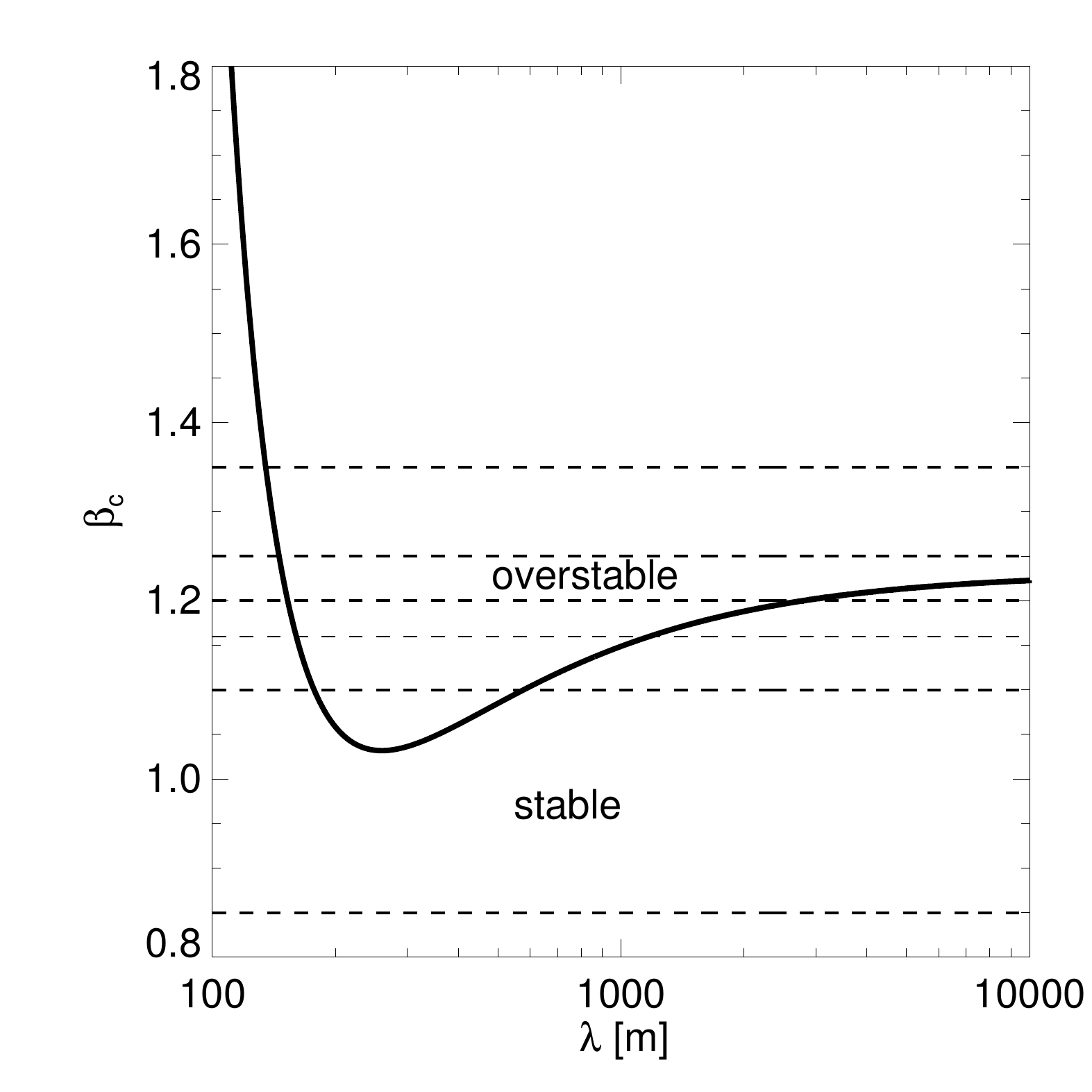}
\caption{Linear stability curve [Equation (\ref{eq:betc})] corresponding to the $Pr76$-parameters (solid curve). The dashed lines indicate the different values 
of the viscosity parameter $\beta$ [Equation (\ref{eq:shearvis})] that are used for the large-scale integrations of resonantly forced density waves discussed
in Section \ref{sec:osdwaves}. Viscous overstability is expected to develop for all values of $\beta$ larger than the minimal value 
$\text{min}[\beta_{c}]_{\lambda} = 
1.03$ which occurs for $\lambda\sim 260\,\text{m}$. For $\beta \gtrsim 1.03$ only a narrow band of wavelengths is unstable. For 
$\beta \gtrsim 1.22 $ all wavelengths larger than a critical one are unstable, implying instability of the resonantly forced density wave.}
\label{fig:lambet}
\end{figure}

We now turn to our hydrodynamical integrations of forced spiral density waves in a model ring which is subjected to viscous overstability such that 
$\beta > \beta_{c}(\lambda)$ [Equation (\ref{eq:betc})] for a non-zero range of wavelengths $\lambda$.
Figure \ref{fig:lambet} displays the linear stability curve  for the $Pr76$-parameters (Table \ref{tab:hydropar}) along with the different values adopted for 
the viscosity parameter $\beta$ [Equation (\ref{eq:shearvis})] in the integrations 
discussed 
in the following. These values are $\beta=0.85,1.10,1.16,1.20,1.25\,\text{and} \, 1.35$. Viscous overstability is expected to develop for all but the 
smallest of these values, resulting in wavetrains which are believed to produce parts of the observed periodic micro-structure in Saturn's A and B 
rings (\citet{thomson2007,collwell2007,latter2009}).
\noindent
For the values $\beta=1.10,1.16\, \text{and} \,1.20$ linear viscous 
overstability is restricted to a relatively narrow band of wavelengths and the forced spiral density wave is stable. 
%
%
%
%
%
%
%
%
%
%
In contrast, for the two 
largest values $\beta=1.25 \,\text{and} \, 1.35$  all wavelengths larger than a critical one are 
unstable. For these two cases the forced spiral density wave itself is unstable and it is expected from existing models that it retains a 
finite amplitude [i.e.\ a finite nonlinearity parameter $q$] at large distance from resonance 
(see BGT86 and LSS2016 for details).

However, these models do not take into account the presence of the 
waves which are spontaneously generated by viscous overstability and which do not depend on the resonant forcing by an external gravitational potential. In 
this section we study the 
interplay of both types of structure in a qualitative manner. All large-scale integrations presented in this section were conducted using a grid with 
$L_{r}=450\,\text{km}$, $h=25\,\text{m}$ and applied \emph{Method A} for the azimuthal derivatives (Section \ref{sec:azima}) as well as the \emph{Straight 
Wire} self-gravity model 
(Section \ref{sec:sgnum}).
\subsubsection{Hydrodynamical Integrations without Forcing}\label{sec:osnofor}

\begin{figure*}
\centering
\includegraphics[width = 0.47 \textwidth]{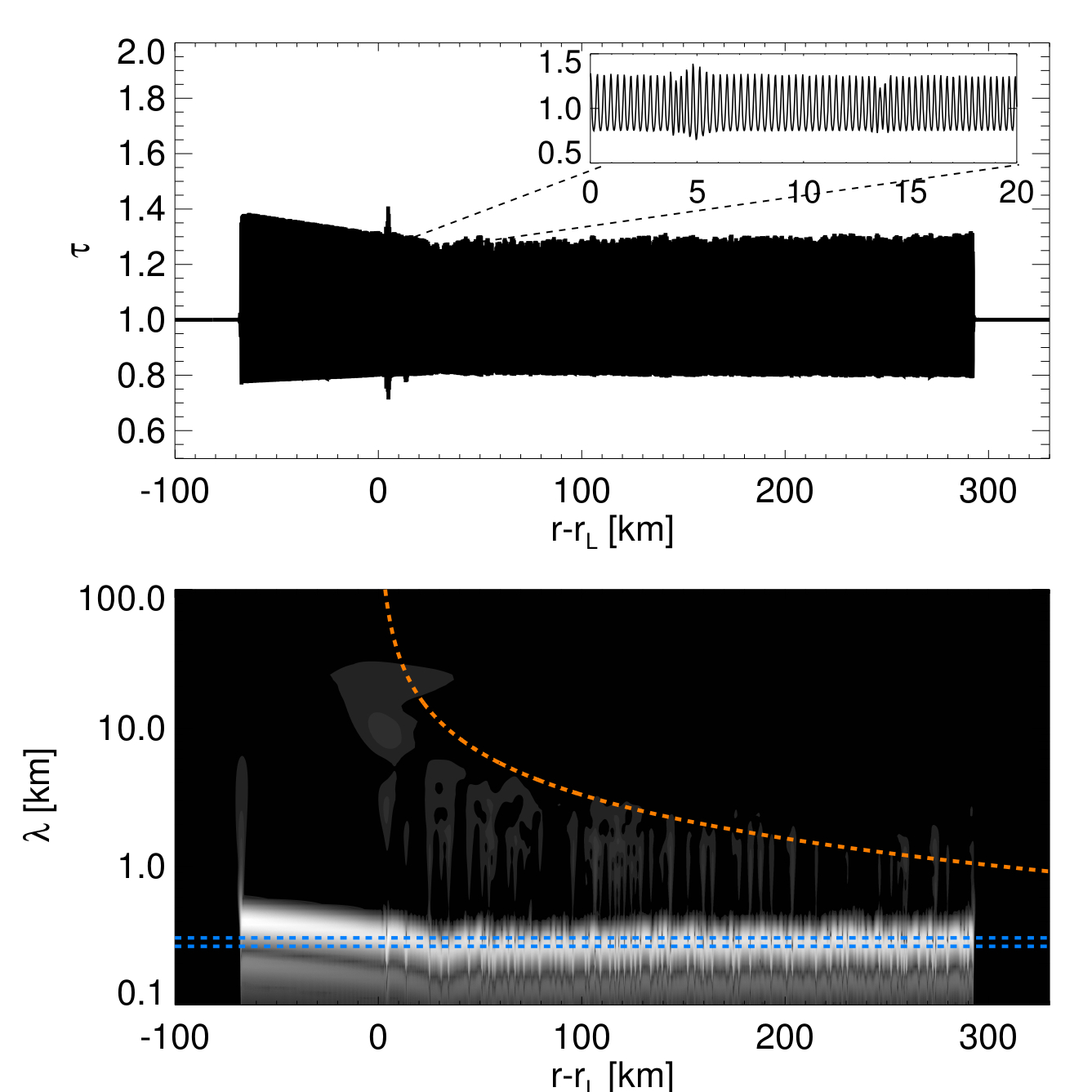}
\includegraphics[width = 0.47 \textwidth]{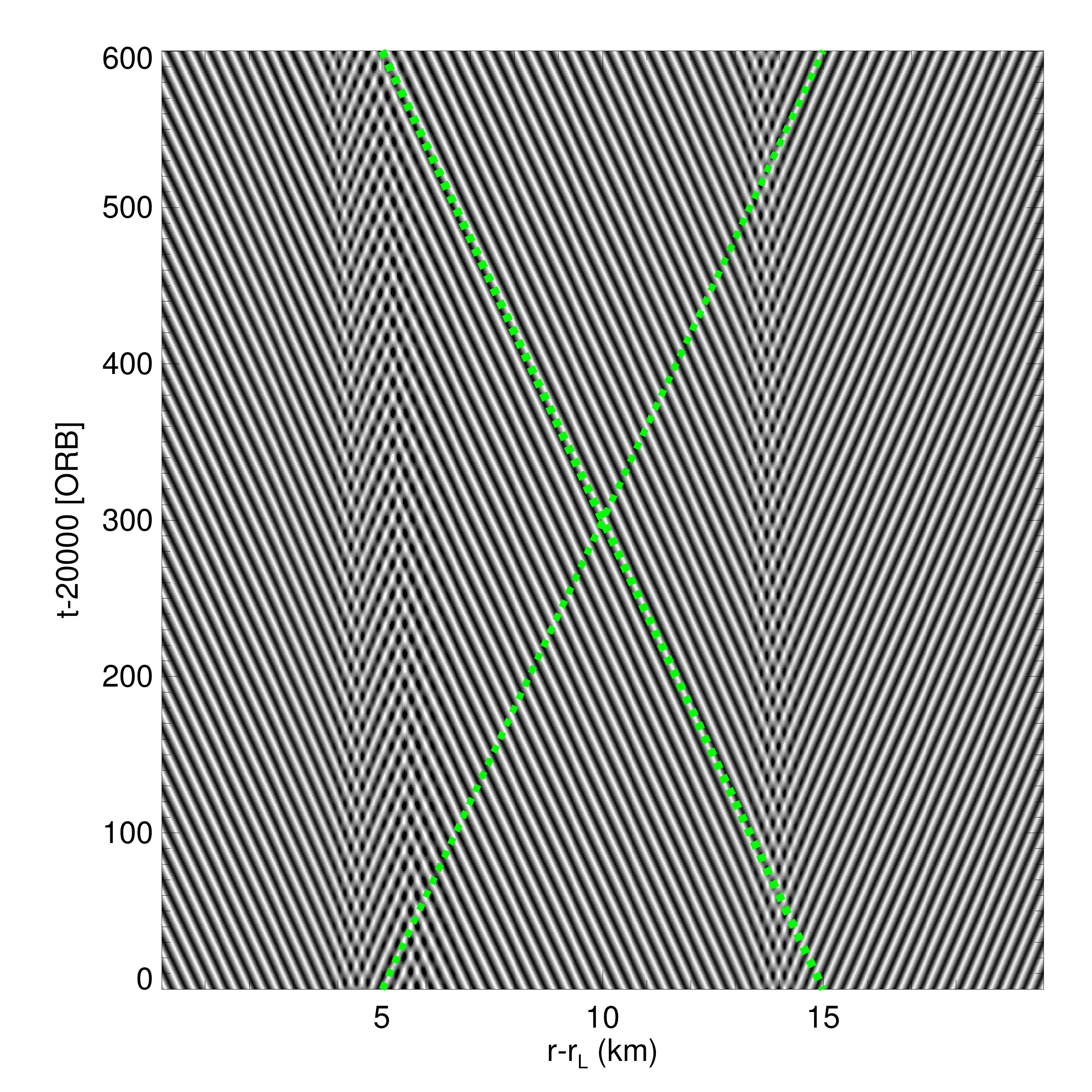}
\caption{\textit{Left}: Radial surface density ($\tau$) profile and its wavelet-power at $t\sim 20,000\,\text{ORB}$ of an 
hydrodynamical integration using the $Pr76$-parameters with $\beta=1.25$ and no satellite forcing ($\tilde{T}^{s}=0$). The short-wavelength structures are due 
to viscous overstability. The red dashed curve represents the linear density wave dispersion relation (\ref{eq:disprel}), which some persistent small amplitude 
undulations, resulting from the azimuthal derivative terms [Section \ref{sec:azideriv}], seem to follow. The blue dashed lines indicate the expected 
nonlinear 
saturation wavelength of viscous 
overstability by margins $\pm 20\,\text{m}$ (see the text). The initial state of the integration is a small amplitude linear combination of left and 
right traveling linear overstable waves on all wavelengths down to about $200\,\text{m}$. \textit{Right}: Stroboscopic space-time diagram showing the 
evolution 
near $t= 
20,000\,\text{ORB}$ of a small radial section at the nominal resonance location. Two source structures are located at $x\sim 4\,\text{km}$ and $x\sim 
14\,\text{km}$, respectively, sending out traveling waves both radially inward and outward. In between the sources (at $x\sim 5\,\text{km}$) 
counter-propagating 
wave patches collide in a sink. The green dashed lines indicate the expected phase velocity $\omega_{I}/k$ of nonlinear overstable waves, 
obtained from 
Figure \ref{fig:pr76osnoforp2} (left panel), for a wavelength of $\lambda=300\,\text{m}$. Since the space-time diagram is stroboscopic, the apparent 
phase velocity of the waves in this plot is reduced (in absolute value) by $\Omega_{L}/k$, compared to the value obtained from the curve in Figure 
\ref{fig:pr76osnoforp2}.}\label{fig:pr76osnofor}
\end{figure*}

\begin{figure*}
\vspace{-0.5cm}
\centering
\includegraphics[width = 0.46 \textwidth]{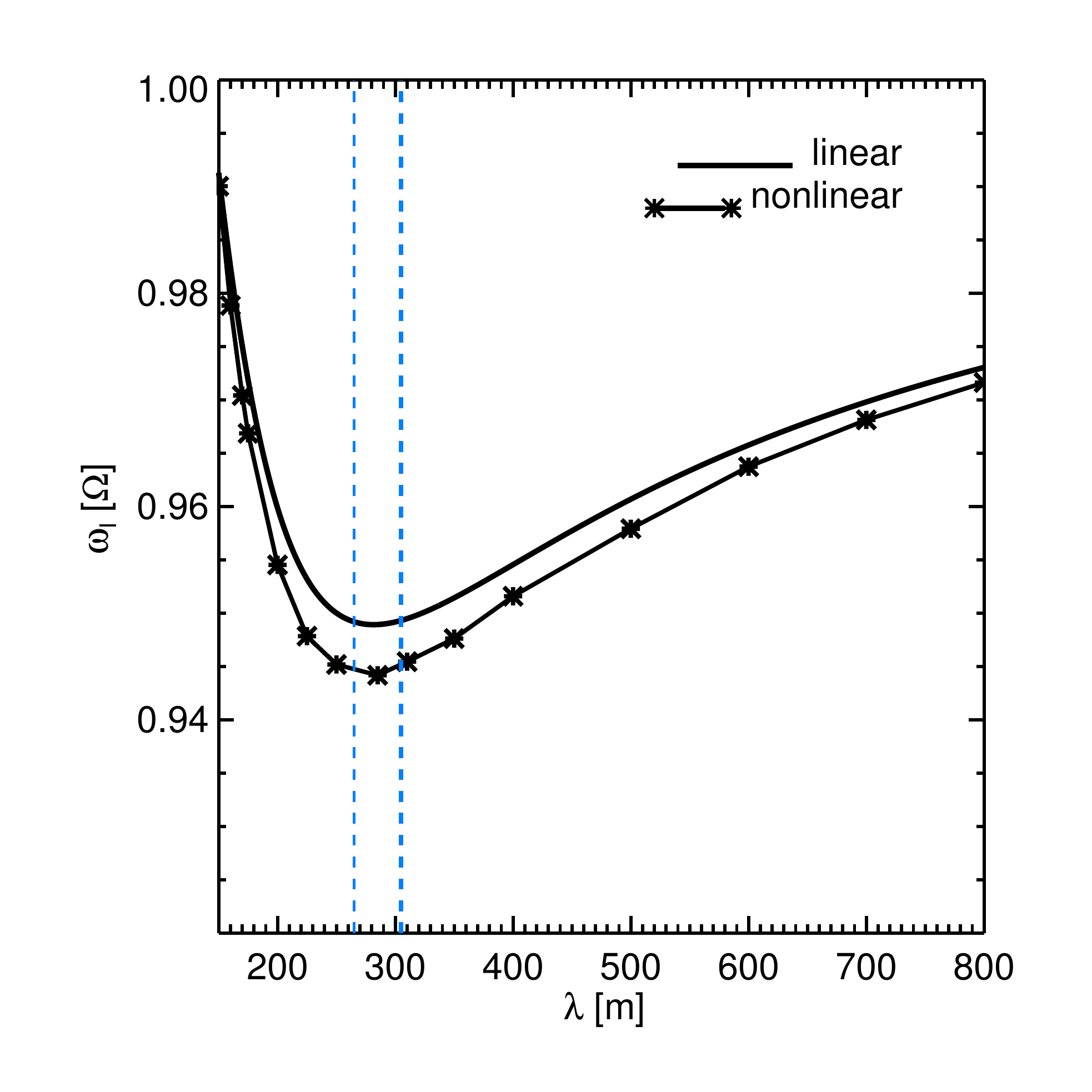}
\includegraphics[width = 0.46 \textwidth]{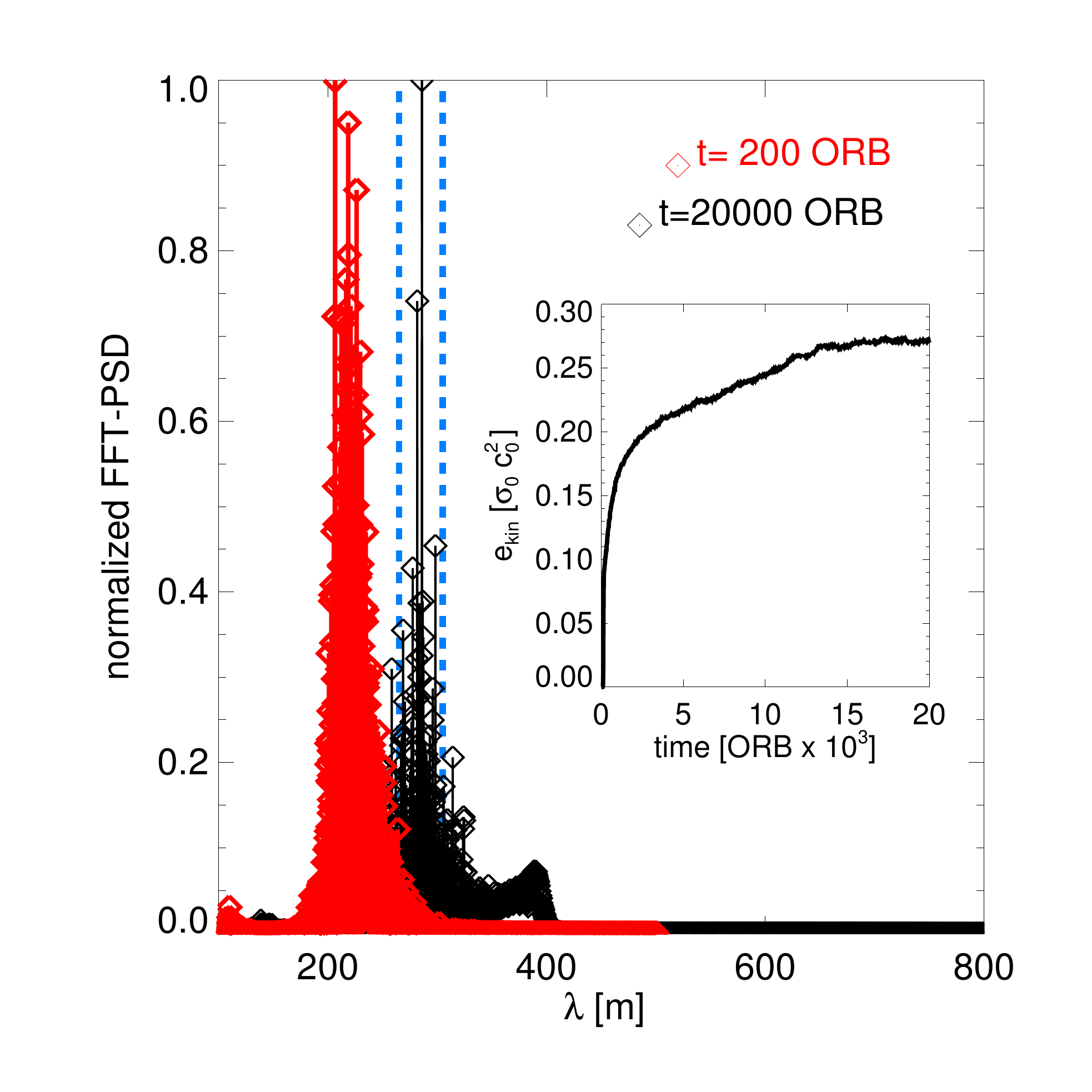}
\caption{\textit{Left}: Linear and nonlinear frequencies of overstable waves adopting the $Pr76$-parameters with $\beta=1.25$. The linear 
curve was obtained from numerical solution of (\ref{eq:evp}). The nonlinear frequencies were obtained from small-domain hydrodynamical integrations of 
saturated overstable waves as described in detail in LSS2017 (see their Section 6.1). The minima of both curves appear at equal 
wavelengths, which is a consequence of the (linear) ideal gas relation (\ref{eq:pres}) for the hydrostatic pressure. For more realistic equations of 
state the nonlinear frequency minimum occurs at larger wavelengths than the linear one (LSS2017). \textit{Right}: Power 
spectral density of $\tau$ from the same integration as in Figure \ref{fig:pr76osnofor} during the linear growth phase ($t=200\,\text{ORB}$) and in the 
saturated state ($t \sim 20,000\,\text{ORB} $) of viscous overstability. The insert displays the evolution of the kinetic energy density (\ref{eq:ekin}).
The blue dashed lines in both panels indicate the expected nonlinear saturation wavelength of viscous overstability with margins $\pm 20 \, \text{m}$ (see 
the text).
}\label{fig:pr76osnoforp2}
\end{figure*}

For reference, Figures \ref{fig:pr76osnofor} and \ref{fig:pr76osnoforp2} describe an integration using $\beta=1.25$, without forcing by the satellite 
($\tilde{T}^{s}=0$). The seed for this integration consists of a small amplitude superposition of linear left and right traveling overstable modes on 
all wavelengths down to about $200\,\text{m}$.
Note that without any seed and in the absence of satellite forcing, no perturbations develop. 
Figure \ref{fig:pr76osnofor} (left panel) shows a profile of $\tau$ (top) after 
about 20,000 orbital periods, along with its wavelet power 
(bottom). The structure on wavelengths $\lambda\sim 200-400\,\text{m}$ represents the nonlinear saturated state of viscous overstability.
This state consists of left- and right traveling wave patches, separated by source and sink structures 
(\citet{latter2009,latter2010};~LSS2017).
This can also be seen in the stroboscopic space-time diagram (right panel), showing the 
evolution of $\tau$ over 600 orbits in the saturated state within a small portion of the 
computational domain near the nominal resonance location. The green dashed lines represent the expected nonlinear phase velocity $v_{ph}=\omega_{I} /k$ of  
overstable modes (Figure \ref{fig:pr76osnoforp2}, left panel) of wavelength $\lambda=300\,\text{m}$, with $\omega_{I}$ and 
$k$ denoting 
the 
nonlinear oscillation frequency and wavenumber of the wave. 
Although the modes seen in Figure \ref{fig:pr76osnofor} are in fact
non-axisymmetric with azimuthal periodicity $m=7$ (see Section \ref{sec:azideriv}), their phase velocity [cf.\ Equation (\ref{eq:omcrit})] is practically the 
same as for axisymmetric modes ($m=0$) since we are in a frame rotating with the orbital frequency at resonance $\Omega_{L}$.
Note that in Section \ref{sec:theo} the symbol $\omega_{I}$ was used to describe the 
\emph{linear} oscillation frequency of overstable waves. 
The sharp decay of the density pattern near the domain boundaries is due to the inclusion of buffer-regions where 
$\beta<\text{min}[\beta_{c}]_{\lambda}$, 
so that the condition for linear viscous overstability is not fulfilled for any wavelength within these regions. We included such buffer-regions in all 
large-scale integrations 
presented in the following. 
Furthermore, Figure \ref{fig:pr76osnoforp2} (right panel) displays for the same 
integration as in Figure \ref{fig:pr76osnofor} the power spectral 
density of $\tau$ at two different times, as well as the evolution of the kinetic energy density (the insert). At the early time ($200\,\text{ORB}$) the 
overstable waves are still in the linear growth phase and the power spectrum corresponds directly to the linear growth rates 
%
%
%
%
$\omega_{R}(\lambda)$ (cf.\ Section \ref{sec:theo} and the curve corresponding to $\beta=1.25$ with $q=0$ in Figure \ref{fig:grqvar}). During this stage the 
kinetic 
energy density increases rapidly.
 At later times nonlinear effects slow down the evolution and the power spectrum at $t=20,000\,\text{ORB}$ reflects the 
nonlinear saturation of the overstable waves.

In LSS2017 we have shown that \emph{axisymmetric} viscous overstability in a self-gravitating disk evolves towards a state of 
minimal nonlinear oscillation frequency $\omega_{I}$ (Figure \ref{fig:pr76osnoforp2}, left panel),
or equivalently, towards a state of vanishing nonlinear 
group velocity $\mathrm{d}\omega_{I} / \mathrm{d} k$.
The dashed blue lines in Figure \ref{fig:pr76osnofor} (lower left panel) and Figure \ref{fig:pr76osnoforp2} indicate the wavelength 
corresponding to this frequency minimum of the nonlinear dispersion relation (by margins  $\pm 20 \,\text{m}$).
The wavelet power in Figure \ref{fig:pr76osnofor} reveals that in the region $r>r_{L}$ the saturation wavelength of viscous overstability is very close to the 
expected value. The overstable waves in this region are responsible for the sharp peak in the power spectrum for $t=20,000\,\text{ORB}$ at $\lambda\lesssim 
300\,\text{m}$ (Figure \ref{fig:pr76osnoforp2}).
On the other hand, the region $r<r_{L}$ contains a left traveling wave with a wavelength that gradually departs from the expectation value towards the left, 
measuring $\lambda\sim 400\,\text{m}$ at the edge of the buffer-region.
We observe the presence of weak 
long-wavelength undulations on top of the overstable waves in the region $r>r_{L}$.
These mild, persistent undulations seem to adhere to the (long) density wave dispersion relation and result from the azimuthal derivative terms 
in the hydrodynamic equations (Section 
\ref{sec:azideriv}). 
They seem to prevent the saturation wavelength of overstable waves in the region $r>r_{L}$ 
to exceed the nonlinear frequency minimum.
As such, 
the azimuthal derivatives seem to effectively remove the artificial influence of the periodic boundary conditions
on the long-term nonlinear saturation of the viscous overstability in that they sustain mild perturbations on the wave trains, at least in the region 
$r>r_{L}$ (see \citet{latter2010} and LSS2017 for more details). This is further illustrated in Figure \ref{fig:oscomp} 
(Appendix \ref{sec:appendixd}) where we compare these results with those of an integration 
%
%
%
%

%
\noindent
without the azimuthal derivative terms. These plots confirm our 
finding in Section \ref{sec:theo} that the linear behavior of small-scale axisymmetric and non-axisymmetric overstable modes is identical in our model. 
Furthermore, the nonlinear saturation behavior is very similar as well. Due to the lack of persistent perturbations in the axisymmetric ($m=0$) integration 
all but one of the source and sink pairs will eventually merge and disappear so that the entire box will be filled out by a single wavetrain  
which originates from the left buffer-region and whose wavelength increases with increasing distance from its origin. At the time $t=35,000\,\text{ORB}$ its 
wavefront, which travels with a group velocity of several meters per orbital period, has reached a radial distance of $r-r_{L} \sim 60\,\text{km}$.
\citet{latter2010} have pointed out that this long-term behavior is actually an artifact of the applied periodic boundary conditions. 

%
%

Note that due to the relatively low grid-resolution the wave profiles are not fully developed since the higher harmonics are diminished. This, and 
also the 
effect of the azimuthal derivatives should, however, not affect our qualitative discussion of the interaction between the density 
wave and viscous overstability in the following. 
 \vspace{0.5cm}
 
\subsubsection{Co-Existence of Density Waves and Viscous Overstability}\label{sec:coex}
In Figure \ref{fig:pr76waveletbetvarhr} (Appendix \ref{sec:appendixd}) we compare integrations with a fixed forcing strength $\tilde{T}^{s}=9\cdot 10^{-2}$ and 
varying value of the 
viscosity parameter $\beta$. The first integration shows the same case as considere in Section \ref{sec:modelcomp} with $\beta<\text{min}[\beta_{c}]_{\lambda}$
so that no viscous overstability develops. The integrations in rows 2-4 use $\text{min}[\beta_{c}]_{\lambda} < \beta < \beta_{c}(\lambda\to\infty)\equiv 
\beta_{c}^{\infty}$, while the 
integration 
shown 
in the bottom panel adopts $\beta > \beta_{c}^{\infty}$. From top to bottom the results show an increasing saturation amplitude of the viscous overstability in
%
%
%
the evanescent 
region of the density wave ($r<r_{L}$), as well as for large distances $r\gg r_{L}$ from resonance, where the density wave is already strongly damped.
Furthermore, as a reaction on the increased value of $\beta$ the amplitude of the density wave shows a mild increase as well, particularly at larger distances.
These trends are expected from existing models for the nonlinear saturation of viscous overstability (\citet{schmidt2003,latter2009}), as well
as the BGT and the WNL models for nonlinear density waves.
However, the behavior seen in the integration with $\beta=1.35$ is not correctly described by the latter models. Since in this case $\beta>  
\beta_{c}^{\infty}$, i.e.\ all wavelengths should be overstable, it is expected from these models that the density wave does not damp but retains a finite 
(saturation) amplitude at large distances from resonance (BGT86;~LSS2016). In contrast, our 
%
%
%
%
%
integration shows a damping of the wave very 
similar to the cases with $\beta\lesssim \beta_{c}^{\infty}$. In this case the viscous overstability possesses a 
sufficiently large amplitude to withstand the perturbation by the density wave at all distances $r-r_{L}$, albeit with strongly 
diminished amplitude in the region of largest density wave amplitude.
In contrast, in the first three integrations viscous overstability is fully damped for a range of distances where the density wave amplitude takes the largest 
values.

Figure \ref{fig:pr76waveletbet125hr} (Appendix \ref{sec:appendixd}) shows a series of integrations with increasing forcing strength 
$\tilde{T}^{s}=10^{-4}-0.16$ 
and fixed value $\beta=1.25 > 
\beta_{c}^{\infty}$. The first wave, excited by a small torque $\tilde{T}^{s}=10^{-4}$ is a linear wave. The development of the viscous overstability is very 
similar to the case without forcing (Figure \ref{fig:pr76osnofor}). With increasing torque, the overstable waves become increasingly distorted by the 
density wave, showing many similarities to those in Figure \ref{fig:pr76waveletbetvarhr}. Eventually, the density wave in the bottom panel is sufficiently 
strong to
suppress viscous overstability in the far wave zone, and the former wave attains a finite saturation amplitude at large distance from resonance until it 
hits the 
buffer-zone near $x=300\,\text{km}$. At times $t\gtrsim 20,000 \,\text{ORB}$ there remain small distortions in the wave profile. It is possible that these 
result from the approximative treatment of the azimuthal derivatives (Section \ref{sec:azideriv}). The saturation amplitude 
$\tau\sim 1.35$ is slightly larger than what is predicted by the BGT and WNL models, which is 
$\tau\sim 1.15$. It is possible that this is a consequence of our approximation for the azimuthal derivatives. 

\begin{figure*}
\centering
\includegraphics[width = 0.95 \textwidth]{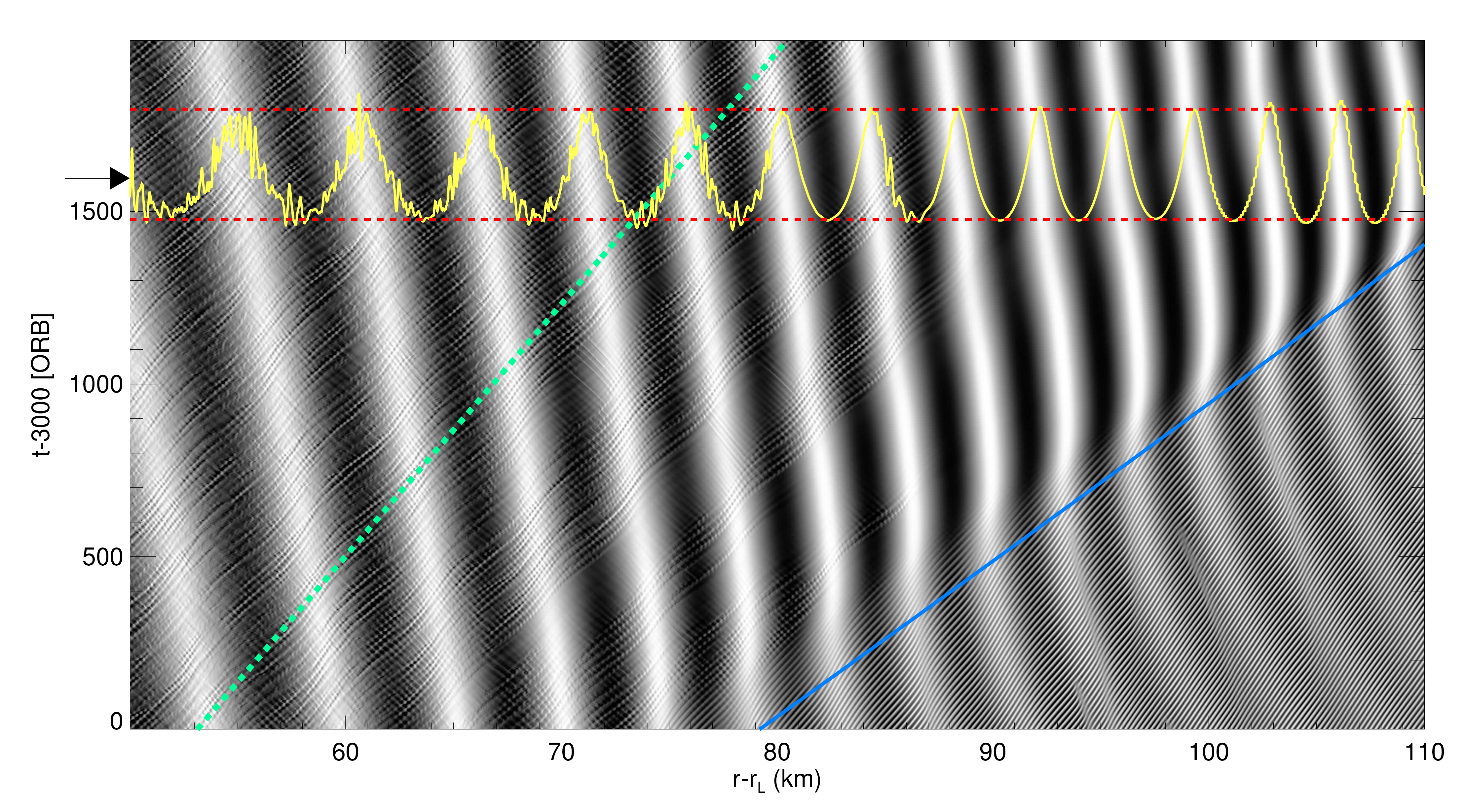}
\caption{Stroboscopic space-time diagram of a $60\, \text{km}$-section of $\tau$ resulting from the 
same hydrodynamical integration using $\beta=1.20$ and scaled torque $\tilde{T}^{s}=9\cdot 10^{-2}$ as displayed in Figure \ref{fig:pr76waveletbetvarhr}
for times $t= 3,000-5,000\,\text{ORB}$. The over-plotted $\tau$-profile describes the state at time $t= 1,600\,\text{ORB}$ and the red dashed lines indicate 
the maximum and minimum values of $\tau$ predicted by Equation (\ref{eq:taupro}) for $q=0.25$.
The blue solid 
line describes the expected location of the density wave front based on the group velocity (\ref{eq:vg}). Overstable waves at radial distances 
$r-r_{L} \gtrsim 90\,\text{km}$ are fully damped once the density wave front has passed this region. At considerably larger radial distances where the density 
wave has damped substantially, overstability reappears (see Figure \ref{fig:pr76waveletbetvarhr}, fourth row). The green dashed line 
indicates the nonlinear phase velocity $\omega_{I}/k$ of overstable waves with $\lambda=250\,\text{m}$ (Figure \ref{fig:pr76osnoforp2}). Overstable 
waves existing at distances $r-r_{L} \sim 80-110\,\text{km}$ before the density wave front arrives possess small amplitude perturbations (the slowly 
left-traveling narrow features). These are expected to propagate with the nonlinear group velocity $\mathrm{d} \omega_{I}/\mathrm{d} k$ of the overstable waves 
(\citet{latter2010};~LSS2017), which is small since the wavelength $\lambda$ of the waves is close to the 
(nonlinear) frequency minimum (Figure \ref{fig:pr76osnoforp2}).}
\label{fig:bet120detail1}
\end{figure*}

\begin{figure*}
\centering
\includegraphics[width = 1. \textwidth]{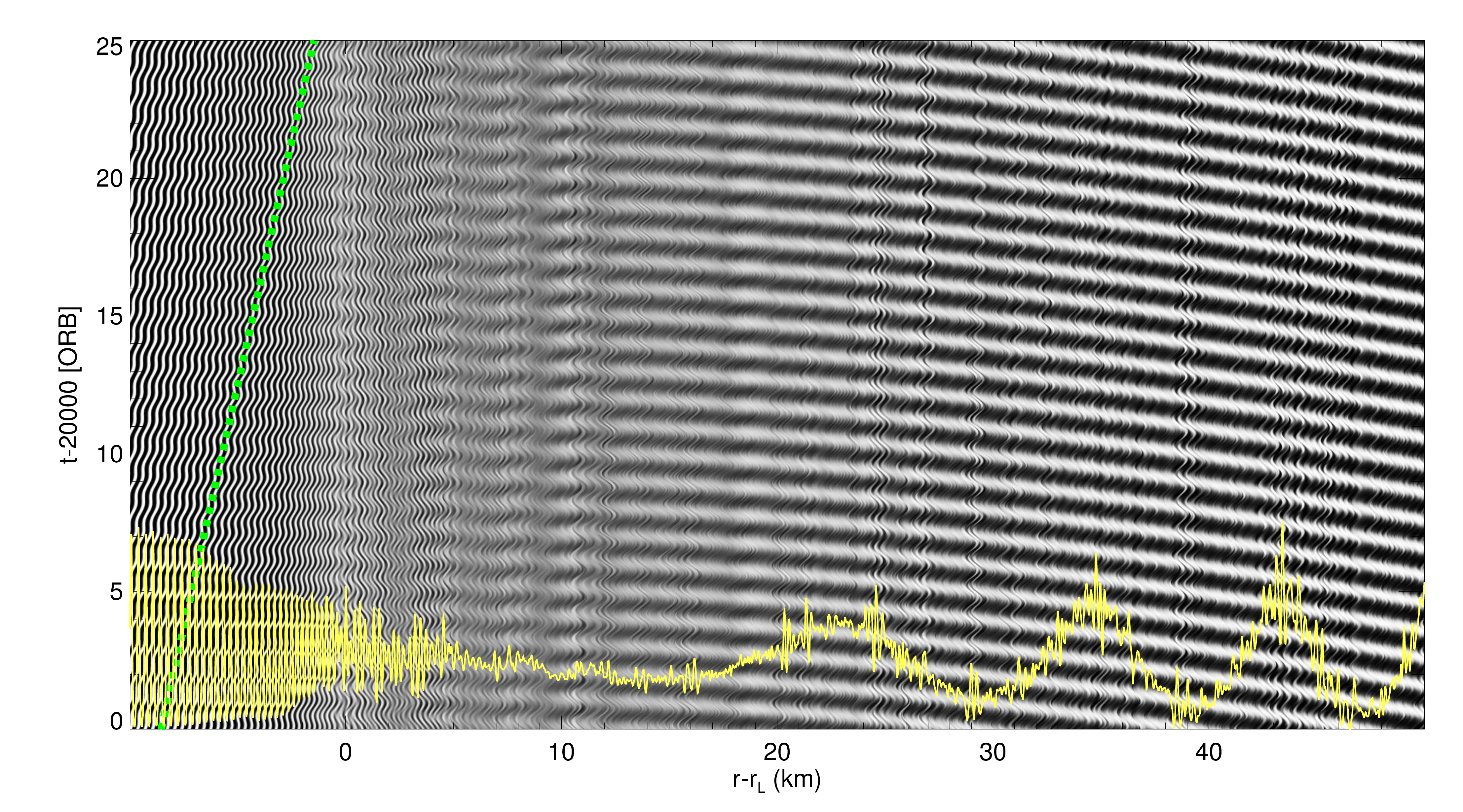}
\caption{Orbit-resolved space-time diagram of $\tau$ for a region near the density wave resonance resulting from the integration shown in Figure 
\ref{fig:pr76waveletbetvarhr} with $\beta=1.20$ for times $t\gtrsim 20,000\,\text{ORB}$. The nearly horizontal pattern which becomes increasingly pronounced 
with increasing $r > r_{L}$ represents the density wave. The smaller-scale structures are overstable waves which are perturbed by the satellite resonance, 
causing the ''wiggles'' in their appearance, in contrast to the waves displayed in Figure \ref{fig:pr76osnofor}. The green dashed line is the expected phase 
speed $\omega_{I}/k$ of (unperturbed) overstable waves 
with $\lambda=300\,\text{m}$ obtained from Figure \ref{fig:pr76osnoforp2}.
A profile of $\tau$ is drawn for the time indicated by the arrow, showing how the amplitude of overstable waves is reduced in approaching the resonance from 
smaller radii. Visible as well are the first wave-cycles of the density wave.  }
\label{fig:bet120detail2}
\end{figure*}

\vspace{0.4cm}
Some details of the wave patterns encountered in our integrations
 are illustrated in Figures \ref{fig:bet120detail1} and 
\ref{fig:bet120detail2}, which describe the 
%
%
%
%
%
%
%
%
%
%
%
%
%
\noindent
integration with $\beta=1.20$ and $\tilde{T}^{s}=9\cdot 10^{-2}$  (same as in Figure \ref{fig:pr76waveletbetvarhr}, 
third row). 
Figure \ref{fig:bet120detail1} shows a stroboscopic space-time plot of a section of the radial $\tau$-profile for times $t = 3,000-5,000\,\text{ORB}$.
During this time the density wave front traverses the considered region (indicated by the blue solid line) and clears overstable waves past a 
radial distance $r-r_{L}\gtrsim 90\,\text{km}$ (see also Figure \ref{fig:pr76waveletbetvarhr}, fourth row). The density wave corresponds to the nearly vertical 
pattern with radially decreasing wavelength $\lambda \sim 10\,\text{km} - 5 \,\text{km}$  (cf.\ Figures \ref{fig:sptdpr76f001}-\ref{fig:sptdpr76f20}), while 
(apparently right-traveling) overstable waves are represented as short-wavelength structure. 
The green dashed line indicates the expected (unperturbed) nonlinear phase velocity 
$\omega_{I}/k$ of these waves, assuming a wavelength $\lambda =250\,\text{m}$ (Figure \ref{fig:pr76osnoforp2}, left panel). 
Note that the frequencies drawn in Figure \ref{fig:pr76osnoforp2} correspond to 
$\beta=1.25$, but the dependence of the overstable frequency on $\beta$ is weak so that the corresponding curves for $\beta=1.20$ are almost identical.
The wavelength of overstable waves is modulated as they traverse the peaks and troughs of the density wave.
That is, the green dashed line matches quite well the phase velocity of the overstable waves within the density wave peaks. In the troughs the phase velocity 
is notably increased, which follows from the decreased wavenumber of the overstable waves in these regions.  
Furthermore, a profile of $\tau$ at $t= 1,600\,\text{ORB}$ (as marked by the arrow) is overplotted.
If we assume that the density wave at a given time can be described through Equation (\ref{eq:taupro}), 
we can 
estimate $q\sim 0.25$ in the region where overstability is damped. The red dashed lines in Figure \ref{fig:bet120detail1} indicate minimum and maximum 
values of $\tau$ resulting from (\ref{eq:taupro}) for  $q=0.25$. In a similar manner it follows that  values of $q\sim 0.23$ and $q\sim 0.20$ lead to a damping
of viscous overstability in the cases $\beta=1.16$ and $\beta=1.10$ (Figure \ref{fig:pr76waveletbetvarhr}), respectively. The associated $q$-values where 
overstability 
reappears at larger distances from resonance seem 
to be slightly smaller in all cases.
The mitigation of viscous overstability by a density wave will be discussed 
in more detail in the following section.

Figure \ref{fig:bet120detail2} shows an orbit-resolved space-time plot of $\tau$, illustrating how overstable wavetrains are distorted in direct vicinity of 
the Lindblad resonance in the integration displayed in Figure \ref{fig:pr76waveletbetvarhr} corresponding to $\beta=1.20$.
In the evanescent region $r<r_{L}$ we recognize a right traveling overstable wave whose phase velocity undergoes periodic perturbations on the orbital time 
scale. These perturbations become stronger as the wave approaches the resonance $r=r_{L}$. In the region $r>r_{L}$ the overstable waves seem to be unable to 
travel over 
any notable distance as their phase velocity rapidly changes its sign. In this region the amplitude of the overstable waves becomes strongly diminished (cf.\ 
Figure 
\ref{fig:pr76waveletbetvarhr}, fourth row).
\subsubsection{Viscous Overstability in a Perturbed Ring: Axisymmetric Approximation}\label{sec:pertring}

The hydrodynamical integrations presented above reveal a variety of structures resulting from interactions between a spiral 
density wave and the free short-scale waves associated with spontaneous viscous overstability. We find that a sufficiently strong spiral density wave 
completely 
mitigates the 
growth of viscous overstability. In this section we will consider this aspect in a more simplified, axisymmetric model which, on the one hand, allows us to 
conduct a 
simple hydrodynamic stability analysis, and, on the other hand, can be investigated with local N-body simulations. 
%
%
%
%
%
%
%
%
%
In what follows we assume that the perturbation is due to a nearby ILR. To that end consider the axisymmetric equations
\begin{equation}\label{eq:lineqpert1}
\partial_{t} \tau   =  -u  \partial_{x} \tau - \tau \partial_{x} u,
\end{equation}
\begin{equation}\label{eq:lineqpert2}
\partial_{t} u  =  -u \partial_{x} u + ( \Omega_{L}^2 r + 2\Omega_{L} \,  v) -\partial_{x}\left[\phi^{d} + \phi^{p}\right] - \frac{1}{\sigma} 
\partial_{x} \hat{P}_{xx},
\end{equation}
\begin{equation}\label{eq:lineqpert3}
\partial_{t} v   =  - u \partial_{x}v  -2 \Omega_{L} u  - \frac{1}{\sigma} \partial_{x} \hat{P}_{xy},
\end{equation}
in the shearing sheet approximation (\citet{goldreich1965}), using a rectangular frame $(x,y=0)$ rotating with $\Omega_{L}$ where $x$ is given by (\ref{eq:xx})
and where, in contrast to 
Equations (\ref{eq:nleq}), $\Omega=\Omega_{L}$ is now a constant. 
Note that the components of the pressure tensor $\hat{P}_{xx}$ and $\hat{P}_{xy}$ are 
identical to $\hat{P}_{rr}$ and $\hat{P}_{r\theta}$ 
given by 
(\ref{eq:pten}), respectively, since we had already neglected 
curvature terms in the latter expressions. Moreover, $v$ denotes here the total azimuthal velocity and $\phi^{p}$ is the gravitational potential due to the 
planet.
We now introduce the perturbed oscillatory ground 
state (\citet{mosqueira1996})
\begin{equation}\label{eq:gspert}
 \begin{split}
 \tau_{0} & = \frac{1}{1- q \cos \left(m \phi + m \Delta \right)},  \\[0.1cm]
u_{0} & =\Omega_{L} x \, \frac{q \sin \left(m \phi + m \Delta  \right)}{1-q \cos \left(m \phi + m \Delta  \right)},\\[0.1cm]
v_{0} & = -\frac{3}{2}\Omega_{L} x  \, \frac{1-(4/3) q \cos \left(m \phi + m \Delta  \right)}{1-q \cos \left(m \phi + m \Delta  
\right)}.
 \end{split}
\end{equation}
As shown in Appendix \ref{sec:gspert} Equations (\ref{eq:gspert}) are valid in the vicinity of a Lindblad resonance where fluid streamlines can be described 
by $m$-lobed orbits
\begin{equation}\label{eq:streamline}
 r(a) = a \left[ 1- e(a) \cos \left(m \phi + m \Delta \right) \right],
\end{equation}
in a cylindrical frame $(r,\phi,z=0)$ rotating with the satellite's mean motion frequency $\Omega^{s}= \hat{\omega}^{s} /m$ [Equation (\ref{eq:omfin})] 
in the 
%
%
%
%
%
%
%
%
%
%
%
%
present context.
The quantities $a$ and $e$ denote a streamline's semi-major axis and eccentricity, respectively and $\Delta$ is a phase angle. 
%
Furthermore, $q$ is the nonlinearity parameter (\citet{Borderies83}) fulfilling
\begin{equation}\label{eq:qpar}
q = \sqrt{  \frac{\mathrm{d}(a e)}{\mathrm{d}a} +  m a e \frac{\mathrm{d}\Delta}{\mathrm{d}a}}.
\end{equation}
In the limit $q \to 0$ Equations (\ref{eq:gspert}) describe the usual homogeneous unperturbed ground state as in Section \ref{sec:theo}.
If we now adapt to the frame ($r,\theta,z=0$) which rotates with the local Kepler frequency $\Omega_{L}$ at the resonance, 
we have
\begin{equation}\label{eq:coordtrans}
m\phi = m\theta - m \left(\frac{\hat{\omega}^{s}}{m} -\Omega_{L} \right) t =m\theta + \Omega_{L} t.
\end{equation}

Using (\ref{eq:gspert}) and (\ref{eq:coordtrans}) the Equations (\ref{eq:lineqpert1})-(\ref{eq:lineqpert3}) are identically fulfilled if one assumes a 
consistently expanded 
planetary potential
\begin{equation*}
  \phi^{p} = -\Omega_{L}^2 \left[ r_{L}^2 - r_{L} x +x^2 \right]
\end{equation*}
at $y=0$ and if one neglects the radial dependencies of the phase angle $\Delta$ and the nonlinearity parameter $q$.
Note that the terms arising from orbital advection [the azimuthal derivatives (Section \ref{sec:azideriv})] are neglected in the axisymmetric equations 
(\ref{eq:lineqpert1})-(\ref{eq:lineqpert3}). These terms would scale relative to the other terms as $x/r_{L}\lesssim 10^{-4}$ in the present situation.
Since we assume that we are close to the Lindblad resonance ($|x|\to 0$) we can ignore the radial variation of $\Delta$ in the arguments of the sine 
and cosine functions appearing in (\ref{eq:gspert}).
That is, in the evanescent region close to the resonance ($x\lesssim 0$) one can approximate $q\sim  a \mathrm{d}e /\mathrm{d} a$ since the eccentricity 
increases steeply 
towards the 
resonance and the disk's response to the perturbation is not wavelike, i.e.\ $\mathrm{d}e/\mathrm{d}a \gg  e \mathrm{d}\Delta/\mathrm{d}a$ (see for 
instance \citet{hahn2009}). 
For $x\gtrsim 0$ one usually adopts the approximation that a non-vanishing $q$ [Eq.\ (\ref{eq:qpar})] arises 
only from the radial variation of the phase angle $\Delta$. 
This is the tight-winding approximation for the disk's response in form of a long spiral density wave propagating outward with \emph{radial wavenumber} $m  
\mathrm{d}\Delta / \mathrm{d}a \gg 1/a$.
However, even in 
this region we can approximate 
$\Delta$ as a constant in 
(\ref{eq:gspert}), as long as the wavenumber of the density wave is much smaller than that of the overlying periodic micro structure that we wish 
to analyze. Thus, the neglect of the radial variation of $\Delta$ restricts the applicability of the above model to a small region at the resonance, since 
for 
sufficiently large $x>0$ the wavelength of the density wave is \emph{not} much greater than that of the overstable waves (cf.\ Figures 
\ref{fig:pr76waveletbetvarhr}, \ref{fig:pr76waveletbet125hr}). 
As for the necessary approximation of a constant $q$ we rely on previous studies which imply that for typical 
length scales of overstable wavetrains (several kilometers) $q$ varies slowly (see for instance Figure 3 of \citet{bgt1986}, Figure 2 of 
\citet{hahn2009}, as well as \citet{longaretti1986} and \citet{rappaport2009} on the Mimas 5:3 wave). 
Note that the tight-winding approximation applied in \citet{bgt1986} inevitably assumes $q(x=0)=0$).

Thus, in what follows we assume that the phase angle $\Delta$ is a constant and we assume (without loss of generality) that the ring is in the 
uncompressed 
state at initial time $t=0$. Then Equations 
(\ref{eq:gspert}) together with (\ref{eq:coordtrans}) yield
\begin{equation}\label{eq:gspertnew}
 \begin{split}
 \tau_{0} & = \frac{1}{1- q \sin \left(\Omega_{L} t \right)},  \\[0.1cm]
u_{0} & =-\Omega_{L} x \, \frac{q \cos \left( \Omega_{L} t \right)}{1-q \sin \left( \Omega_{L} t \right)}, \\[0.1cm]
v_{0} & = -\frac{3}{2}\Omega_{L} x  \, \frac{1-(4/3) q \sin \left( \Omega_{L} t \right)}{1-q \sin \left( \Omega_{L} t \right)}.
 \end{split}
\end{equation}
To the ground state (\ref{eq:gspertnew}) we now add axisymmetric perturbations
\begin{equation}\label{eq:linper}
 \mathbf{\Psi}(x,t) = \begin{pmatrix} \tau^{'}(x,t)  \\ u^{'}(x,t) \\  v^{'}(x,t)  \end{pmatrix} = \begin{pmatrix} 
 \hat{\tau}(x,t) \\ \hat{u}(x,t)  \\ \hat{v}(x,t) \end{pmatrix} \exp\left[i k(t) x\right]
\end{equation}
with time-dependent wavenumber
\begin{equation}\label{eq:kpert}
k(t) = \frac{k_{0}}{1 - q \sin \Omega_{L} t }.       
\end{equation}
The time dependence in (\ref{eq:kpert}) stems from the periodic variation of the radial width of a streamline resulting from the perturbation by 
the density wave.
This ansatz is chosen since Equations (\ref{eq:gspertnew}) contain only the kinematic effect of the density wave on the considered ring region.
That is, (\ref{eq:gspertnew}) describe how a single fluid streamline behaves in the presence of a density wave whose wavelength is assumed much larger than
the extent of the streamline. For a nonlinear study of viscous overstability (implying longer time scales) in a perturbed ring region, the dynamical 
evolution of a streamline due to neighboring streamlines should be considered as well, which requires a more sophisticated treatment than the one adopted 
here. The behavior of the wavenumber according to Equation (\ref{eq:kpert}) is also seen directly in N-body simulations (cf.\ Figure \ref{fig:nbodyevo}).

We assume that the $x$-dependency of the quantities $\hat{\tau}(x,t)$, $\hat{u}(x,t)$ and $\hat{v}(x,t)$ in (\ref{eq:linper}) is only weak so that
\begin{equation}\label{eq:kapprox}
 \partial_{x} \begin{pmatrix} 
 \hat{\tau}(x,t) \\ \hat{u}(x,t)  \\ \hat{v}(x,t) \end{pmatrix} \ll k(t) \begin{pmatrix} 
 \hat{\tau}(x,t) \\ \hat{u}(x,t)  \\ \hat{v}(x,t) \end{pmatrix} 
\end{equation}
and can be ignored.

We insert the resulting expressions for $\tau=\tau_{0}+\tau^{'}$, $u = u_{0}+ u^{'}$, $v = v_{0} + v^{'}$ into (\ref{eq:lineqpert1})-(\ref{eq:lineqpert3}) and 
linearize with 
respect to the perturbations $\tau^{'}$, $u^{'}$ and $v^{'}$.
This procedure yields the linear system 
\begin{equation}\label{eq:pde}
 \partial_{t} \mathbf{\Psi}(x,t) = \hat{M}(x,t) \mathbf{\Psi}(x,t)
\end{equation}
where the radial location $x$ is a parameter and
\begin{equation}\label{eq:matni}
 \renewcommand*{\arraystretch}{1.}
\begin{array}{@{}*{22}{l@{}}}
\hat{M}(x,t) = \begin{pmatrix} M_{11}   \hspace{0.2 cm} & M_{12} \hspace{0.2 cm} & M_{13} \hspace{0.2 cm}   \\[0.4 cm] 
M_{21}   \hspace{0.2 cm}  & M_{22}  \hspace{0.2 cm}  & M_{23} \hspace{0.2 cm} \\[0.4 cm] 
M_{31}  \hspace{0.2 cm} & M_{32} \hspace{0.2 cm} & M_{33}   \hspace{0.2 cm}  \\[0.15cm] 
\end{pmatrix} 
\end{array}
\end{equation}
with
\begin{equation}\label{eq:matnic}
\begin{split}
M_{11} & = q \left(1+i k(t) x\right) J^{-1} \cos  t  ,\\
M_{12} & = -ik(t) J^{-1} ,\\
M_{13} & = 0 ,\\
M_{21} & = i \left[2g -k(t) \left( J + \left(\beta+1\right) \alpha  \nu_{0}q J^{-\beta} \cos  t  \right)   \right], \\
M_{22} & = J^{-1} q \cos  t \left[ 1 +  i k(t)  x  \right] - \alpha \nu_{0} k^2(t) J^{-\beta} , \\
M_{23} & =2 ,\\
M_{31} & =\frac{1}{2} i \left(\beta+1\right) k(t) \nu_{0} J^{-\beta} \left(4 q \sin  t -3 \right) ,\\
M_{32} & = -\frac{1}{2}J,\\
M_{33} & = -k(t) \left[k(t) \nu_{0} J^{-\beta} -i q J^{-1} x \cos  t  \right].
\end{split}
\end{equation}
To arrive at (\ref{eq:matnic}) we also used (\ref{eq:wkbsgos}) and (\ref{eq:gg}) where $k_{x}$ has been replaced by (\ref{eq:kpert}).
Here we apply scalings so that time and length are scaled with $1/\Omega_{L}$ and $c_{0}/\Omega_{L}$, respectively.
We also define the dimensionless quantities
\begin{equation}
\label{eq:shorteq}
\begin{split}
 J & =1-q \sin  t,\\
  \alpha & = \frac{4}{3}+\gamma,
  \end{split}
\end{equation}
for notational brevity.
To illustrate the procedure for obtaining (\ref{eq:matnic}) let us consider the linearization of the continuity equation (\ref{eq:lineqpert1}).
The latter can be written as
\begin{equation*}
\begin{split}
\partial_{t} \left(\tau_{0}+\tau^{'}  \right)  = & -\left(u_{0}+ u^{'}\right)  \partial_{x} \left(\tau_{0}+\tau^{'}  \right)\\
\quad & - \left(\tau_{0}+\tau^{'}  \right) \partial_{x} \left(u_{0}+ u^{'}\right). 
\end{split}
\end{equation*}
Since the ground state quantities are an exact solution in the current approximation, we end up with
\begin{equation*}
\begin{split}
\partial_{t} \tau^{'}   = - u_{0} \partial_{x} \tau^{'}  - \tau^{'} \partial_{x} u_{0} - \tau_{0} \partial_{x} u^{'} ,
\end{split}
\end{equation*}
where we used $\partial_{x} \tau_{0}=0$ [Equation (\ref{eq:gspertnew})].
Using (\ref{eq:linper}) and (\ref{eq:kapprox}) yields 
\begin{equation*}
 \partial_{t} \tau^{'}   =  -\left[ i k(t) u_{0} + \partial_{x} u_{0}\right]  \, \tau^{'}  - i k(t) \tau_{0} \,u^{'}.
\end{equation*}
By applying (\ref{eq:gspertnew}), (\ref{eq:kpert}) and (\ref{eq:shorteq}), as well as the aforementioned scalings, we obtain $M_{11}$, $M_{12}$ and $M_{13}$ as 
given in (\ref{eq:matnic}).
All other matrix components are derived in the same fashion.

The aim is now to investigate whether a seeded overstable wavetrain (\ref{eq:linper}) will decay or grow in amplitude by integrating (\ref{eq:pde}) over a 
given time 
range. 
In the case $q=0$ one can assume
\begin{equation*}
 \mathbf{\Psi}(x,t)_{q=0} = \begin{pmatrix} \tau^{'}(x,t)  \\ u^{'}(x,t) \\  v^{'}(x,t)  \end{pmatrix} = \begin{pmatrix} 
 \hat{\tau} \\ \hat{u}  \\ \hat{v} \end{pmatrix} \exp\left[\omega t + i k_{0} x\right],
\end{equation*}
i.e.\ the solution is a traveling wave with constant growth (or decay) rate, determined by the imaginary part of $\omega$ 
(\citet{schmit1995,schmidt2001b,latter2009}).
For $q> 0$, the behavior is more complicated.

We integrate the complex-valued system of equations (\ref{eq:pde}) numerically with a 4th-order Runge-Kutta method 
on a grid of radial size $L_{x}=2\,\text{km}$.
As initial state we use an eigenvector of (\ref{eq:matni}) in the limit $q\to 0$ at marginal stability $\beta=\beta_{c}$ which reads (\citet{schmidt2003})
\begin{equation}\label{eq:istat}
 \mathbf{\Psi}(x,t_{0}) =  \begin{pmatrix}  2 k_{0} (i k_{0}^2 \nu_{0} + s) \\ 2 s (i k_{0}^2 \nu_{0} + s)  \\ - i s + k_{0}^2 (\nu_{0} + \alpha 
\nu_{0} s^2) + \alpha \nu_{0}^3 k_{0}^6 \end{pmatrix} \exp\left[i k_{0} \, x\right],
\end{equation}
where
\begin{equation*}
s=\sqrt{1- 2 g k_{0} + \alpha k_{0}^4 \nu_{0}^2 + k_{0}^2 }
\end{equation*}
is the unperturbed frequency of the overstable mode at marginal stability [c.f.\ Equation (\ref{eq:omcrit})]. The initial state corresponds to 
a right traveling wave.

In order to obtain the growth rate of a seeded mode $\lambda_{n} = L_{x}/n$ where $n$ denotes the radial mode number, we write
\begin{equation*}
  \mathbf{\Psi}(x,t) =  \mathbf{A}_{n}(t) \exp\left[i k_{n}(t) x\right]
\end{equation*}
 where $k_{n}(t) = (2\pi/\lambda_{n})\,  J^{-1}(t)$. 
The complex amplitude is then obtained by numerical solution of
\begin{equation}\label{eq:ampcalc}
 \mathbf{A}_{n}(t)= \frac{1}{L_{x}}\int\limits_{-L_{x}/2}^{L_{x}/2} \mathrm{d}x \mathbf{\Psi}(x,t)  \exp\left[- i k_{n}(t) x\right]
\end{equation}
for each time step. Since (\ref{eq:istat}) is not an exact eigenvector of (\ref{eq:matni}) for $q>0$, the first orbital periods of integrations with $q>0$ are 
excluded from the computation of the growth rates as the system is yet to settle on an exact eigensolution.

The growth rates of $\mathbf{A}_{n}$ for different radial modes $n$ resulting with the $Pr76$-parameters adopting different values of $\beta$ and $q$ are 
%
%
%
%
\begin{figure*}
\centering
\includegraphics[width = 0.4 \textwidth]{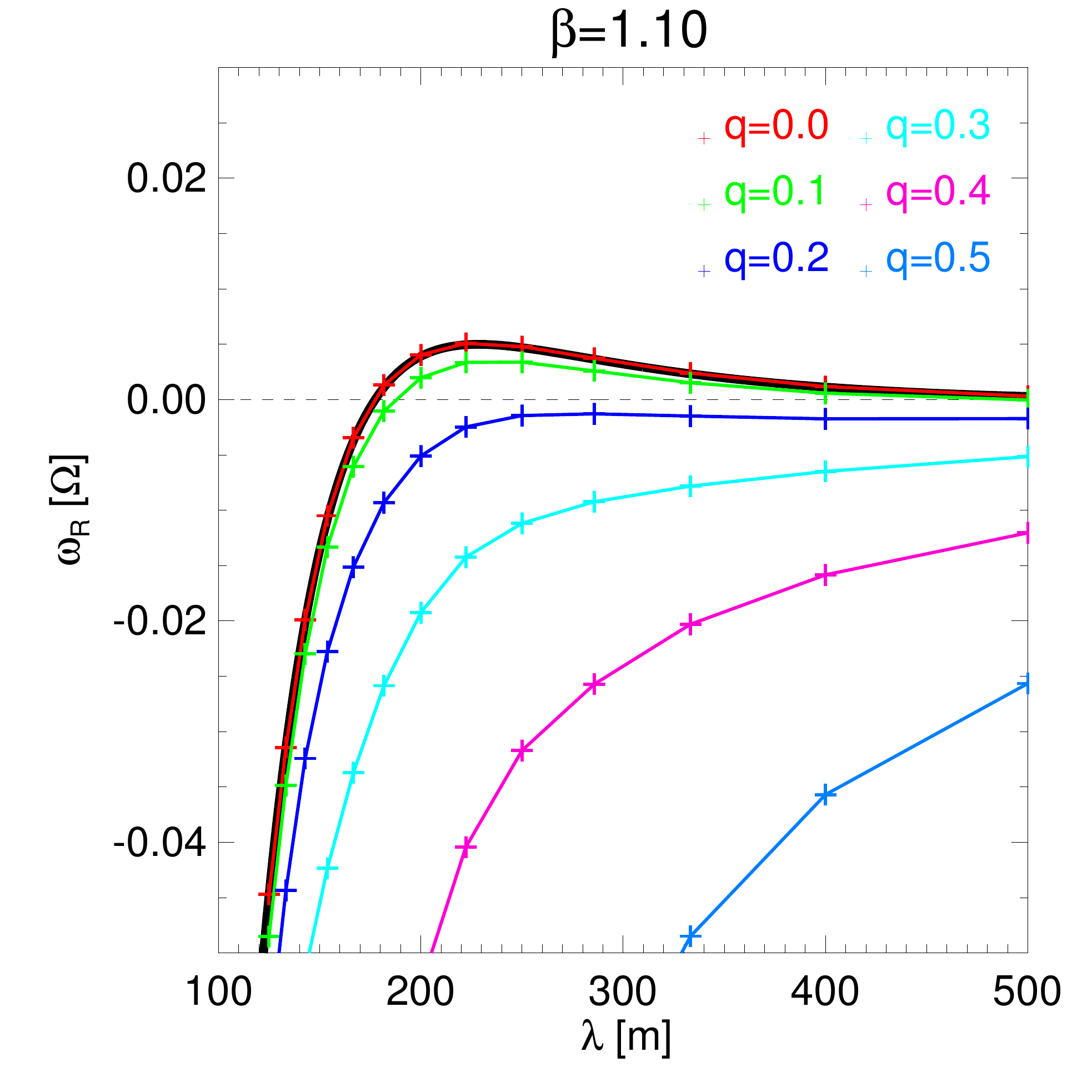}
\includegraphics[width = 0.4 \textwidth]{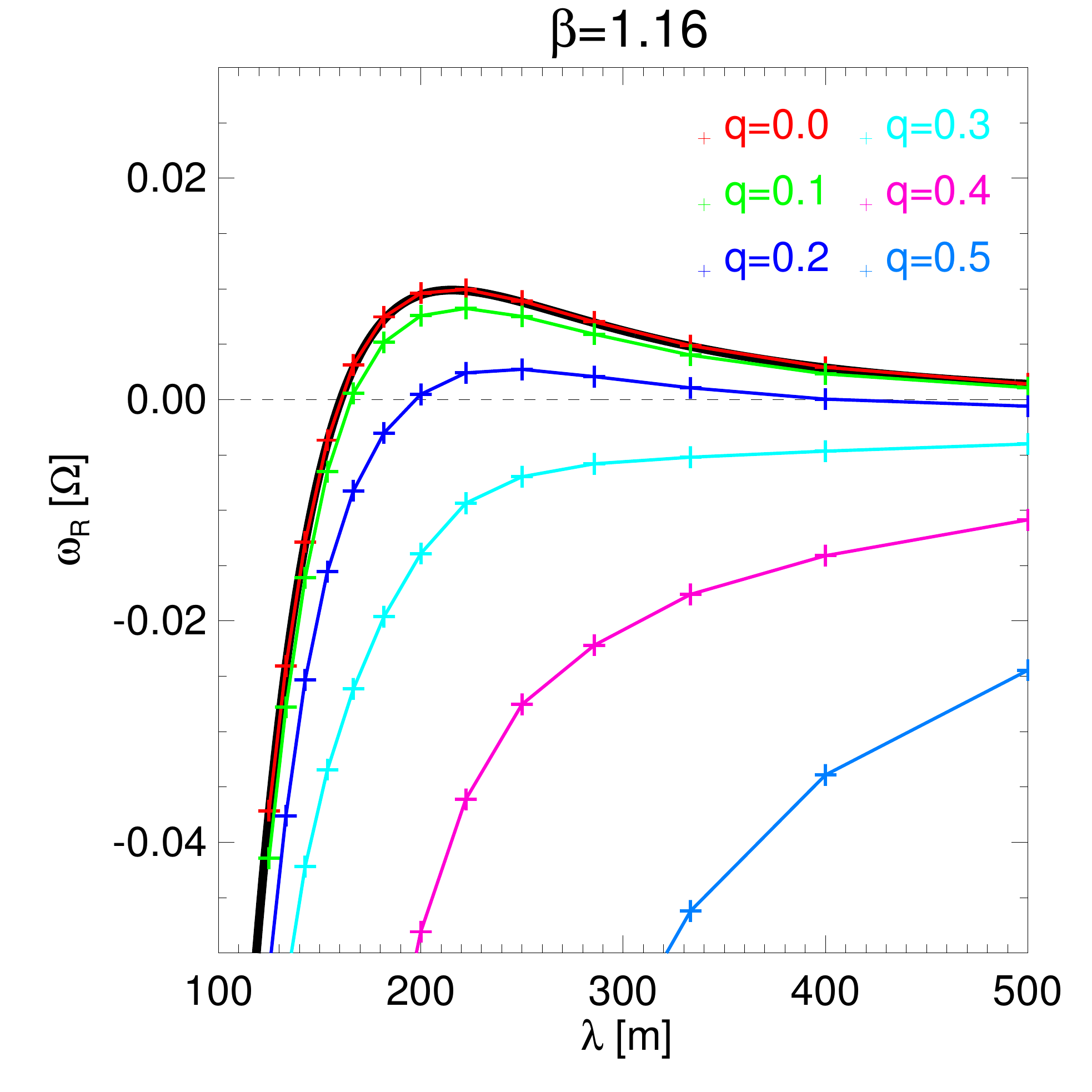}
\includegraphics[width = 0.4 \textwidth]{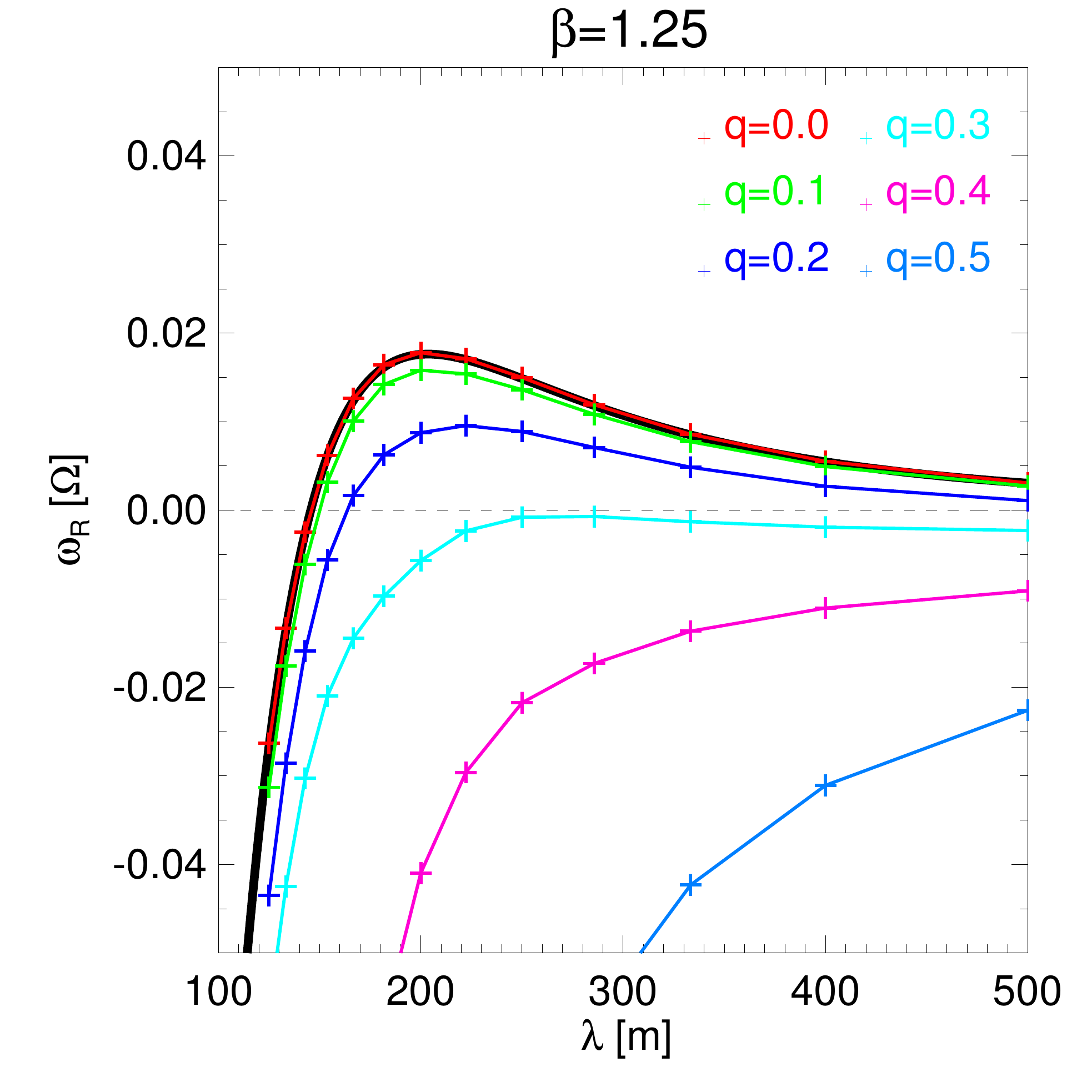}
\includegraphics[width = 0.4 \textwidth]{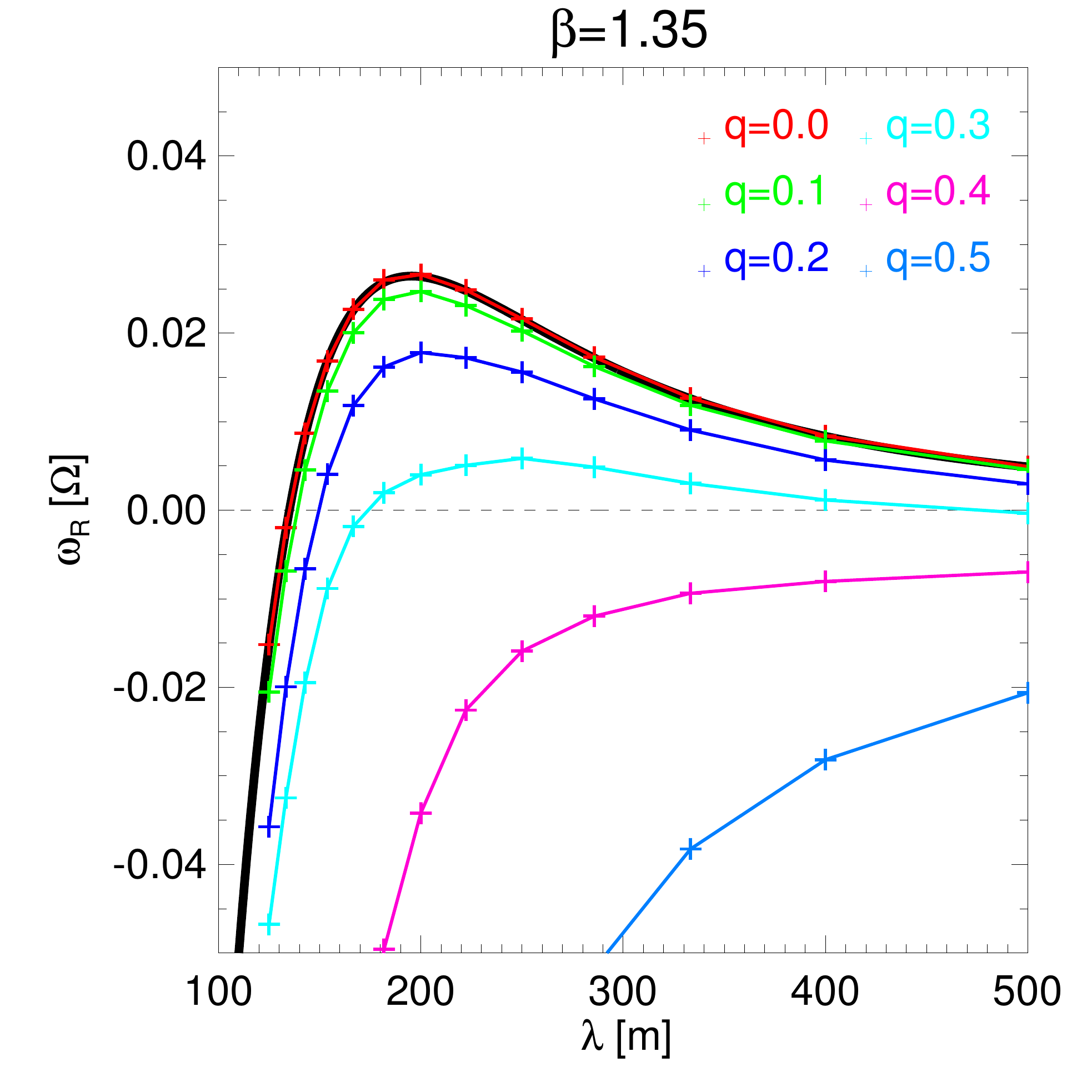}
\caption{Linear hydrodynamic growth rates of overstable modes in a perturbed ring for different values of 
the nonlinearity parameter $q$ describing the satellite perturbation. The wavelengths $\lambda$ correspond to the uncompressed state of the model ring adopted 
at times $t= l \pi /\Omega_{L}$, with non-negative integer $l$ [cf.\ Equation (\ref{eq:kpert})].
The used parameters are the $Pr76$-parameters with varying value of $\beta$ and all growth rates are scaled with $\Omega=\Omega_{L}$. In all panels the growth 
rates for $q=0$, obtained by numerical solution of (\ref{eq:evp}), are plotted additionally as black solid curves.}
\label{fig:grqvar}
\end{figure*}
%
%
%
%
%
drawn in Figure 
\ref{fig:grqvar}.
These plots show a monotonic decrease of the growth rates with increasing nonlinearity parameter $q$ on all wavelengths. At the same time the maxima of the 
curves shift towards 
larger wavelengths. From these plots we can estimate for given $\beta$ the critical values $q_{c}$ that yield negative growth rates on all wavelengths. 
In the presence of a perturbation with $q \geq q_{c}$ no axisymmetric viscous overstability is expected to develop. The so obtained values of $q_{c}$ seem to 
agree quite well with those estimated from the large-scale integrations in Section \ref{sec:coex}.

As an illustration, in Figure \ref{fig:sptdpert} (Appendix \ref{sec:appendixd}) we show space-time diagrams of the radial velocity perturbation $u^{'}$ of the 
mode $n=10$ for the case 
$\beta=1.35$ with different 
values of $q$. While for $q=0$ and $q=0.1$ the wave amplitude grows with time, as indicated by the gradual brightening in upward direction in the 
first two figures, for $q=0.4$ the amplitude diminishes. Furthermore, with increasing $q$ the overstable pattern becomes less of a uniform traveling wave. 
The waves seen in these space-time plots show many similarities to the overstable waves encountered in our large-scale integrations (Figure 
\ref{fig:bet120detail2}).
In terms of the here applied hydrodynamical model the mitigation of overstable oscillations by the satellite perturbation can be explained by an increasing 
de-synchronization of 
specific terms appearing in the dynamical Equation (\ref{eq:lineqpert3}).
That is, the viscous overstability mechanism describes a transfer of energy from the background azimuthal shear into the epicyclic fluid motion through 
a coupling of the viscous stress to the Keplerian shear (\citet{latter2006,latter2009}), resulting in an oscillating angular momentum flux which instigates
the epicyclic oscillation if the following two conditions are met (\citet{latter2006,latter2008}).
On the one hand, the viscous stress needs to possess a sufficiently steep dependence on the surface mass density. This condition is expressed in terms of a 
critical (wavelength-dependent) viscosity parameter $\beta_{c}(\lambda)$ that must be exceeded for a given wavelength $\lambda$.  Figure \ref{fig:lambet} 
shows $\beta_{c}(\lambda)$ for an unperturbed ring ($q=0$), with minimal value $\text{Min}\left[\beta_{c}\right]_{\lambda}=1.03$.
Figure \ref{fig:grqvar} reveals that $\beta_{c}(\lambda)$ increases with increasing $q$ for all $\lambda$. 
For instance, for $q=0.2$ one finds $ 1.16 > \text{Min}\left[\beta_{c}\right]_{\lambda} >1.1$. Furthermore, for $q=0.3$ one can see that 
$\text{Min}\left[\beta_{c}\right]_{\lambda} \sim 1.25$.

The second condition that must be fulfilled for the viscous overstability to operate in a planetary ring is that the oscillation of the angular momentum flux 
must be sufficiently in phase with the epicyclic oscillation associated with an overstable wave.
In Figure \ref{fig:coup} (Appendix \ref{sec:appendixd}) we show snapshots of the term in the equation for the azimuthal velocity perturbation $v^{'}$ that 
describes 
the coupling of 
the viscous stress to the Keplerian shear and which can be written $M_{31} \, \tau^{'}$ [cf.\ (\ref{eq:pde})-(\ref{eq:matnic})]. The used 
parameters are the same as in Figure \ref{fig:sptdpert} and
the snapshots cover one orbital period in equal time-intervals. Over-plotted for the same instances of time is the epicyclic term $M_{32} \, u^{'}$, appearing 
in the same equation. For $q\lesssim 0.3$ these two terms retain a nearly constant phase-shift for all times. In contrast, for larger $q$ 
the phase difference of these terms drifts constantly. Therefore the energy transfer into the epicyclic oscillation is too inefficient, resulting 
in a damping of seeded wavetrains.

\begin{figure}[h!]
\centering
\includegraphics[width = 0.4 \textwidth]{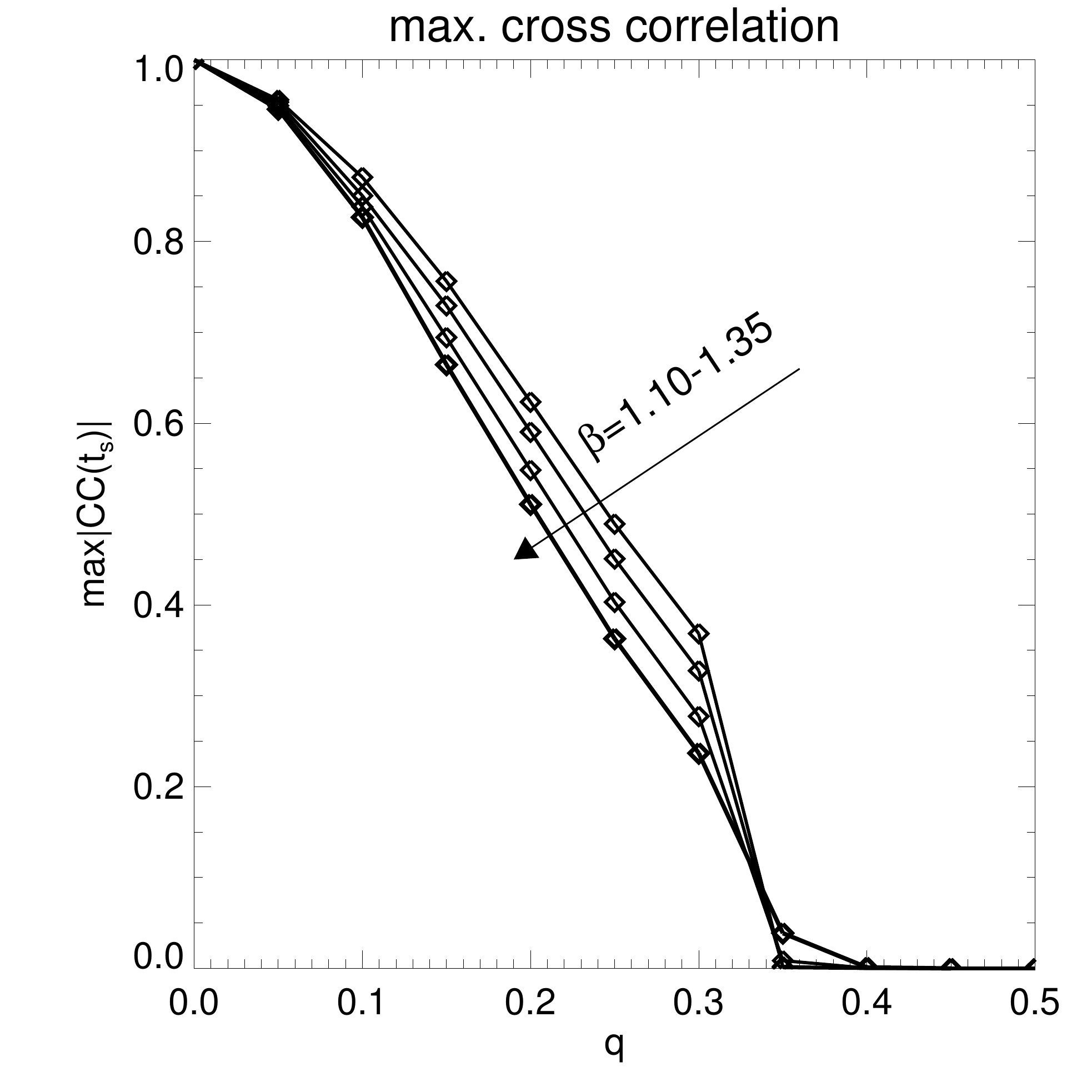}
\caption{Curves showing the maximum absolute value of the cross correlation of the quantities $M_{31}\tau{'}$ and $M_{32} u^{'}$ [Equation (\ref{eq:croscor})] 
associated with an 
overstable mode with $\lambda=200\,\text{m}$ for different values of $q$, and the values of $\beta$ used for the plots in Figure 
\ref{fig:grqvar}. 
The arrow indicates the direction of increasing $\beta$. The monotonic decrease of these curves with increasing $q$ is the reason why the growth 
rates of overstable modes (Figure \ref{fig:grqvar}) become smaller with increasing $q$. All curves 
have been normalized to yield unity for $q=0$. }
\label{fig:croscor}
\end{figure}

To quantify the phase relation between the angular momentum flux and the epicyclic oscillation associated with an overstable wave we define the cross 
correlation of the two aforementioned 
%
%
%
%
%
%
relevant terms
\begin{equation}\label{eq:croscor}
\mathcal{CC}(t_{s}) = \int \limits_{[l 2\pi]}  \left[ M_{31}(t) \, \tau^{'}(t) \, M_{32}(t+t_{s}) \, u^{'}(t+t_{s}) \right] \mathrm{d} t,
\end{equation}
where the shift parameter $t_{s}$ takes values between $0$ and $2\pi$ and the integration is performed over $l$ orbital periods.
If the two quantities $M_{31} \, \tau^{'}$ and $M_{32} \, u^{'}$ are periodic with a constant phase shift, their cross correlation will posses a sharp 
maximum for a specific value of the time shift $t_{s}$. On the other hand, if their relative phase shift varies in a more or less uniform manner during one 
orbital period, the cross correlation will be small for all values of $t_{s}$.
Thus, as a measure for the synchronization of the two oscillatory quantities we take the maximum absolute value of the cross correlation 
$\text{max}|\mathcal{CC}|_{t_{s}}$.
Figure \ref{fig:croscor} displays this value for the parameters as used in Figure \ref{fig:grqvar}. As anticipated, the curves show a steady decrease with 
increasing $q$ and exhibit a fairly sharp drop for $q=0.3-0.4$. This explains why for $q\gtrsim 0.4$ the growth rates in Figure \ref{fig:grqvar} are 
negative for all values of $\beta$ used here. To explain the different critical values $q_{c}$ for which the growth rates become negative for the different 
$\beta$-values (e.g.\ $q_{c}\sim 0.2$ for $\beta=1.1$), one must in addition take into account 
%
%
%
%
the dependence of the growth rates on $\beta$. In the 
unperturbed case ($q=0$) the growth rates depend linearly on the factor $(\beta-\beta_{c})$. 
%

%
%

%
%

%
%

%
Note that in a dilute ring, better described in terms of a kinetic model than a hydrodynamic one, the aforementioned de-synchronization can 
occur already in absence of an external perturbation. In a kinetic model of a dilute ring of sufficiently low dynamical optical depth the viscous stress tensor 
components are subjected to long (collisional) relaxation time-scales. Therefore, these cannot follow the (fast) epicyclic oscillation of the ring flow 
on the orbital time scale, which also prevents viscous overstability (\citet{latter2006}).

We complement these findings with a series of local N-Body simulations of a perturbed ring that include aspects of the vertical self-gravity force in terms of 
an enhancement of the frequency 
%
%
of vertical oscillations (\citet{wisdom1988}) and also the effect of collective radial self-gravity. 
A detailed description of the simulation method can be found in \citet{salo2018} and references therein. In particular, the force method for particle impacts 
as introduced in \citet{salo1995} is used and the radial self-gravity is calculated as in \citet{salo2009}.
The latter is parametrized through a pre-specified ground state surface mass density (see also LSS2017). 
Here we apply modified initial conditions and boundary conditions that account for a perturbed mean flow in the ring 
following the method of \citet{mosqueira1996}.
We perform simulations with meter-sized particles in a periodic box of \emph{uncompressed} radial size $L_{x}=L_{0}/ \sqrt{1-q^2}$ and azimuthal
size $L_{y}=10\,\text{m}$. The number of particles is slightly less than 10,000 in all simulations. The quantity $L_{x}$ is chosen such that the 
time-averaged ground state optical depth of the system is the same for different values of $q$, as shown below.
The radial size of the simulation region changes periodically as 
\begin{equation}\label{eq:lxt}
L_{x}(t,q) = L_{0} \frac{1-q \cos \Omega_{L} t }{\sqrt{1-q^2}}. 
\end{equation}
For a fixed azimuthal width and a fixed number of simulation particles the ground state dynamical optical depth is then given by
\begin{equation}\label{eq:taudyn}
  \tau^{dyn}(t,q) =  \frac{\tau_{0} \sqrt{1-q^2}}{1-q \cos \Omega_{L} t}
\end{equation}
where $\tau_{0}$ is the time-averaged ground state optical depth over one period $2 \pi / \Omega_{L}$, i.e.\ $\tau_{0}  \equiv \langle \tau^{dyn} (t,q) 
\rangle_{t}$ and is independent of $q$. Hence, our choice of $L_{x}$ removes the purely geometrical increase of the time-averaged mean optical depth 
with increasing $q$ and isolates the effect of the perturbation on the evolution of viscous overstability.
Note that the quantity  $\tau_{0}$ is not to be confused with the scaled surface density $\tau$ used elsewhere in this 
paper. In what follows we use $\tau_{0}=1.5$, and 
$L_{0}=2\,\text{km}$. 
The ground state surface mass density $\sigma\left(t,q\right)$ of the simulation region will vary in the same way as the optical depth (\ref{eq:taudyn}) and we 
adopt a time-averaged ground state surface mass density of $\sigma_{0}=300\,\text{kg}\,\text{m}^{-2}$.
Furthermore, the simulations utilize the velocity dependent normal coefficient of restitution by \citet{bridges1984}, while particle spins are neglected.

\begin{figure*}
\centering
\includegraphics[width = 0.83 \textwidth]{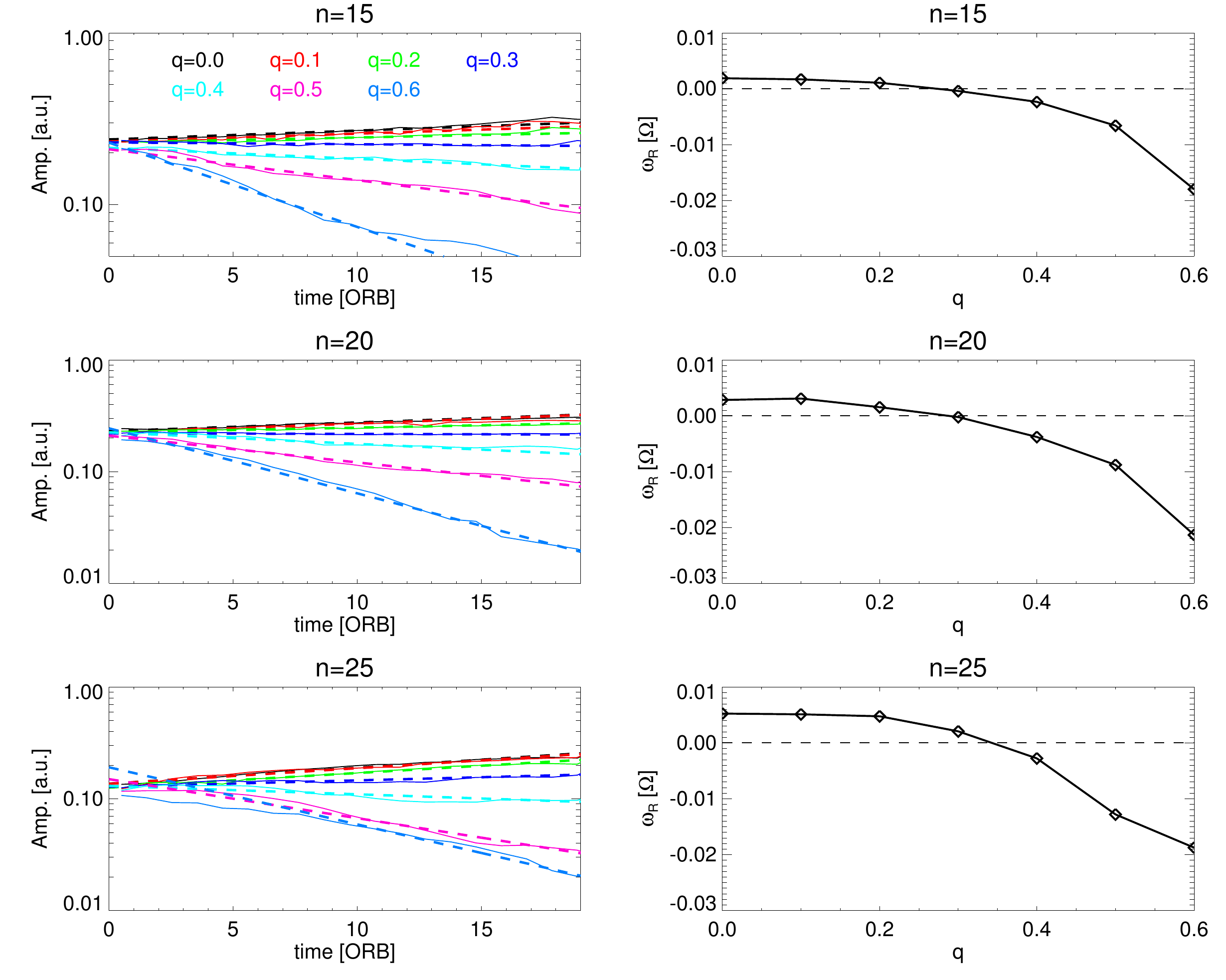}
\caption{Determination of linear growth rates of three different seeded overstable modes (indicated by their radial mode number $n$) in N-body 
simulations for different values of the nonlinearity 
parameter $q$, 
quantifying the amount of perturbation in the ground state that corresponds to a density wave. The left panels display the time evolution of amplitudes $A_{n}$ 
[Equation (\ref{eq:ampcalc})] with a sampling interval of one orbital period.
The right panels show the resulting growth rates, obtained from linear fits as drawn in the left frames. The simulations used time-averaged values of 
the ground state optical depth and surface mass density of $\tau_{0}=1.5$ and $\sigma_{0}=300\,\text{kg}\,\text{m}^{-2}$, respectively, as well as a vertical 
frequency enhancement of $\Omega_{z}/\Omega =2$ to mimic vertical self-gravity.}
\label{fig:mosqgr}
\end{figure*}

\begin{figure*}
\centering
\includegraphics[width = \textwidth]{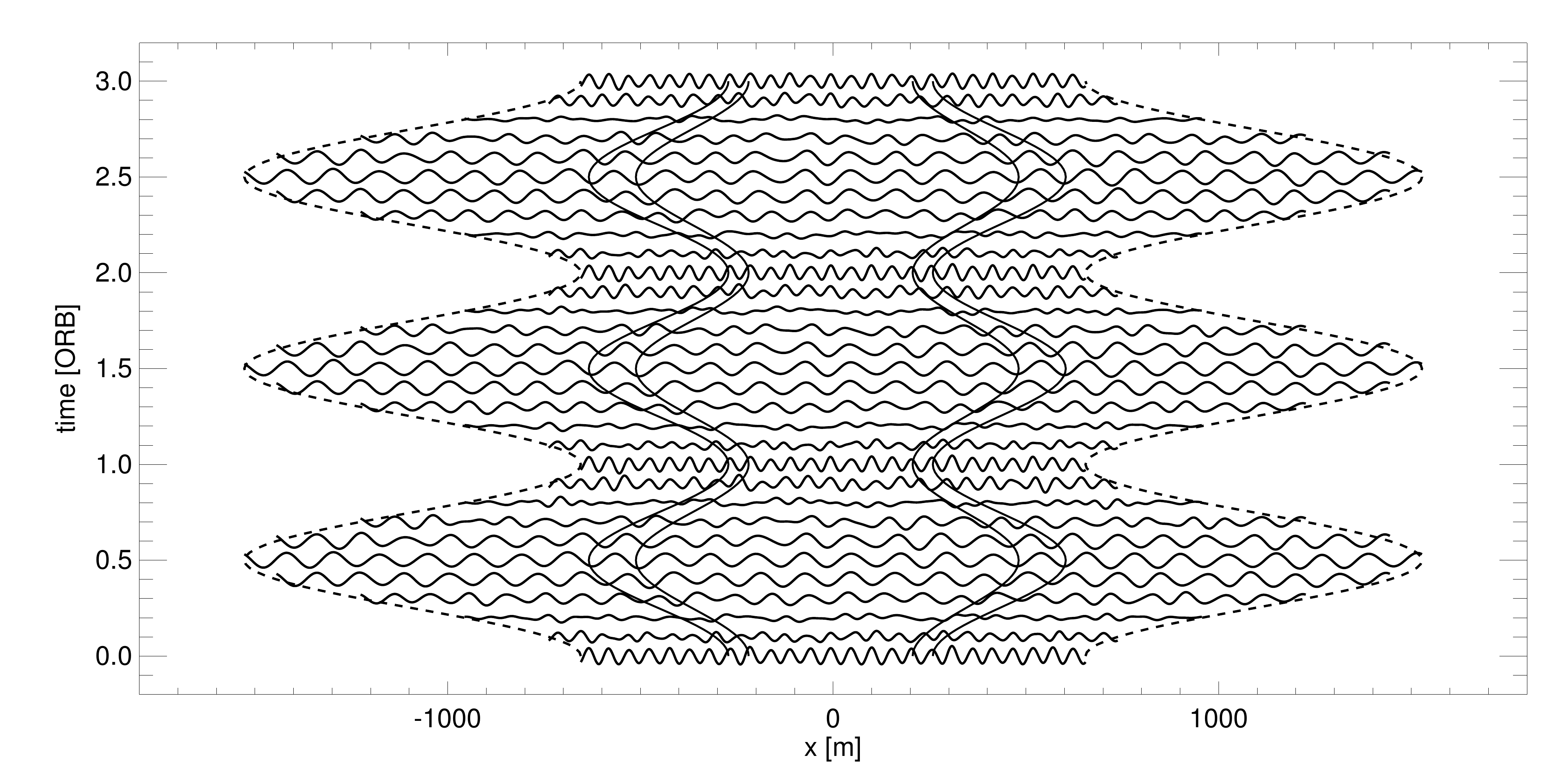}
\caption{The evolution of the radial velocity perturbation $u^{'}$ during the first three orbital periods of the N-body simulation with $n=25$ and 
$q=0.4$ of Figure \ref{fig:mosqgr}. Clearly visible is the periodic variation of the radial size of the simulation region according to 
Equation (\ref{eq:lxt}) which is drawn as dashed curves. Also indicated is the analogous radial variation of one wavelength of the seeded 
mode, represented by the two pairs of solid curves. Since the seed is a standing wave its amplitude undergoes an oscillation with twice the overstable 
wave frequency [cf.\ Equation (\ref{eq:omcrit})].}
\label{fig:nbodyevo}
\end{figure*}

Figure \ref{fig:mosqgr} shows measurements of linear growth rates of three different seeded overstable modes in N-body simulations using different 
values of the nonlinearity parameter $q=0-0.6$. 
The initial state corresponds to a standing linear overstable 
%
%
%
%
%
%
%
%
%
%
wave (see Equation (37) of \citet{schmidt2001b} expanded to the lowest order of the scaled wavenumber $k$) in a phase where only the perturbation in 
the radial velocity $u^{'}$ has a non-zero amplitude. 
When this radial perturbation velocity is seeded with small amplitude,
then the simulation practically starts on an overstable eigenvector of the linear hydrodynamic model. 
The radial modes $n=15-25$ correspond to time-averaged wavelengths $\langle\lambda\rangle_{t}=(L_{0}/n)/\sqrt{1-q^2} = (133\,\text{m}-80\,\text{m}) 
/\sqrt{1-q^2}$ [see Equation (\ref{eq:lxt})]. Figure \ref{fig:nbodyevo} illustrates the evolution the 
the radial velocity perturbation for the case $n=25$ with $q=0.4$ during the first three orbital periods.
The procedure for obtaining the growth rates is similar to that used by \citet{schmidt2001b} and LSS2017. However, in 
the present situation we need to take into account the varying size of the radial domain, i.e.\ we use Equation (\ref{eq:ampcalc}) to obtain the mode 
amplitude, where $\Psi$ is replaced by the tabulated radial velocity field $u^{'}(x)$.

In accordance with the hydrodynamic growth rates (Figure \ref{fig:grqvar}) 
the measured growth rates in N-body simulations (the right panels in Figure \ref{fig:mosqgr}) decrease with increasing magnitude of the perturbation, 
quantified through $q$. 
Note that the overstable modes considered here would be stable in the hydrodynamic model for all used values of $\beta$ (Figure \ref{fig:grqvar}).
Since our N-body simulations do not include particle-particle gravitational forces (but merely the radial component of its mean-field approximation), the 
attained (equilibrium) velocity dispersion takes considerably smaller values than what we adopt for our hydrodynamic integrations (Table \ref{tab:hydropar}). 
This is the main reason why viscous overstability in the N-body simulations considered here occurs on smaller wavelengths than in our hydrodynamic model 
(LSS2017).

\begin{figure*}
\centering
\includegraphics[width = 0.75 \textwidth]{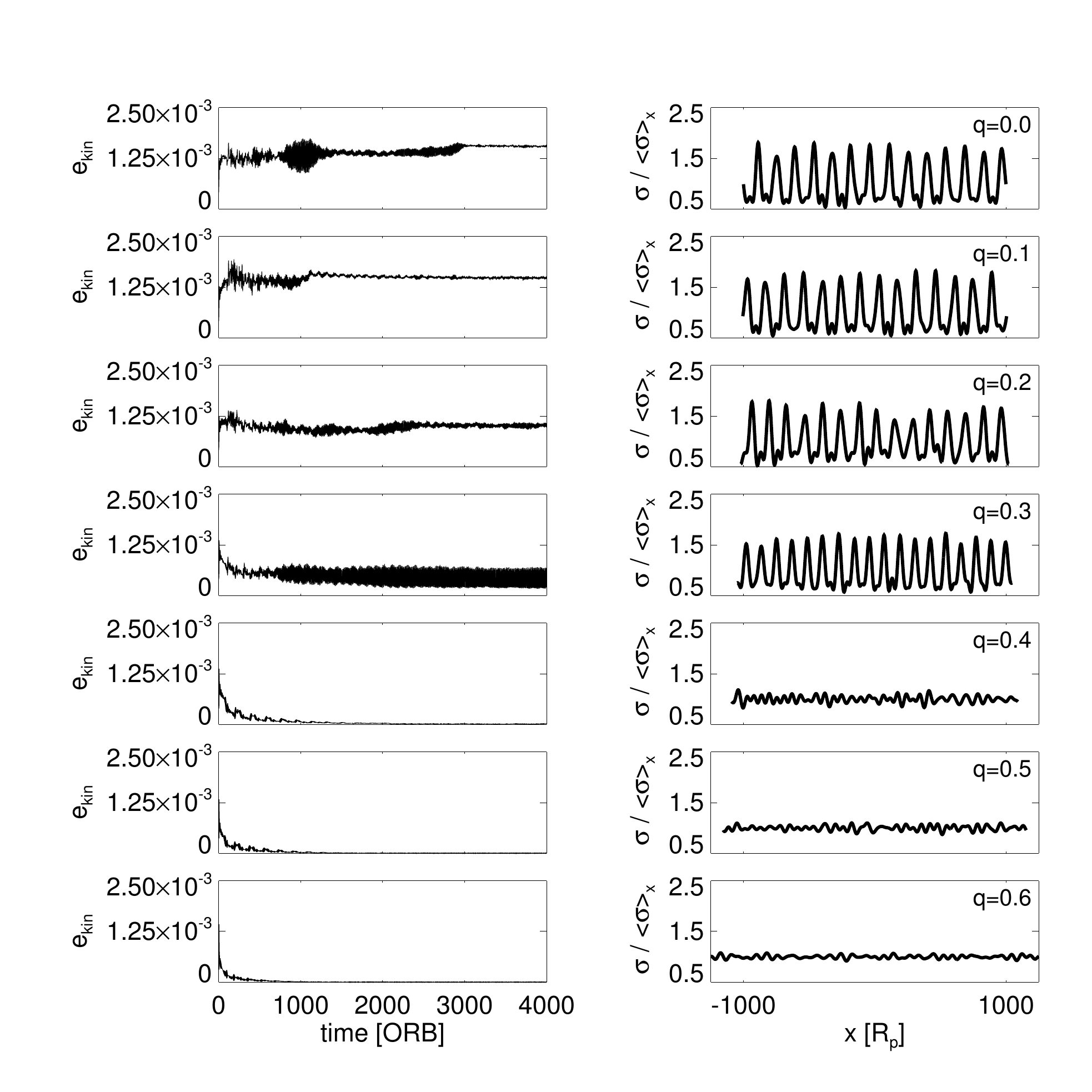}
\caption{Nonlinear evolution of viscous overstability in N-body simulations of a perturbed ring. \textit{Left}: Curves of kinetic energy 
density (\ref{eq:ekin}) of perturbations in units $\sigma_{0} R_{P}^2 \Omega^2$, excluding the ground state velocities (\ref{eq:gspert}). The curves are 
sampled at quadrature such that the radially averaged optical depth takes the value $\langle \tau^{dyn} \rangle_{x}=\tau_{0} \sqrt{1-q^2}$ with time-averaged 
ground state optical depth $\tau_{0}=1.5$.
\textit{Right}: Snapshots of the radial surface density profile for each $q$ taken at the same final time at quadrature where $\langle \tau^{dyn} 
\rangle_{x}=\tau_{0}\sqrt{1-q^2}$ and the radial width of the simulation region takes the value $2\,\text{km} / \sqrt{1-q^2}$. Note that the profiles are 
scaled with the radially averaged surface mass density $\langle \sigma \rangle_{x}$ which oscillates about its 
time-average $\sigma_{0}=300\,\text{kg}\,\text{m}^{-2}$ in the same way as (\ref{eq:taudyn}) .
Furthermore, a vertical frequency enhancement of $\Omega_{z}/\Omega =2$ was used.}
\label{fig:nlsatqvar}
\end{figure*}

For the same parameters Figure \ref{fig:nlsatqvar} illustrates the nonlinear saturation of overstability in simulations that started from white noise. The left 
column displays curves of the kinetic energy density of perturbations that have developed on top of the ground state (\ref{eq:gspert}).
The curves are sampled at quadrature (i.e.\ at times $t$ where $\Omega_{L} t = \pi/2 +  l 2 \pi  $  with integer $l$) where the radially averaged (mean) 
optical depth 
takes the value $\langle \tau^{dyn}\rangle_{x} =\tau_{0} \sqrt{1-q^2}$ so as to mask out the orbital oscillation due to the 
perturbation itself. 
While the curves for $q=0-0.2$ show a clear increase of kinetic energy with time, indicating the formation of nonlinear wavetrains, we observe a 
sharp drop in the kinetic energy densities for $q\geq 0.3$.
The right column shows snapshots of the radial particle number density profile near the end of each simulation where also
$\langle \tau^{dyn}\rangle_{x}=\tau_{0} \sqrt{1-q^2}$. In agreement with the energy curves we find clearly developed nonlinear overstable wavetrains for 
$q=0-0.2$, while for 
$q\geq 0.3$ the density profile develops into noise. Note that due to the particulate nature of the simulations 
a retaining noise level induced by the (strong)
perturbation is inevitable.
From these results we estimate a critical value $q_{c}\sim 0.4$ above which overstability is completely suppressed for the parameters used here.
%
%
%
%
%

%

%
%
%

%
%

%
%

\section{Discussion}\label{sec:disc}

We developed a one-dimensional hydrodynamical scheme to study the excitation of a resonantly forced spiral density wave in a dense planetary 
ring.
Due to the restriction to one space-dimension, the advection caused by orbital shear needs to be approximated. 
We constructed corresponding azimuthal derivative terms from the weakly nonlinear model of LSS2016. 
Profiles of nonlinear density waves in a viscously stable ring computed with our scheme show good agreement with those resulting from the models by 
BGT86 and 
LSS2016. 

We applied our scheme to investigate the damping behavior of spiral density waves in a planetary ring which is subject to viscous overstability.
The results of our large-scale hydrodynamical integrations confirm the observation that resonantly forced spiral density waves can co-exist with short-scale 
waves generated by the viscous overstability (\citet{hedman2014a}), an aspect not taken into account in existing models for the damping of density waves. 
Due to our approximation of the azimuthal derivative terms the free short-scale overstable modes appearing in our hydrodynamical integrations
are also non-axisymmetric with the same azimuthal periodicity $m$ as the spiral density wave. We have shown that the nonlinear evolution of these short-scale 
modes is 
very similar to that of the strictly axisymmetric short-scale overstable modes investigated in earlier studies. 

We find that the damping behavior of a spiral density wave can be very different from what is predicted by existing models, depending on its resonance 
strength. 
A sufficiently strong spiral density wave damps the short-scale viscous overstability. Furthermore, if the density wave is sufficiently 
strong and it is itself overstable it behaves according to the models by BGT86 and 
LSS2016 in that it retains a finite saturation amplitude at large distances from resonance.
If, on the other hand, the density wave is overstable and sufficiently weak the short-scale modes dominate and damp the density wave.

It should be noted that these results are quantitatively (but most likely not qualitatively) affected by the approximation of
the azimuthal derivatives in our numerical scheme. That is, although we have shown that this approximation works well if we consider a nonlinear density wave 
alone, or the nonlinear evolution of viscous overstability in absence of a density wave, it cannot be ruled out that in cases where both wave types co-exist
certain nonlinear terms in the hydrodynamic Equations (\ref{eq:nleq}) would produce \emph{spurious} quasi-resonant higher-order coupling terms between both 
wave types.
Such spurious terms would be quasi-resonant due to the approximation that \emph{all} terms in Equations (\ref{eq:nleq}) are assumed to have 
$m$-fold periodicity and the fact that the wavelength of the density wave is much greater than that of the short-scale overstable modes, at least in close 
vicinity of the resonance radius. It is very unlikely though that such terms would dominate the many \emph{physical} coupling terms. Therefore we believe that 
our findings are qualitatively correct despite the approximation of the azimuthal derivatives.

We verified the damping of viscous overstability by the density wave by performing N-body simulations as well as a linear hydrodynamic stability 
analysis of a simplified axisymmetric model for a 
ring perturbed by a nearby ILR. 
Our N-Body simulations, using modified initial and boundary conditions as 
introduced by \citet{mosqueira1996}, confirm the formation of nonlinear overstable wavetrains if the perturbation by the ILR is not too strong, as 
well as a complete suppression of overstability if the nonlinearity parameter $q$ associated with the perturbation exceeds a certain value.
Critical values of $q$ which result in a damping of viscous overstability obtained from our large-scale integrations compare well with those that follow from 
the linear stability analysis of the axisymmetric model. 
 Based on our results we conclude that the mitigation of 
viscous overstability by a density wave is due to a destruction of the phase relation of the oscillating angular momentum flux and the epicyclic 
oscillation 
associated with overstable waves. 
Note that a quantitative match of $q$-values in the aforementioned comparison should not be expected though. That is, on the one
hand the overstable waves 
found in our large-scale hydrodynamical integrations suffer from the relatively low spatial resolution of the computational grid, which is 
expected to reduce the estimates of $q_{c}$ from these integrations. 
Furthermore, the applied approximation of the azimuthal derivatives could affect these values as well.
On the other hand, the neglect of the variation of the phase angle $\Delta$ due to the 
density wave in the linear stability analysis is most likely not 
justified in the far 
wave region of a density wave. It is not clear how this affects the computed growth rates of overstable modes and associated values of $q_{c}$.

Although we understand the mitigation of viscous overstability by a density wave,
an explanation for the damping of overstable density waves, such as those 
presented in the the last panel of Figure \ref{fig:pr76waveletbetvarhr} and the first three panels of \ref{fig:pr76waveletbet125hr}, remains to be sought for. 
One difficulty is that in this case the nonlinear interaction of the two different modes needs to be considered. It is noteworthy that the 
observed density waves in Saturn's A ring associated with the 7:6 ILR and the 10:9 ILR with the moons Atlas and Pan, respectively, seem to correspond to this 
case (see \citet{hedman2014a}, Figure 5).

Furthermore, it should be noted that due to the neglect of particle-particle self-gravity in our modeling, self-gravitational wakes (\citet{salo1992b}) do not 
form.
In principle their effect on the density wave profile may be described in terms of a gravitational viscosity (\citet{daisaka2001}).
In parts of Saturn's dense rings (particularly the A ring) it is expected that this gravitational viscosity is the dominant mode of viscosity.
However, the wakes will interact with viscous overstability in a more or less complex manner (\citet{salo2001,ballouz2017}) and as such they
will indirectly affect the damping of a density wave.
Moreover, in the regions of strong density waves the size of self-gravitational wakes is expected to be much increased.
That is, the ''straw``-like structures observed in optical Cassini images of strong density waves 
(e.g.\ \citet{tiscareno2017}), are believed to represent self-gravity wakes of kilometer length 
scales. 
These length scales are much greater than the typical length scale of self-gravity wakes that can form in an unperturbed planetary ring 
(\citet{salo1992b}). 
An increased size of self-gravity wakes in the troughs of nonlinear density waves is also expected from N-body simulations (\citet{salo2014})
and theoretical studies (\citet{stewart2017}).
\citet{stewart2017} has shown that the characteristic length scale of self-gravitational perturbations (the
Toomre-wavelength) is greatly enhanced in the troughs of strongly 
nonlinear density waves.
This result suggests that the gravitational viscosity, which scales with the square of the Toomre-wavelength (\citet{daisaka2001}), can be greatly enhanced in 
the wave region,
consequently leading a stronger damping of the density wave.

Our numerical scheme allows for in principle straightforward extensions, such as the inclusion of the energy equation by using the numerical method of 
LSS2017. Ultimately, a two-dimensional scheme should be developed to overcome the necessity to approximate the orbital advection 
terms.

\section*{Acknowledgments}
We thank the reviewer Glen Stewart for helpful and constructive comments. We are grateful to Pierre-Yves Longaretti 
for discussions that greatly improved the manuscript.
We acknowledge support from the Academy of Finland. ML acknowledges funding from the University of Oulu Graduate School 
and the University of Oulu Scholarship Foundation.

%
%

\clearpage

\appendix

\section{Figures of Section 7.2}\label{sec:appendixb}
\begin{figure}[ht!]
\centering
\includegraphics[width = 0.42 \textwidth]{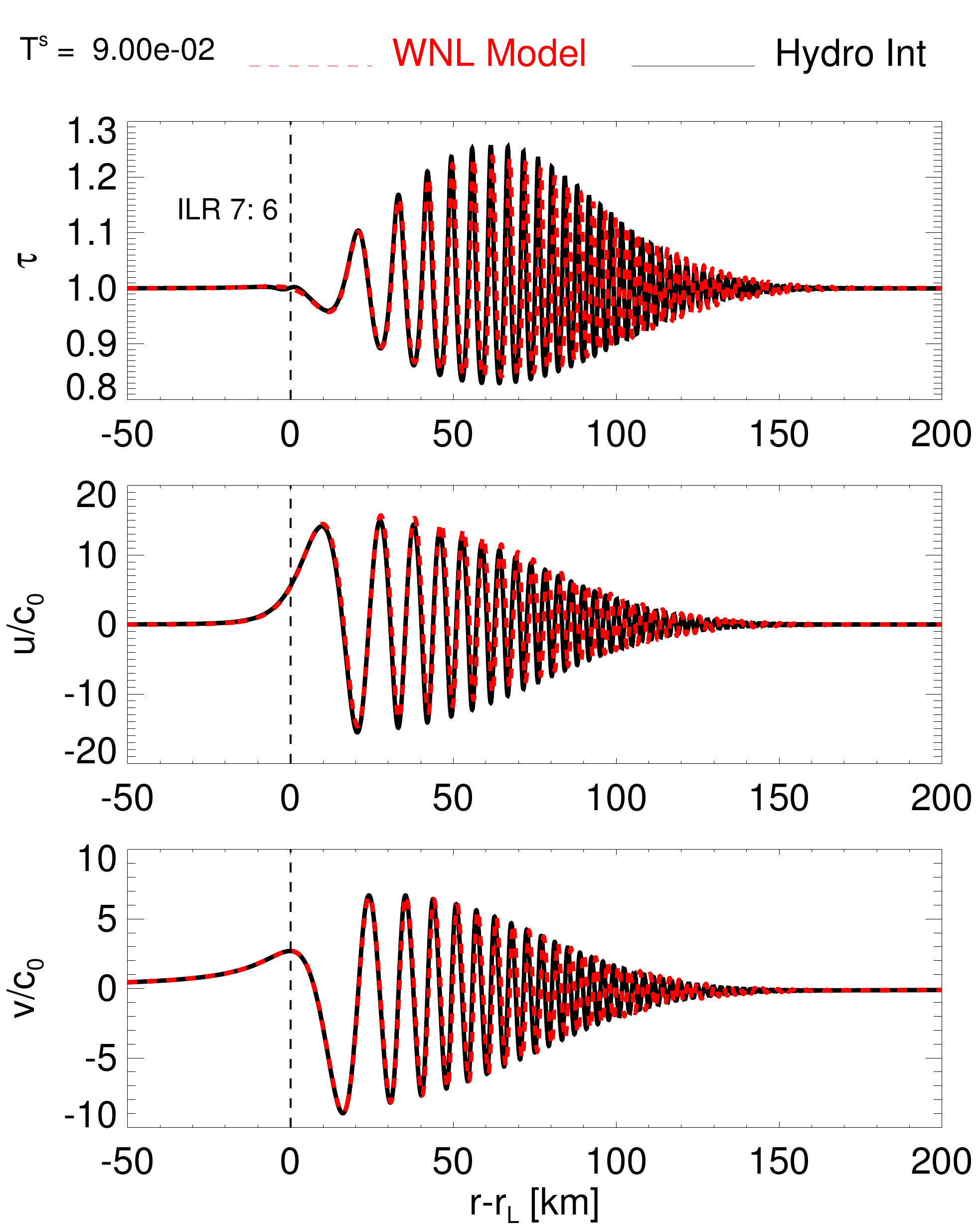}
\includegraphics[width = 0.42 \textwidth]{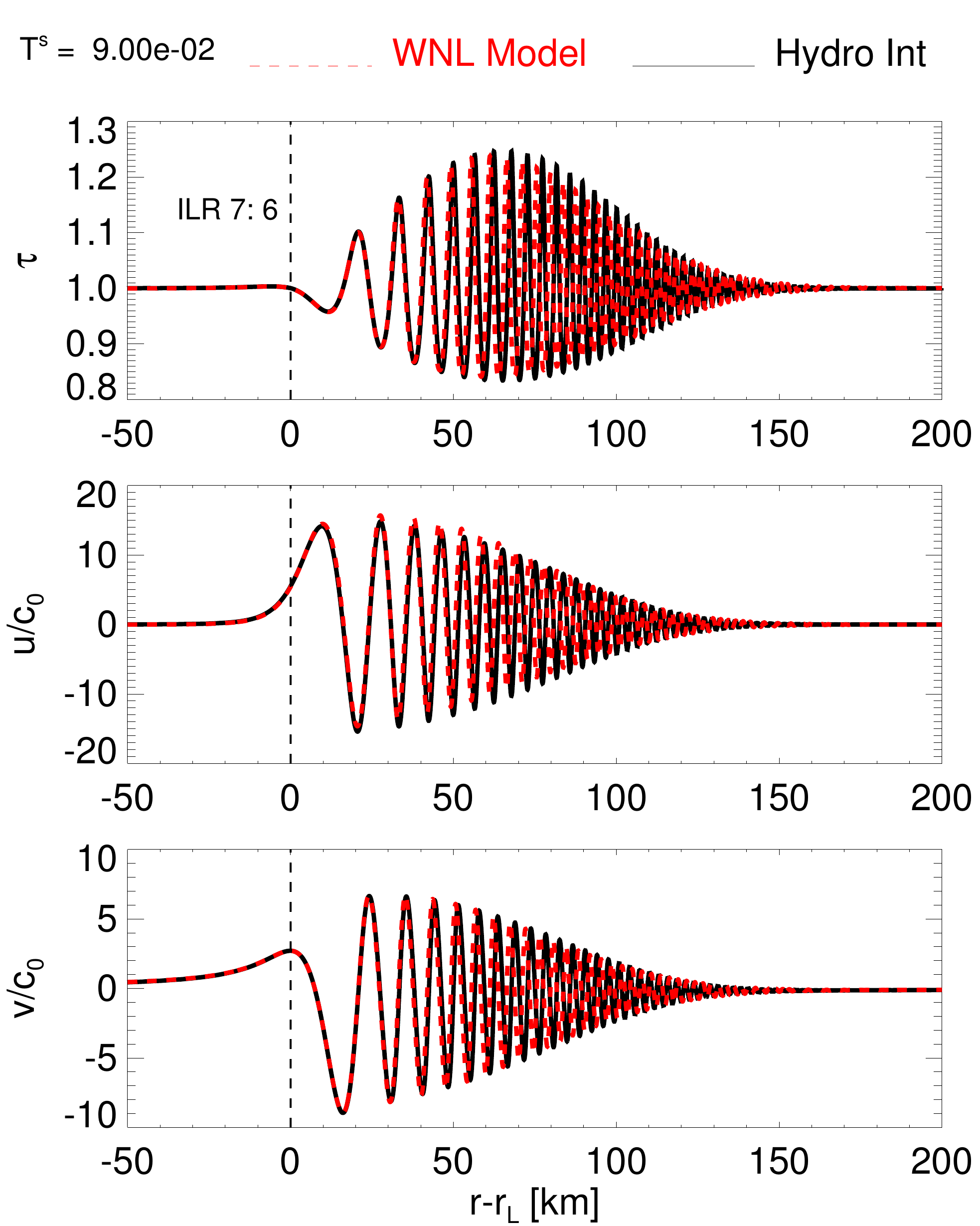}\\
\includegraphics[width = 0.42 \textwidth]{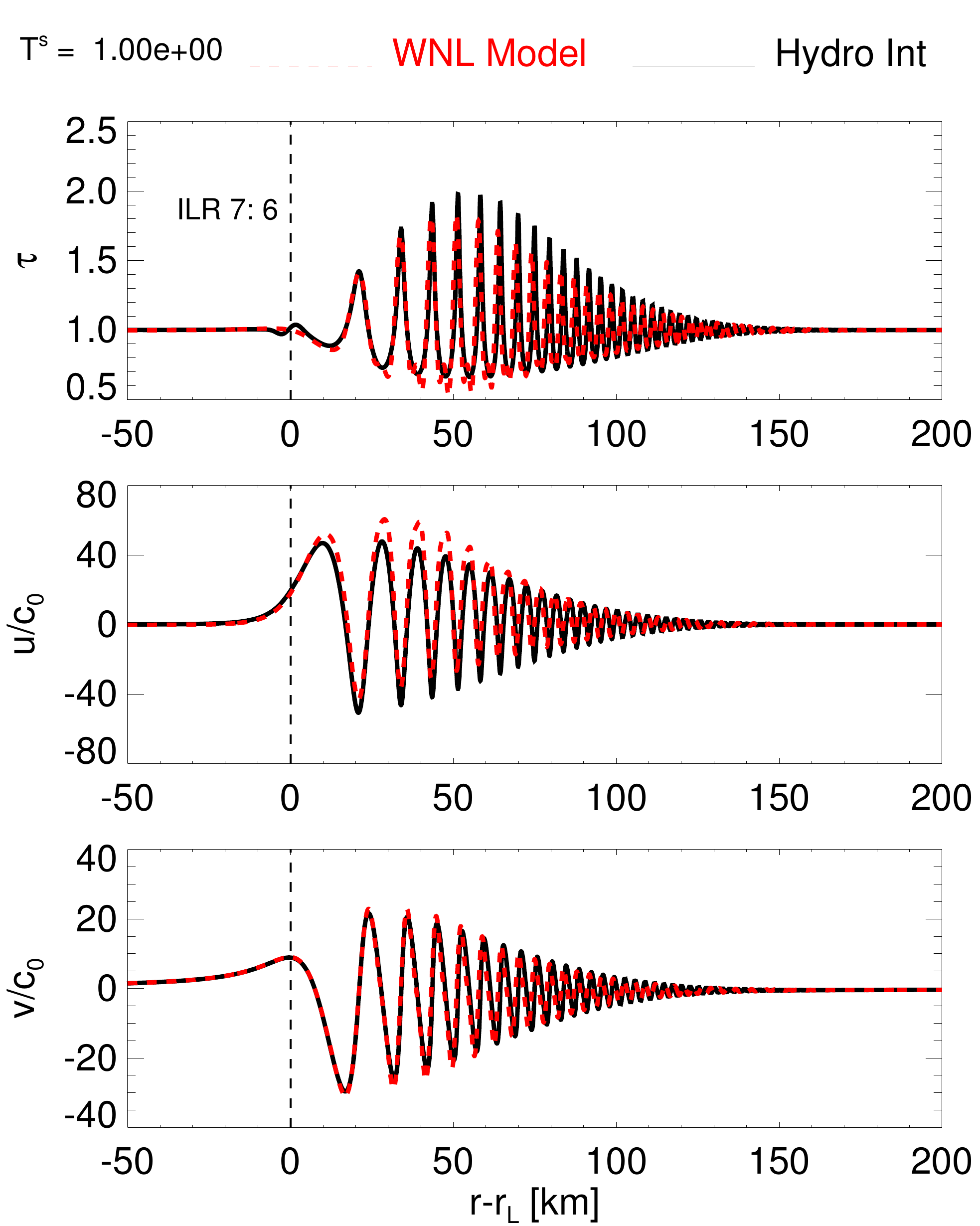}
\includegraphics[width = 0.42 \textwidth]{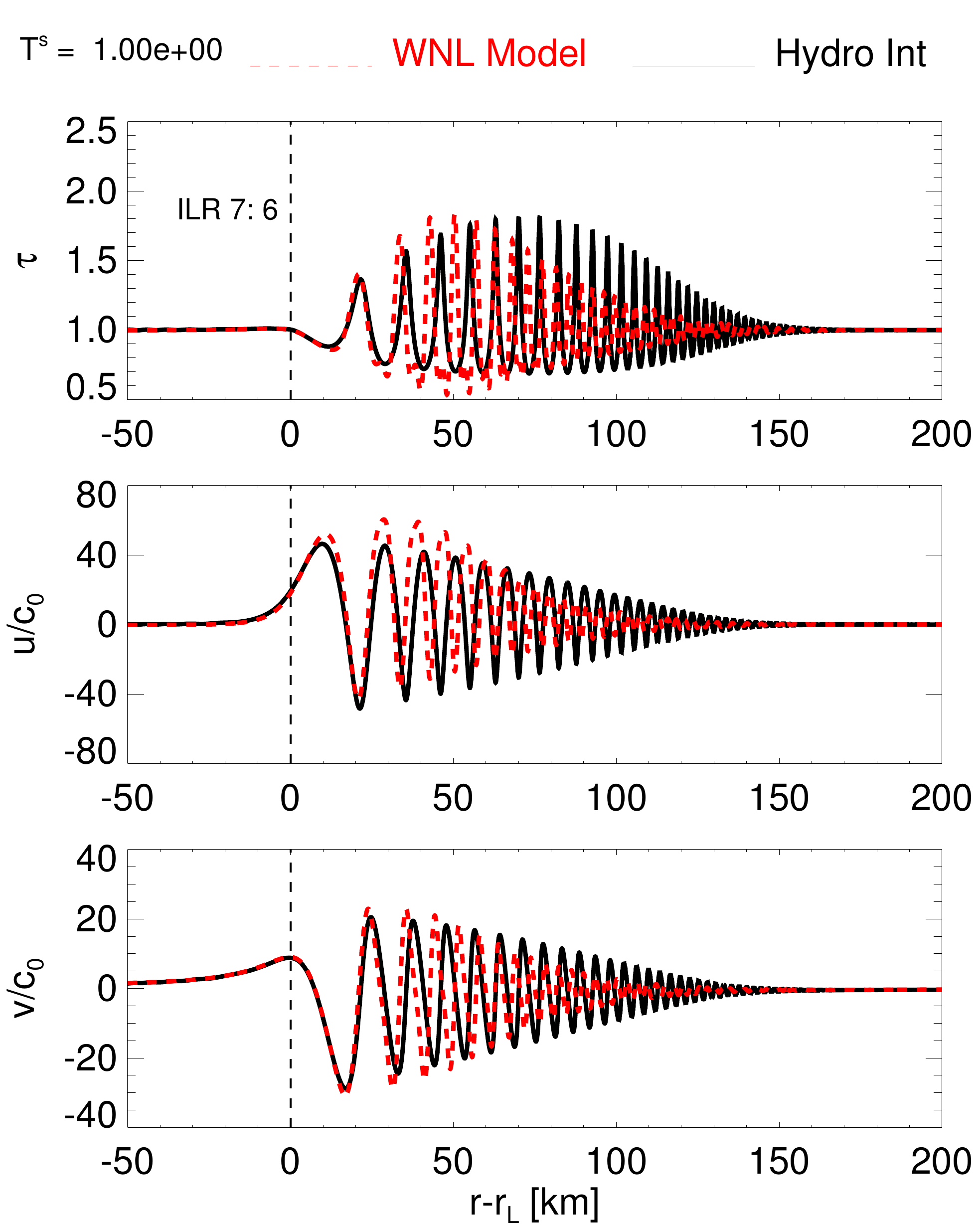}
\caption{Comparison of state variables resulting from hydrodynamical integrations and the WNL model using the $Pr76$-parameters with 
$\tilde{T}^{s}=9\cdot 10^{-2}$ (top panels) and $\tilde{T}^{s}=1$ (bottom panels). The integration shown in the left (right) panels applied \emph{method A} 
(\emph{method B}) for the azimuthal derivatives (Section \ref{sec:azideriv}).}
\label{fig:pr76f0103}
\end{figure}
\FloatBarrier
\clearpage
\begin{figure*}
\vspace{-0.5cm}
\centering
\includegraphics[width = 0.45 \textwidth]{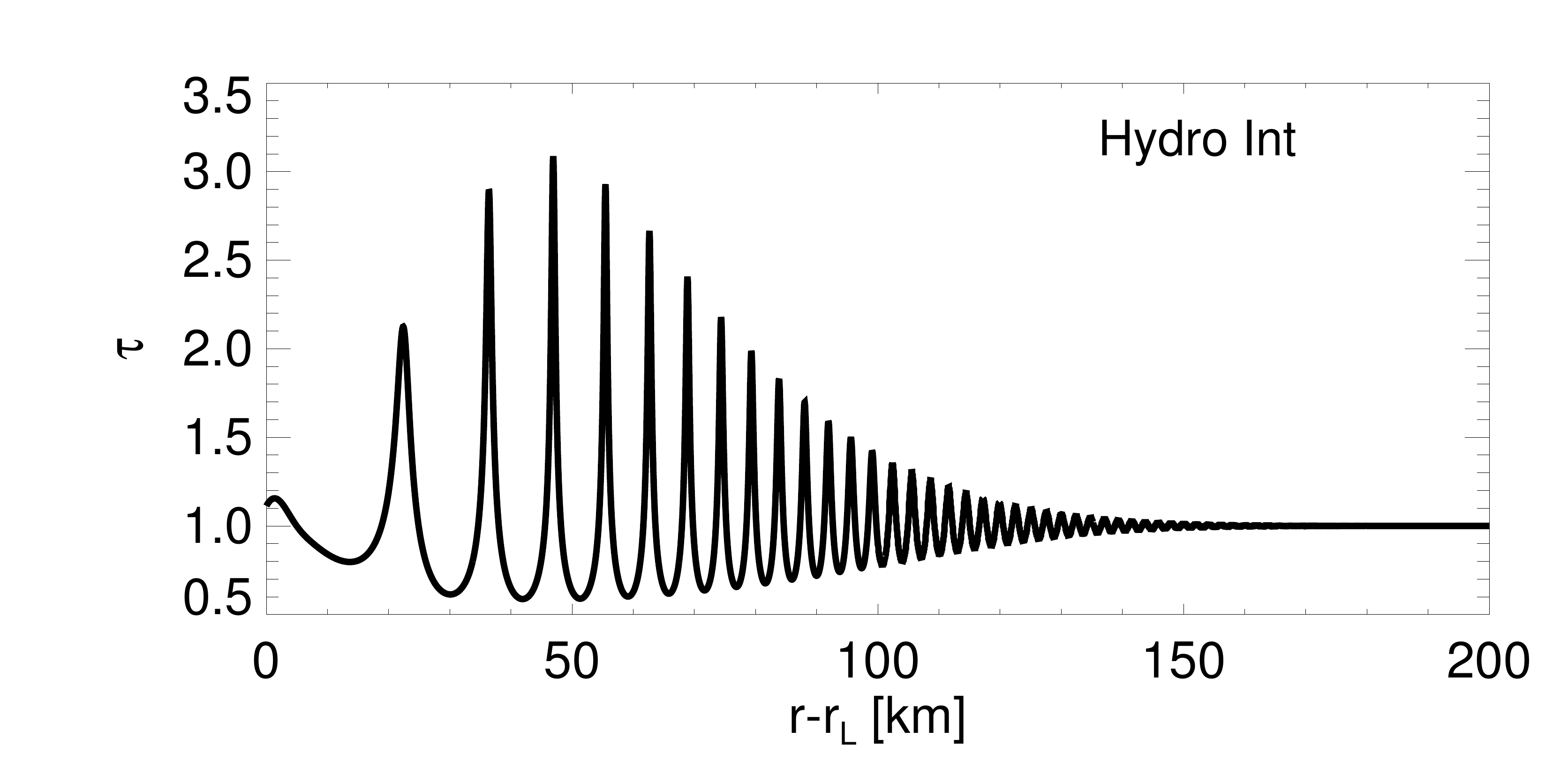}
\includegraphics[width = 0.45 \textwidth]{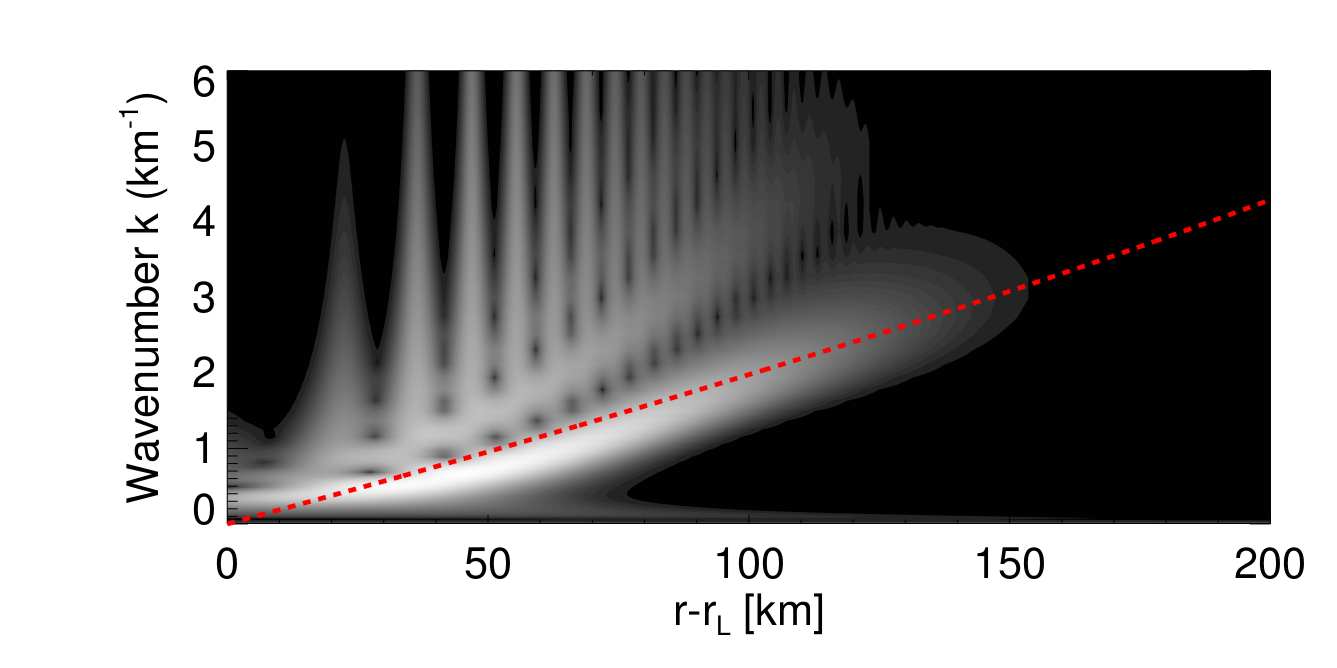}\\
\includegraphics[width = 0.45 \textwidth]{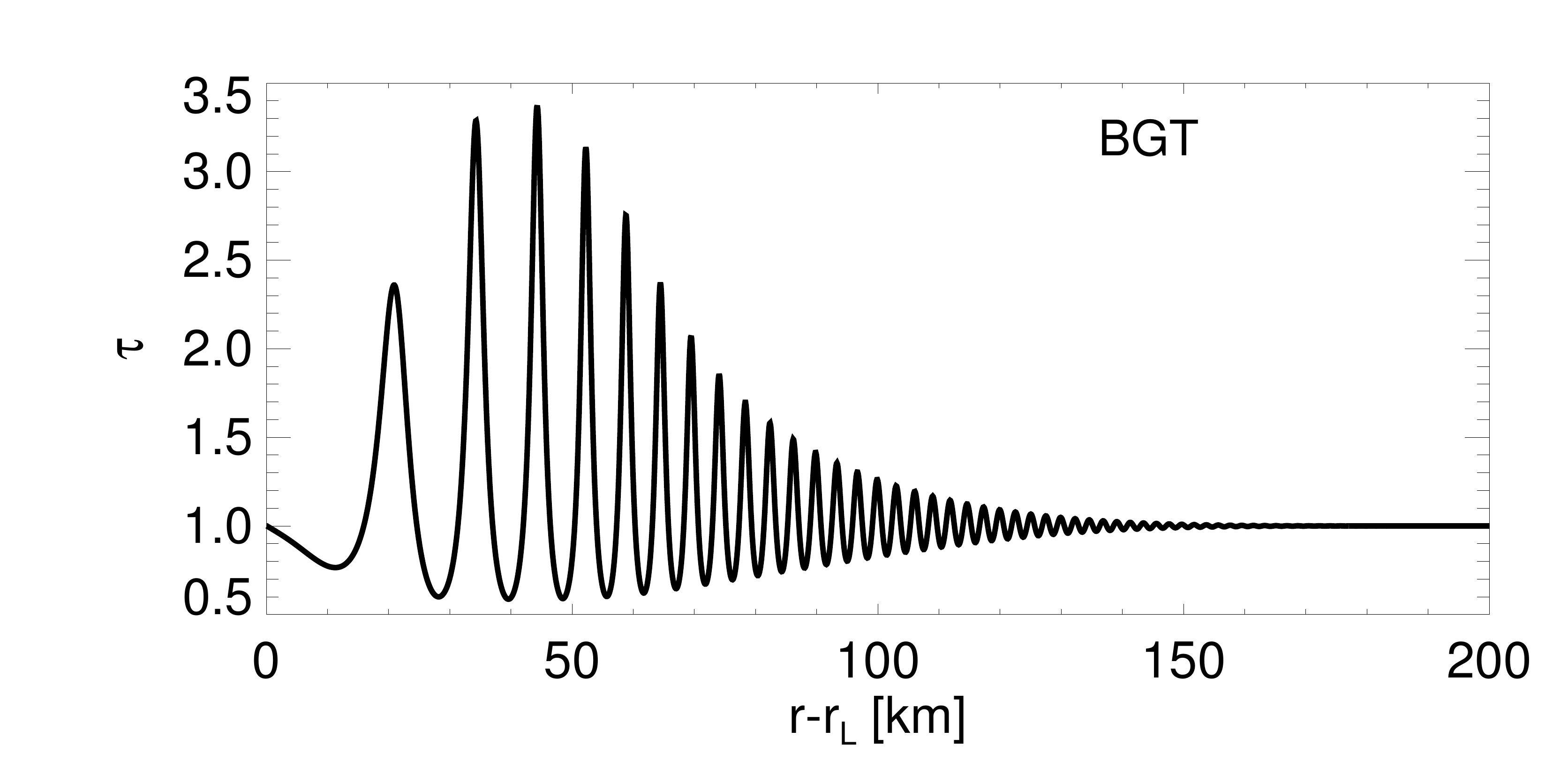}
\includegraphics[width = 0.45 \textwidth]{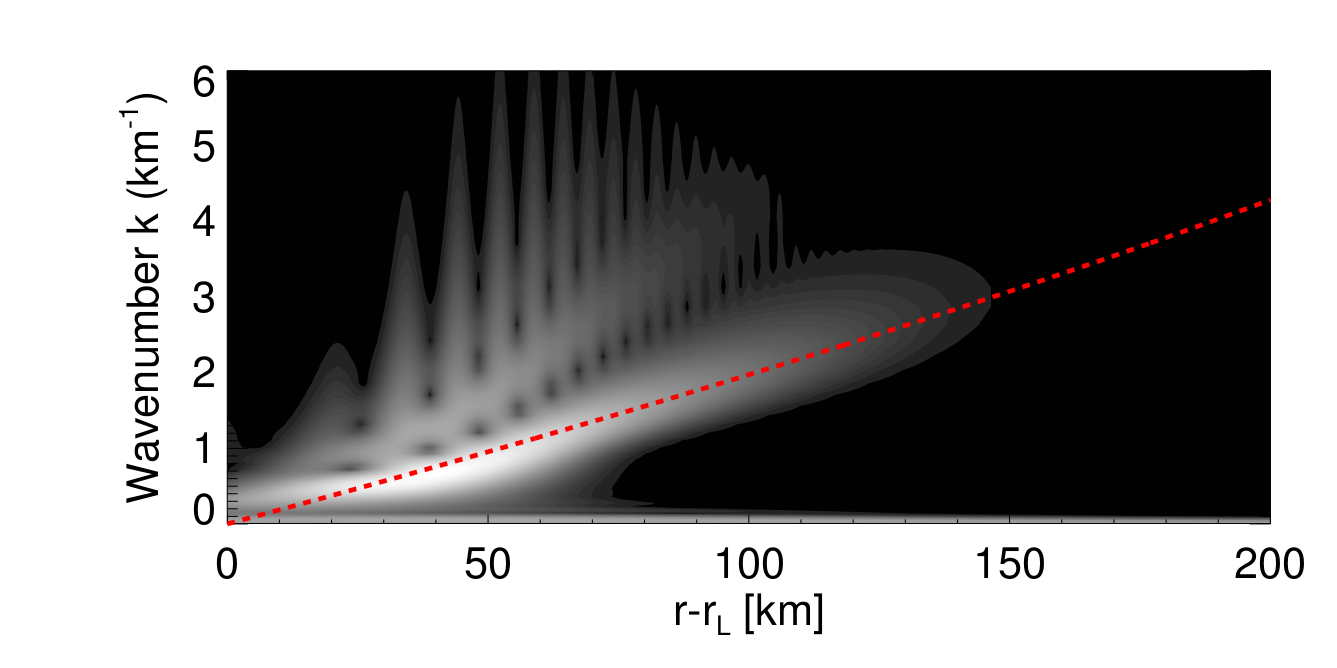}\\
\includegraphics[width = 0.45 \textwidth]{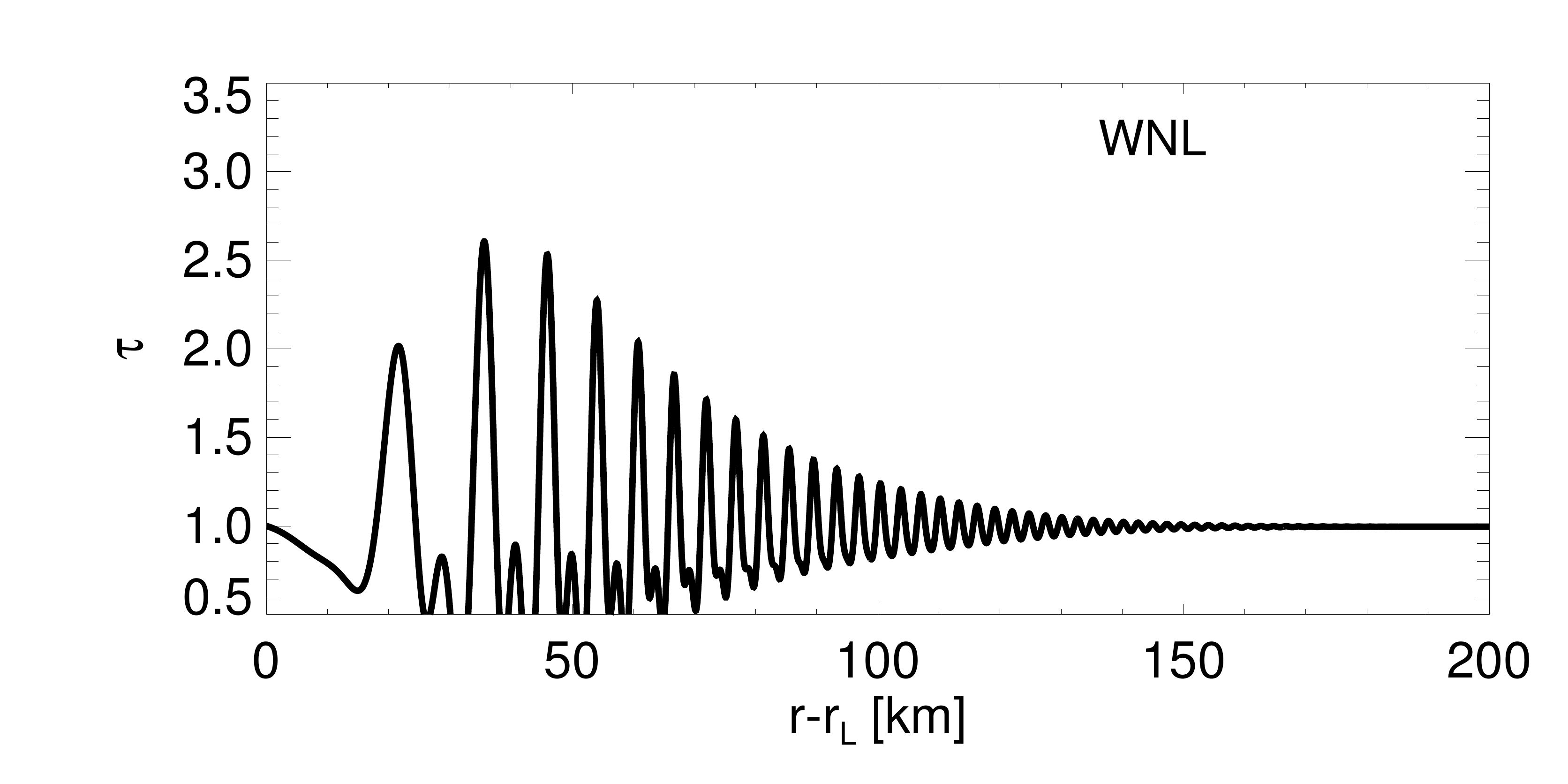}
\includegraphics[width = 0.45 \textwidth]{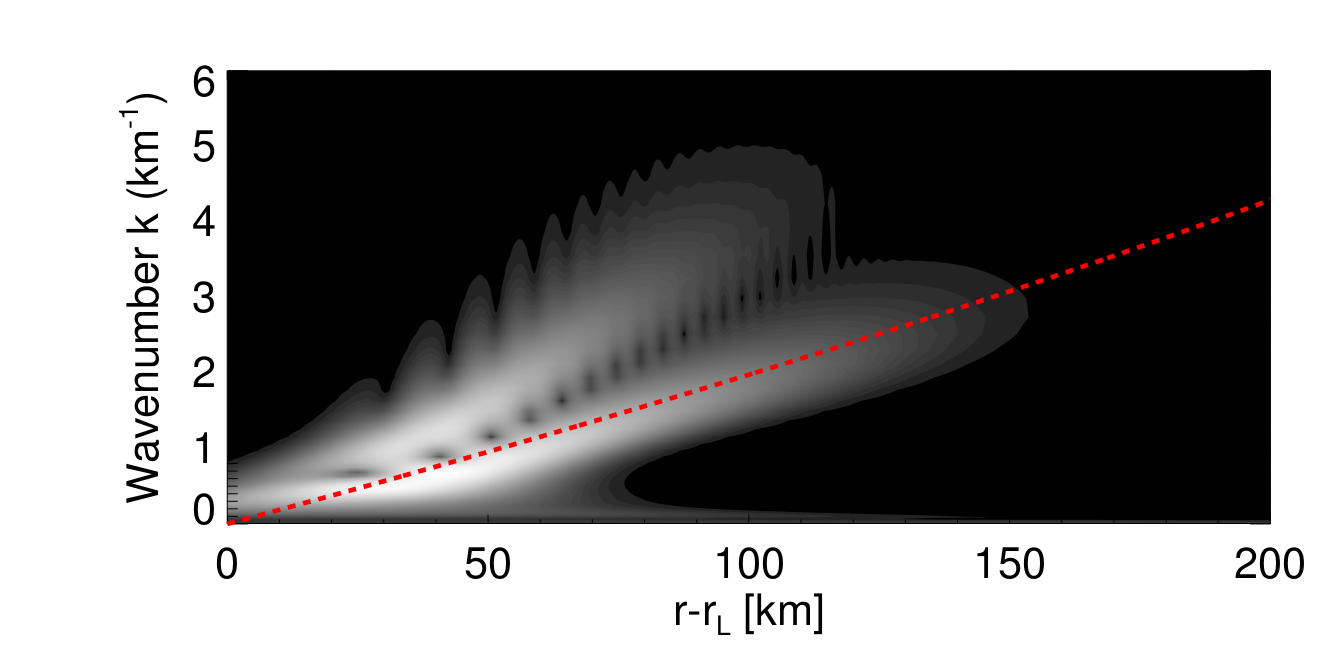}\\
\caption{Comparison of profiles of $\tau$ along with their Morlet wavelet powers resulting from a hydrodynamical integration 
and the WNL and BGT models using the $Pr76$-parameters with $\tilde{T}^{s}=4$. The dashed red lines represent the linear dispersion relation 
(\ref{eq:disprel}). }
\label{fig:pr76waveletf20}
\end{figure*}
\begin{figure*}
\vspace{-0.4cm}
\centering
\includegraphics[width = 0.45 \textwidth]{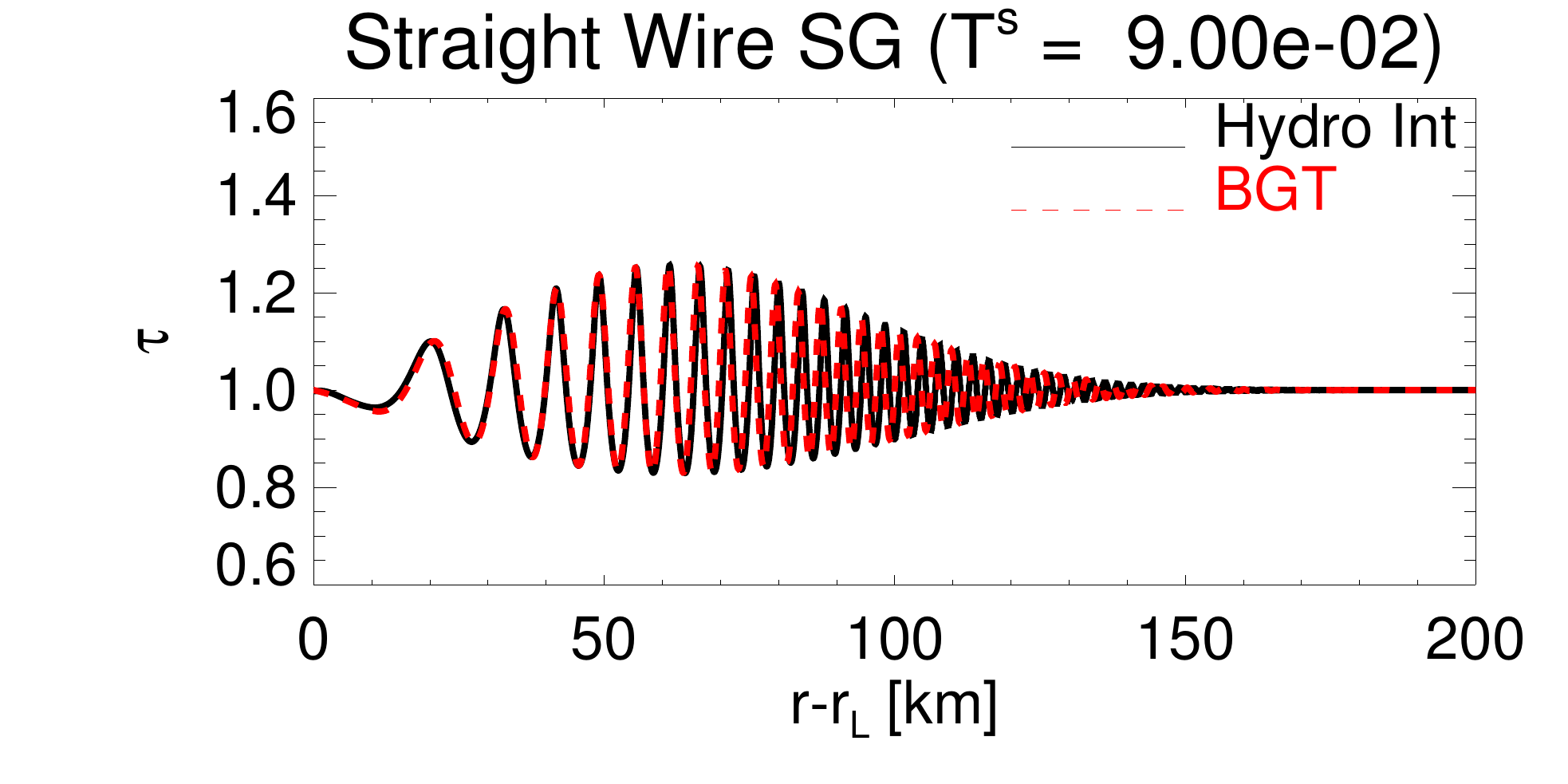}
\includegraphics[width = 0.45 \textwidth]{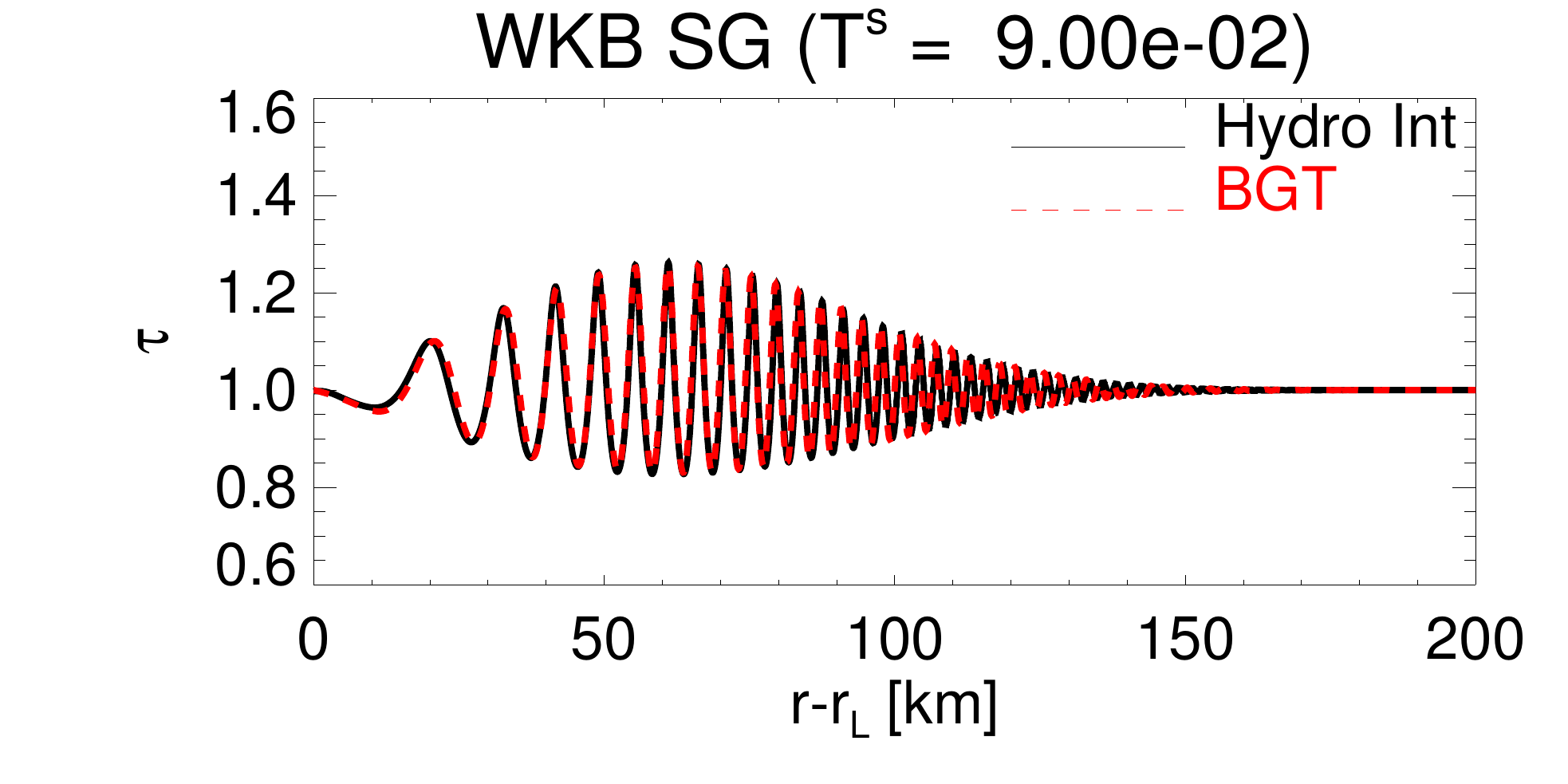}\\
\includegraphics[width = 0.45 \textwidth]{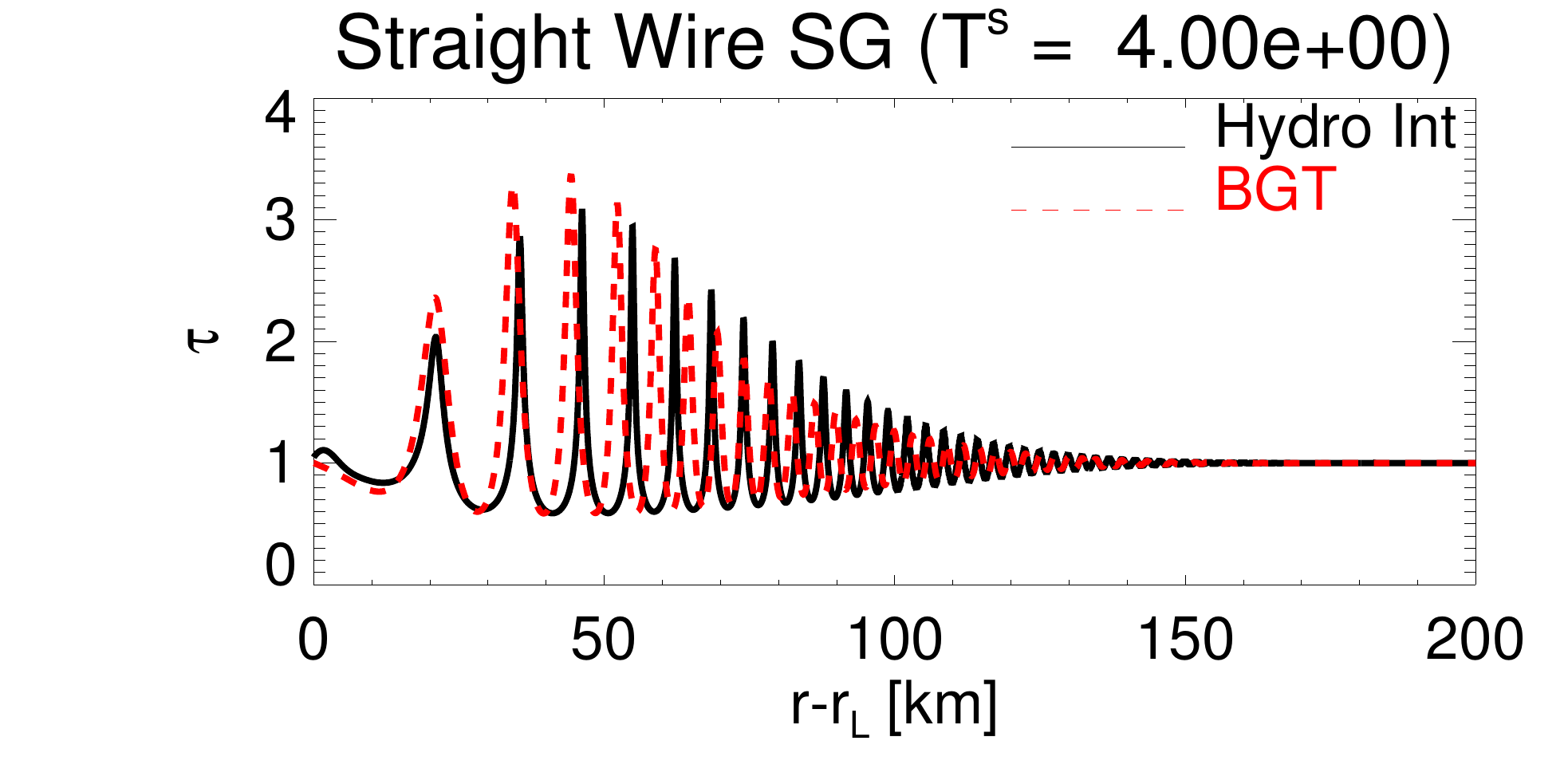}
\includegraphics[width = 0.45 \textwidth]{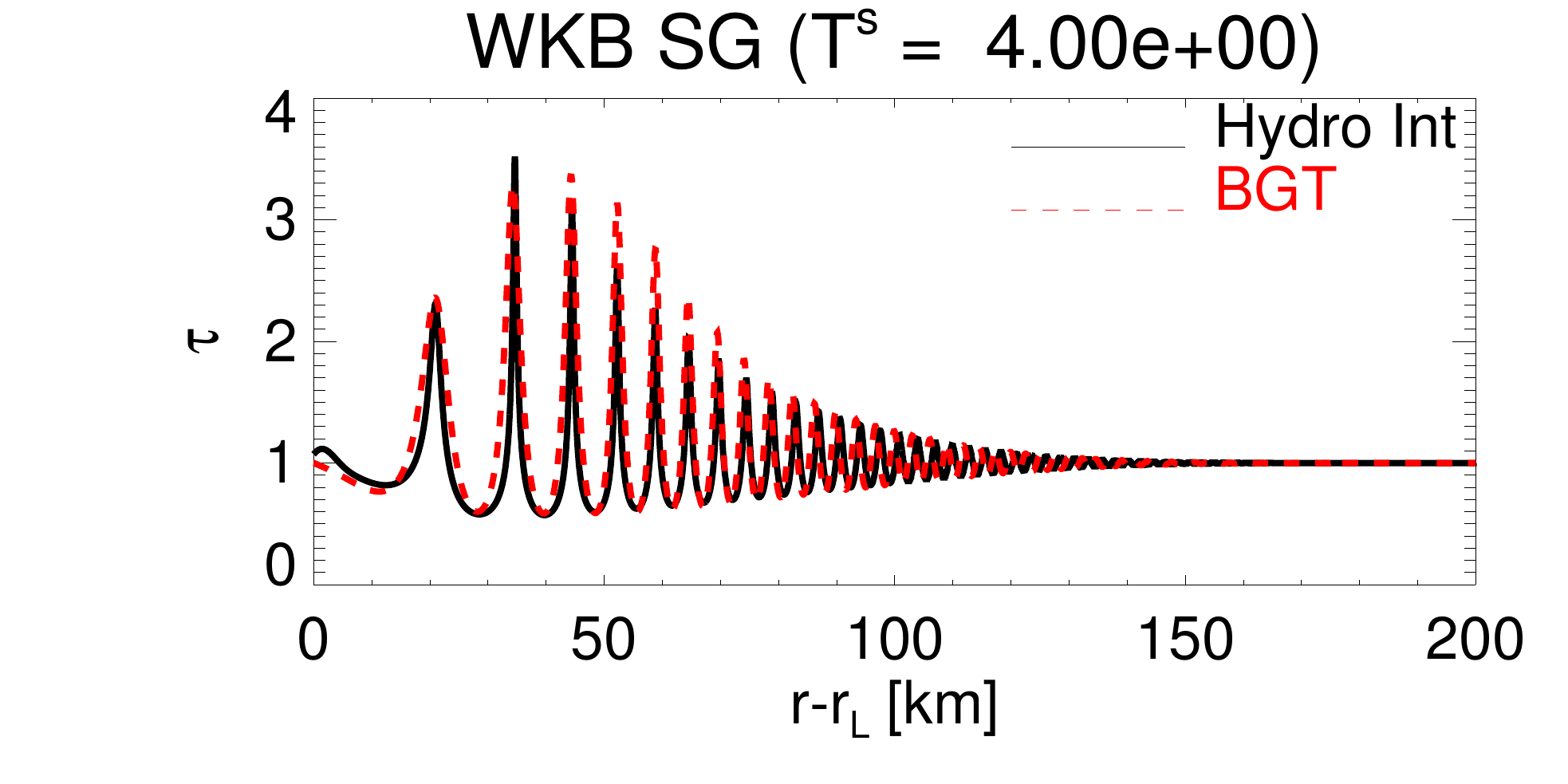}
\caption{Comparison of profiles of $\tau$ resulting from integrations with the \emph{Straight Wire} self-gravity (left column) and the WKB self-gravity (right 
column) with 
corresponding waves of the BGT model. Note that the WKB-approximation is intrinsic to the BGT model.}
\label{fig:sgcompf10}
\end{figure*}
 \FloatBarrier
%
%

%
\section{Figures of Section 7.3}\label{sec:appendixc}
\begin{figure}[h!]  
\vspace{-0.5cm}
\centering
\includegraphics[width = 0.43 \textwidth]{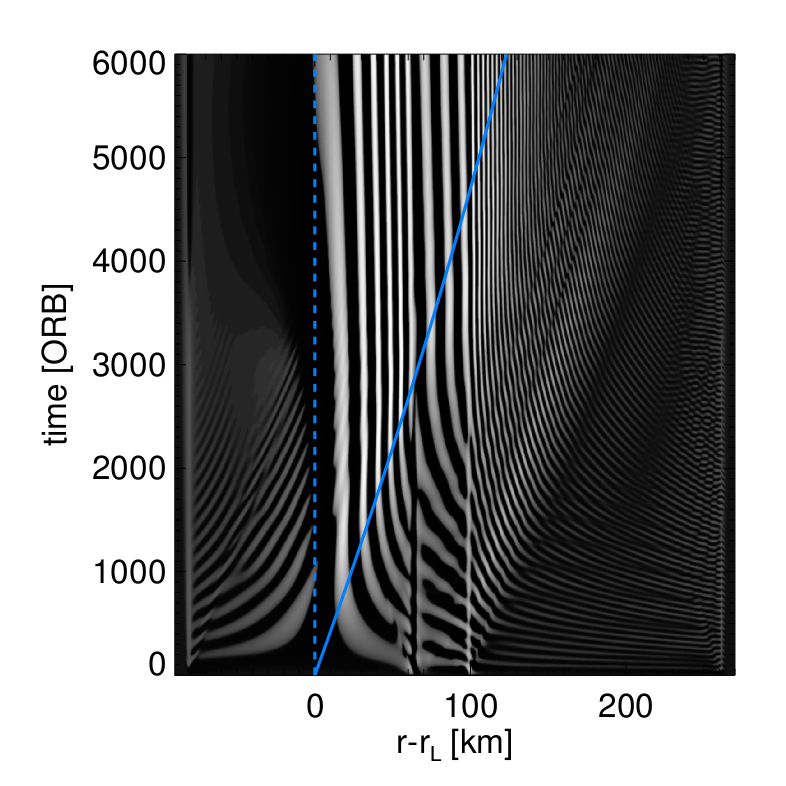}
\includegraphics[width = 0.43 \textwidth]{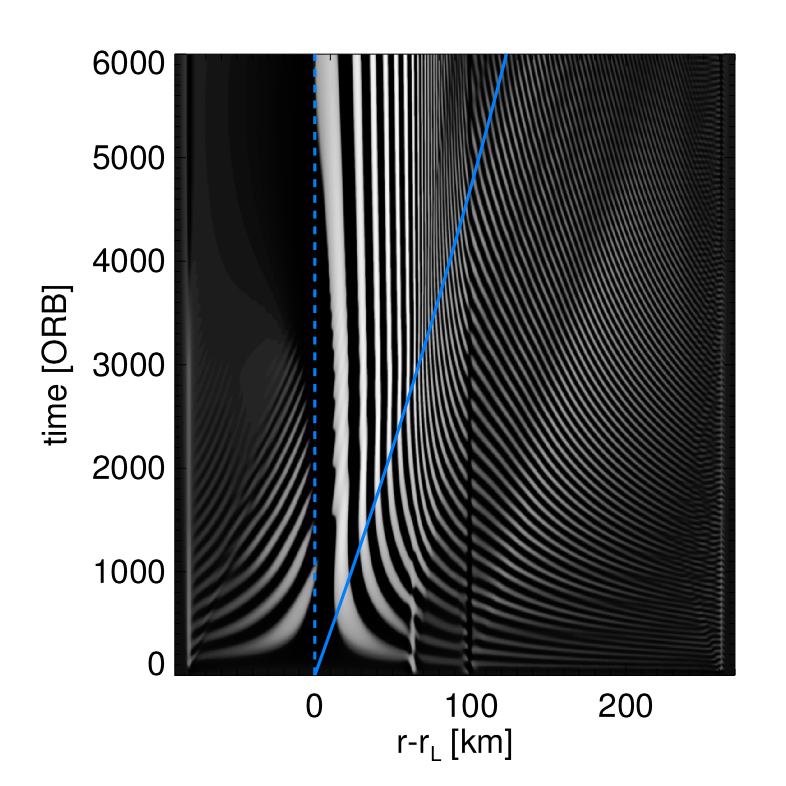}
\caption{Space-time plots of a $m=7$ density wave with scaled torque $\tilde{T}^{s}=9 \cdot 10^{-2}$ passing through a region ($r-r_{L}\sim 60-100\,\text{km}$) 
of increased ($\tau_{0}=3$, left frame) and decreased ($\tau_{0}=0.5$, right frame) equilibrium surface mass density. As in Figures 
\ref{fig:sptdpr76f001}-\ref{fig:sptdpr76f20} the blue solid curve indicates the expected curve of the density wave front in a homogeneous ring [Equation 
(\ref{eq:vg})]. Note that these plots only show the density perturbation on top of the background density.}
\label{fig:sptdbargap}
\end{figure}
\begin{figure}[ht!]  
\centering
\vspace{-1cm}
\includegraphics[width = 0.41 \textwidth]{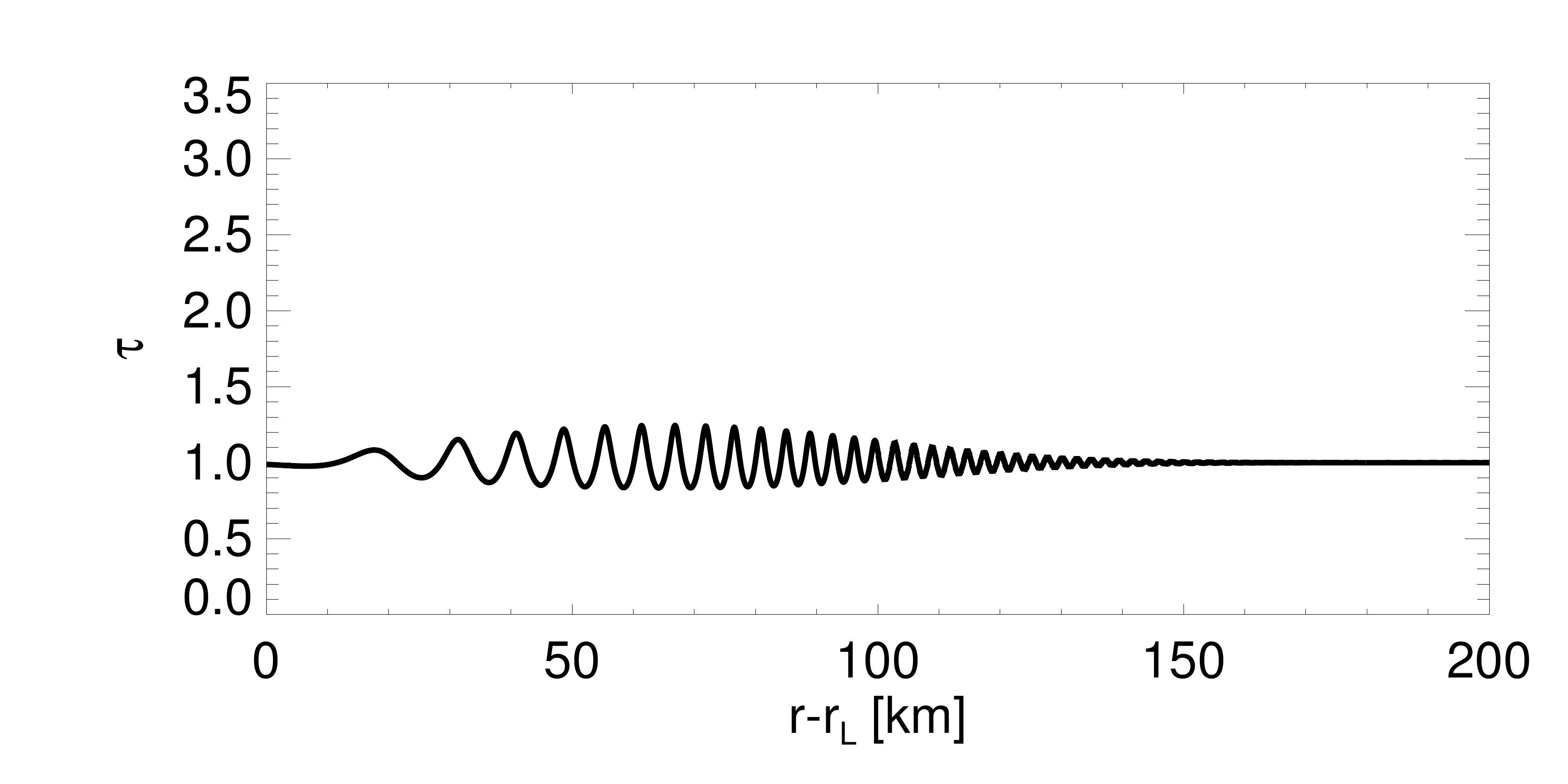}
\includegraphics[width = 0.41 \textwidth]{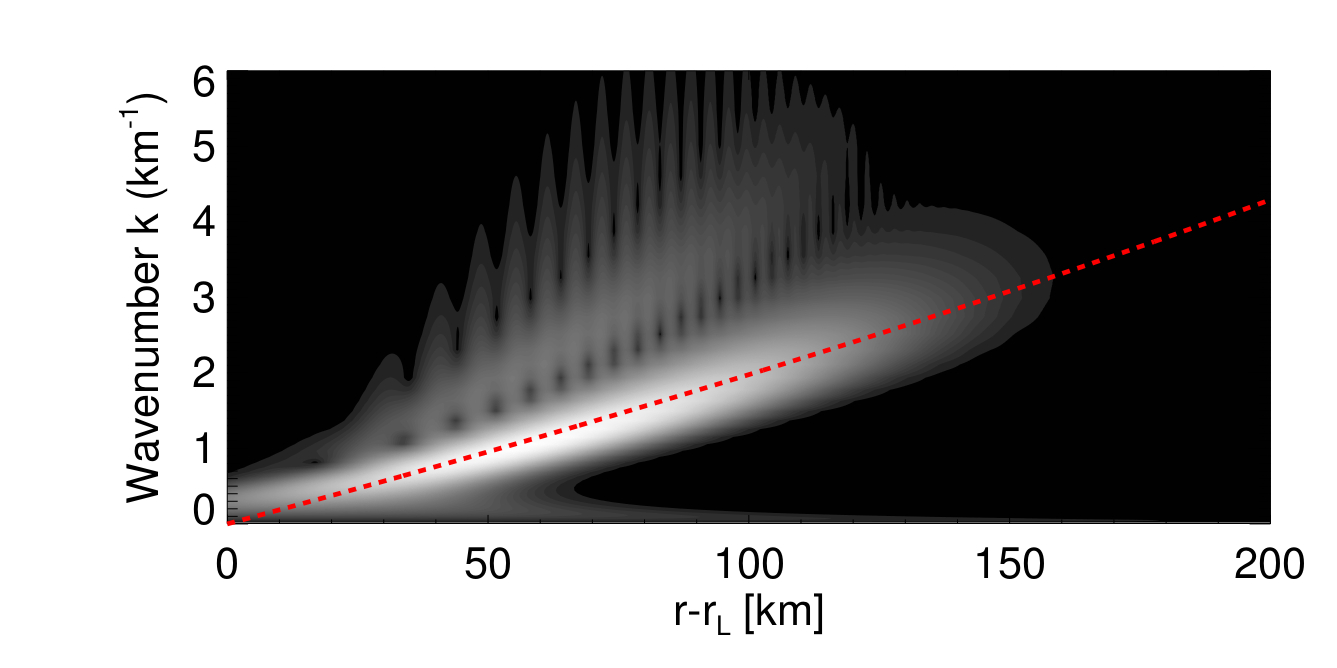}\\
\includegraphics[width = 0.41 \textwidth]{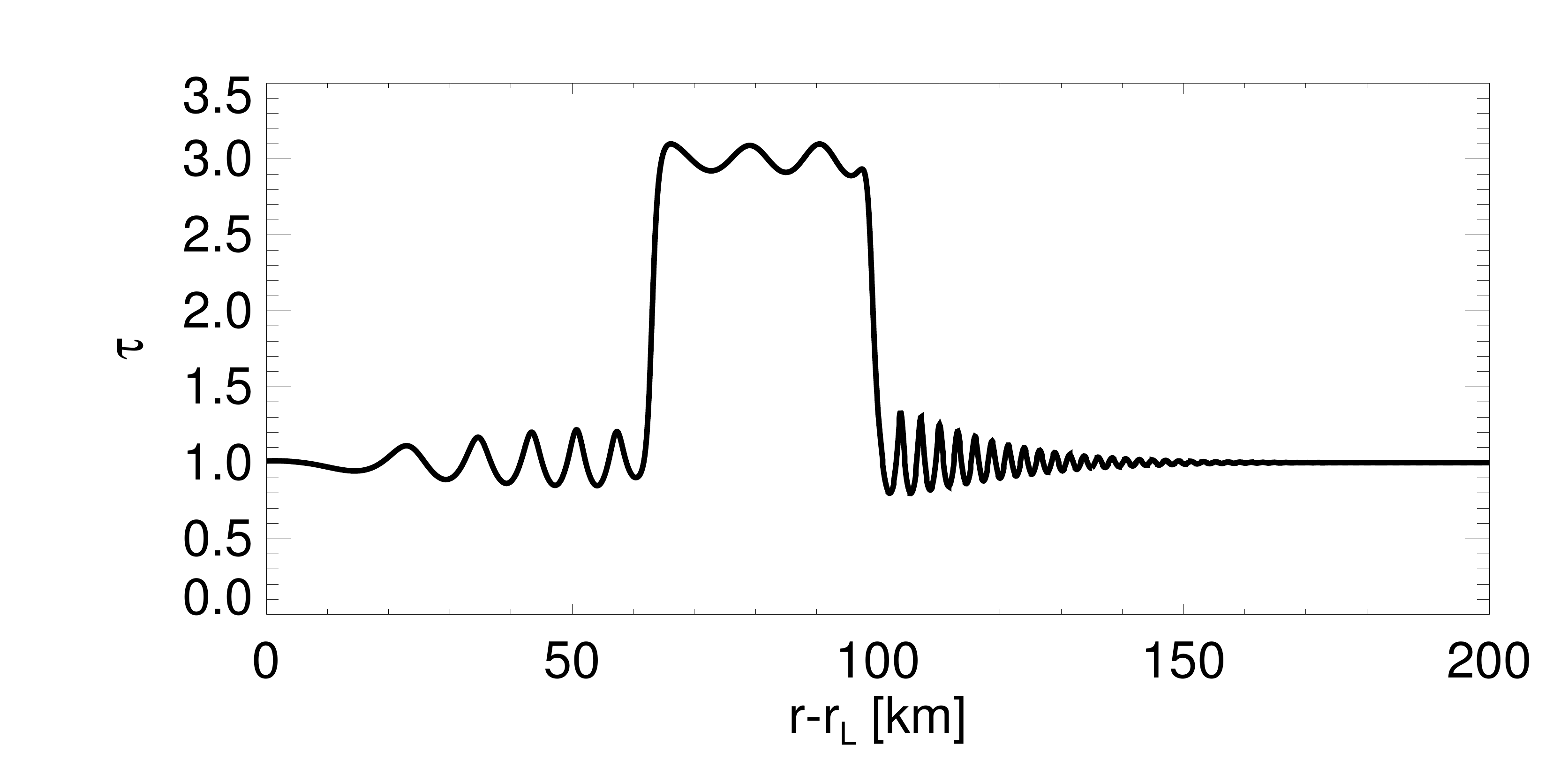}
\includegraphics[width = 0.41 \textwidth]{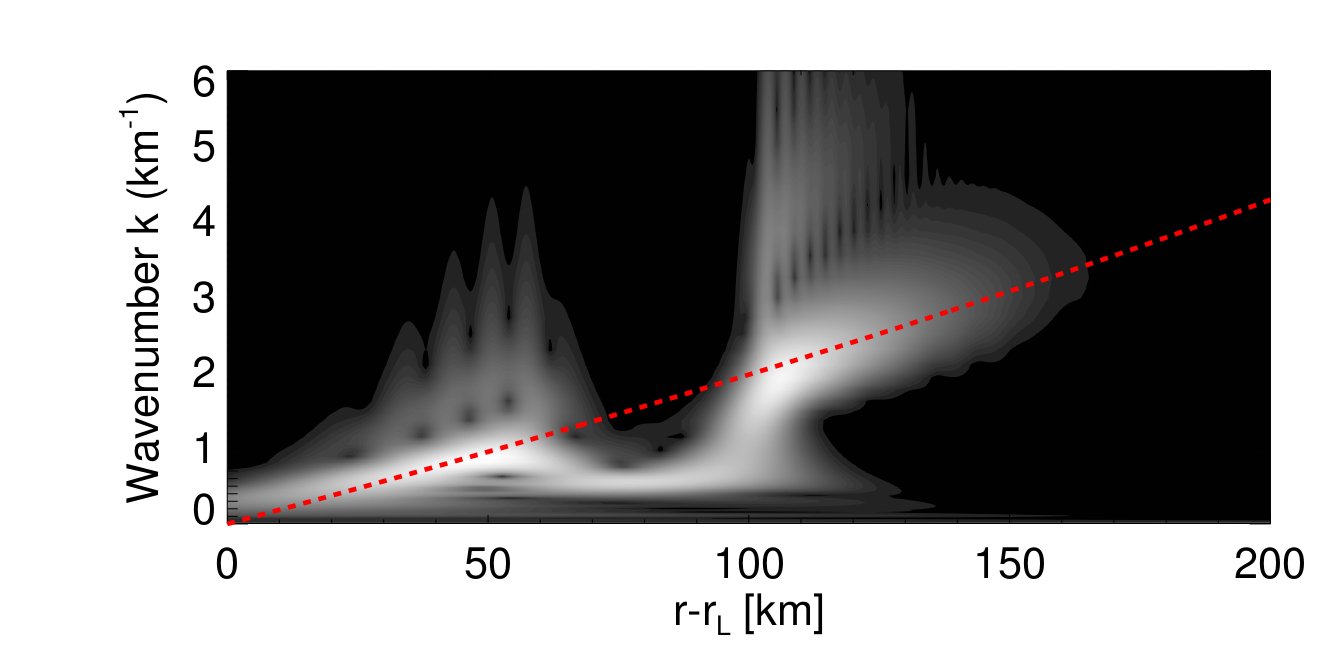}\\
\includegraphics[width = 0.41 \textwidth]{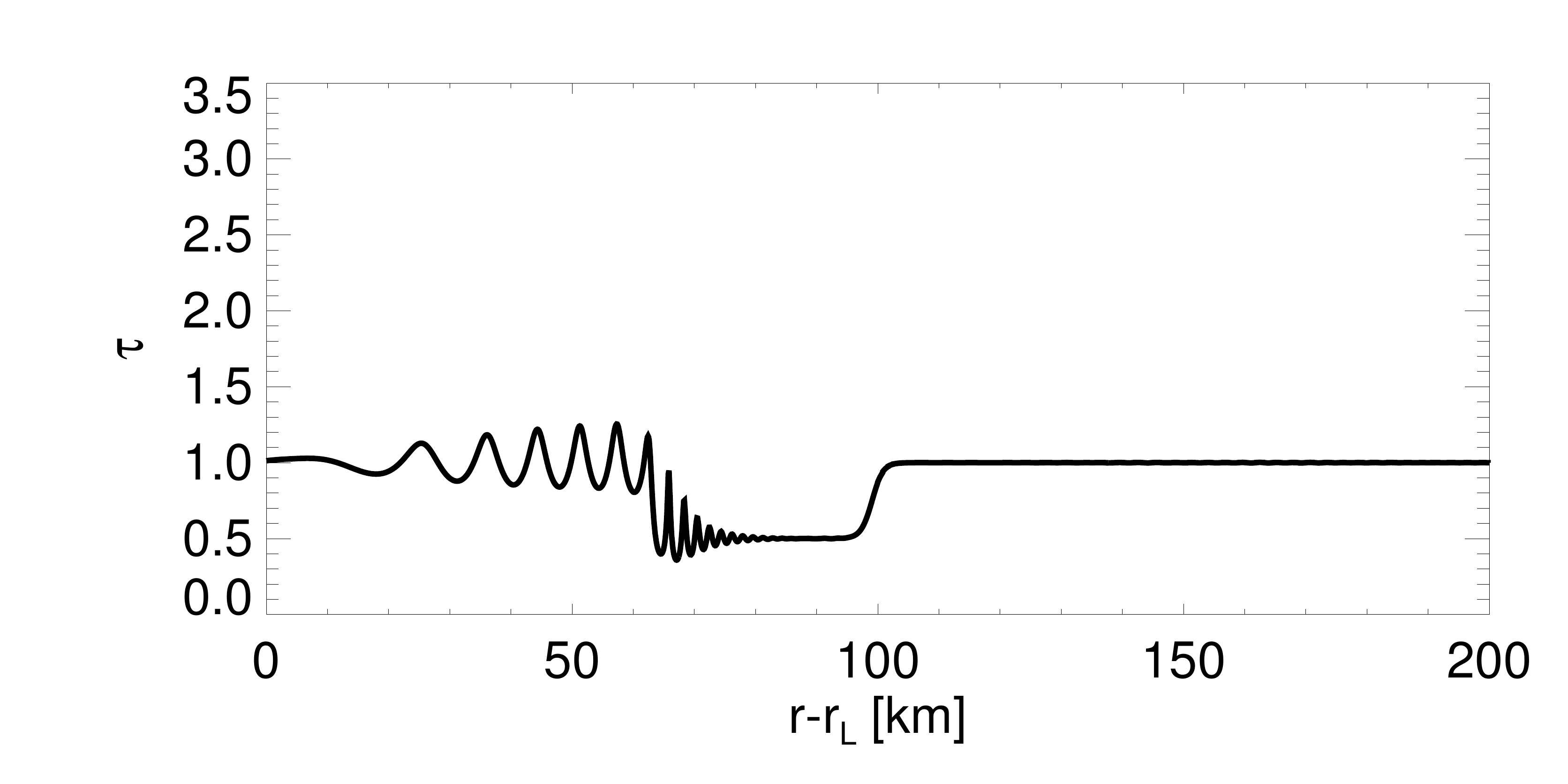}
\includegraphics[width = 0.41 \textwidth]{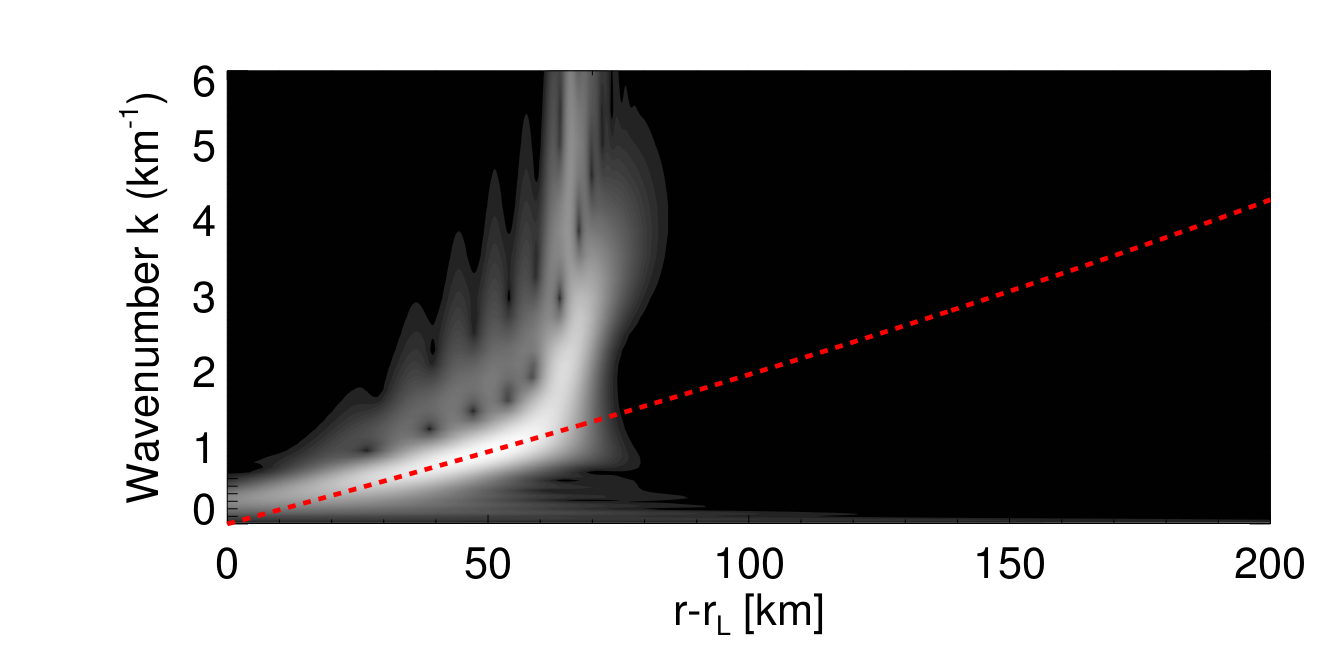}
\caption{Comparison of profiles of $\tau$ along with their Morlet wavelet powers resulting from hydrodynamical integrations using the $Pr76$-parameters and  
scaled torque $\tilde{T}^{s}=9 \cdot 10^{-2}$. From top to bottom the equilibrium surface density $\tau_{0}$ is homogeneous, elevated ($\tau_{0}=3$), as well 
as decreased ($\tau_{0}=0.5$) within regions of radial width $\sim 40\,\text{km}$.}
\label{fig:pr76cleanbargap}
\end{figure}

\section{Figures of Section 7.4}\label{sec:appendixd}

\begin{figure}[ht!]
\vspace{-0.4cm}
\centering
\includegraphics[width = 0.44 \textwidth]{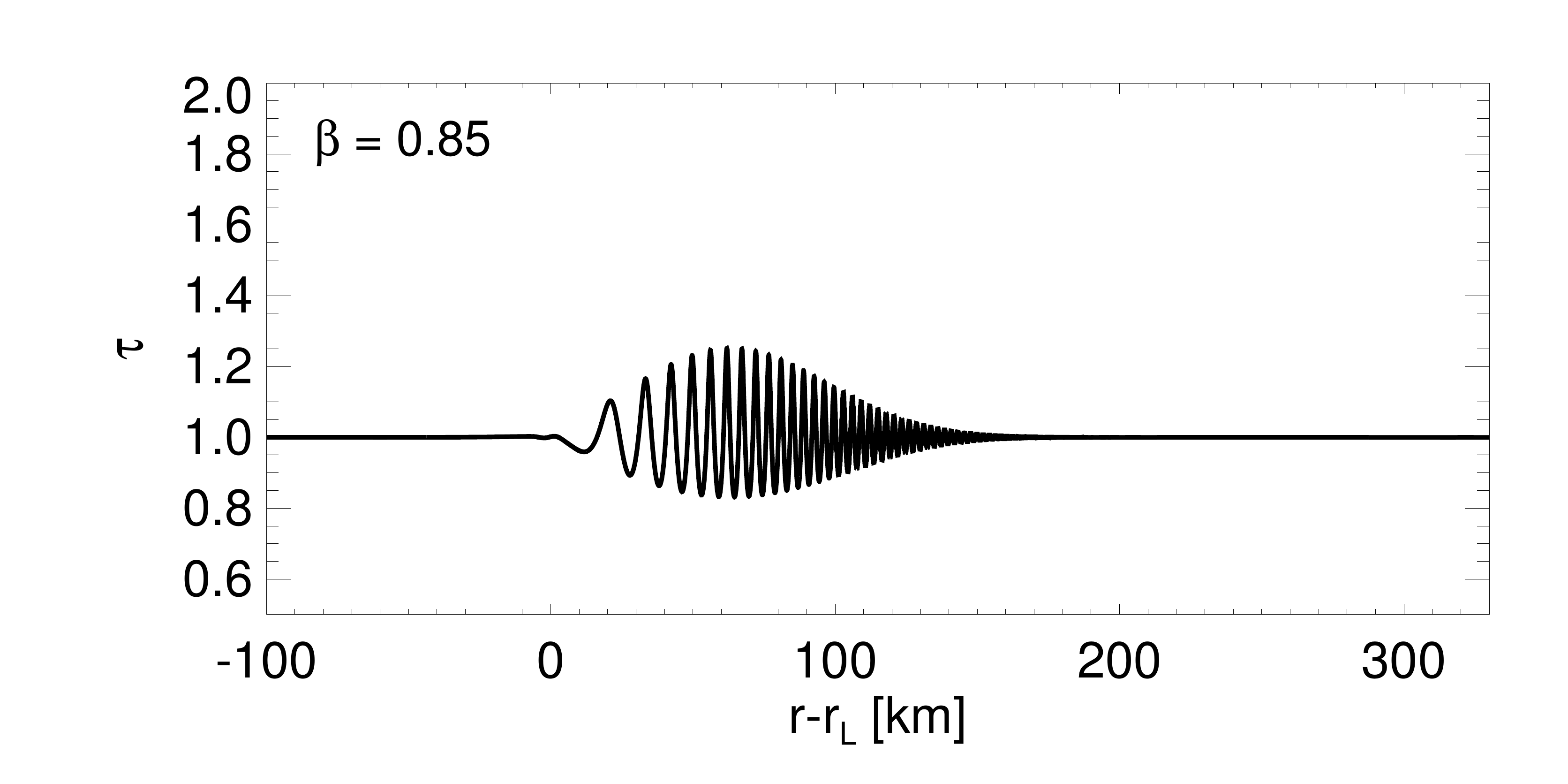}
\includegraphics[width = 0.44 \textwidth]{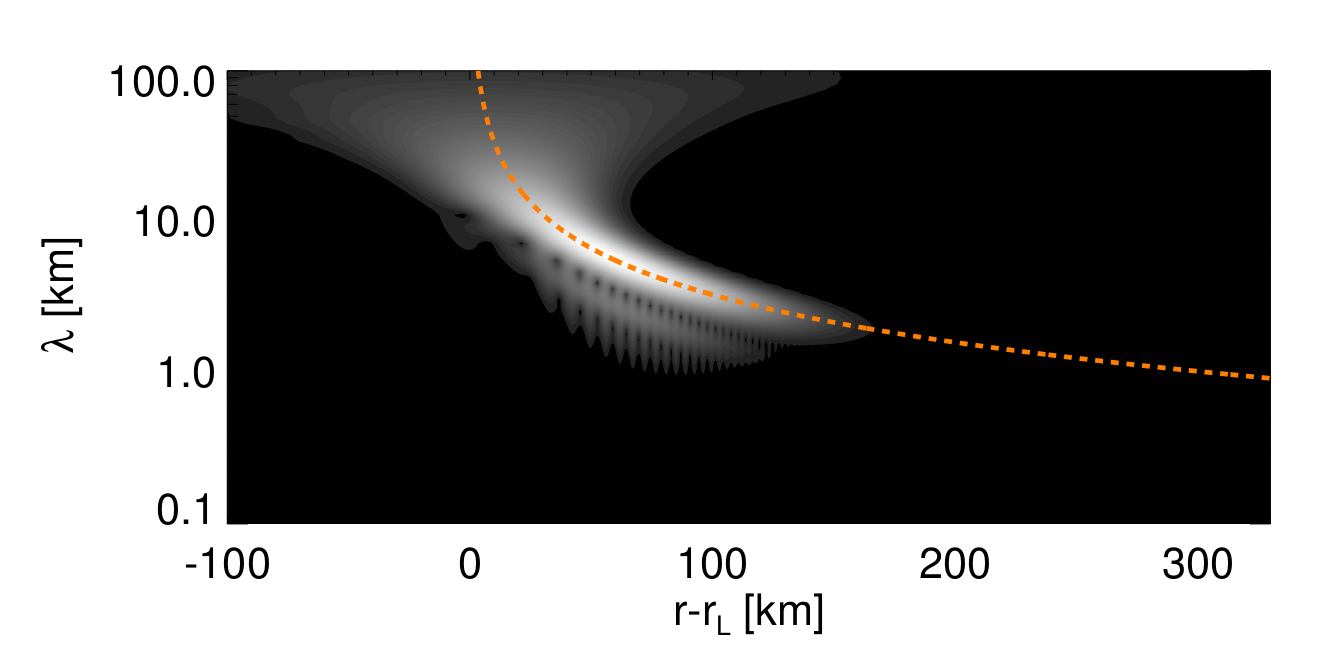}\\
\includegraphics[width = 0.44 \textwidth]{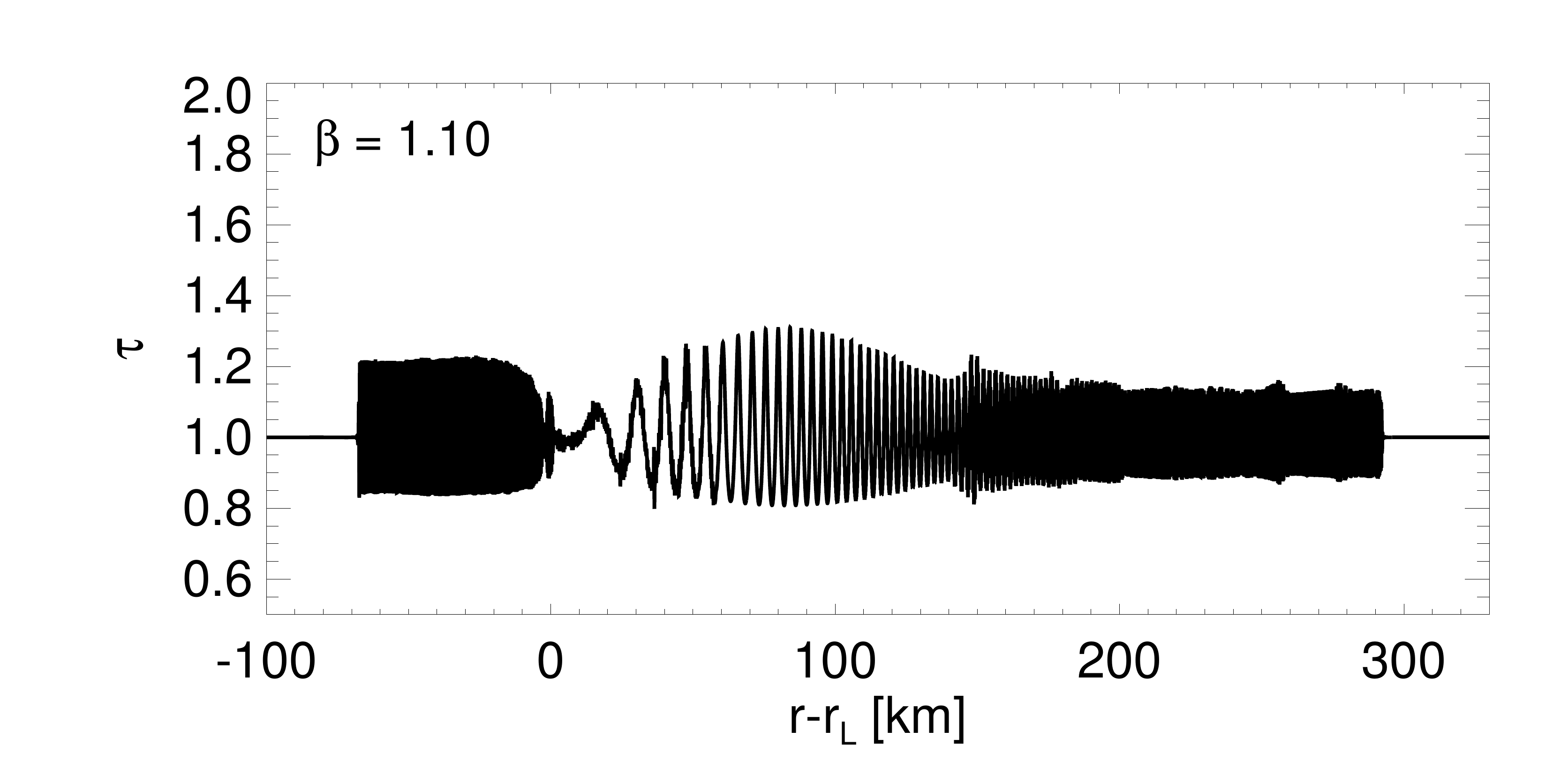}
\includegraphics[width = 0.44 \textwidth]{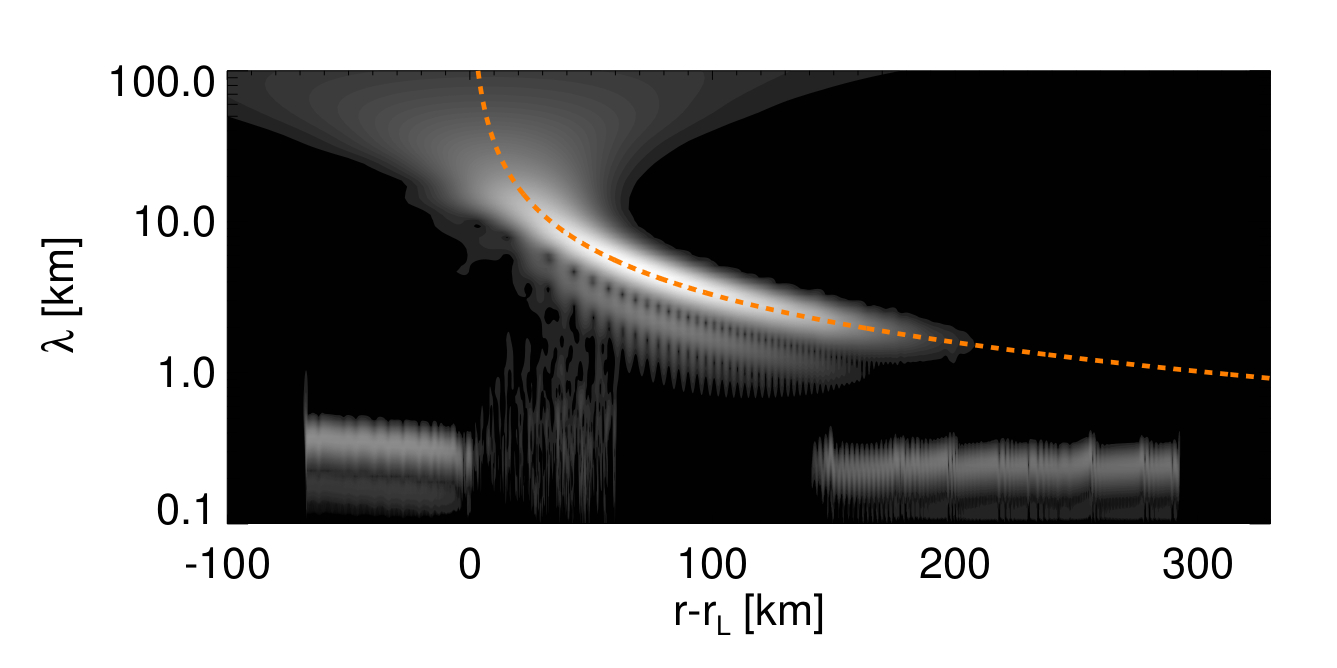}\\
\includegraphics[width = 0.44 \textwidth]{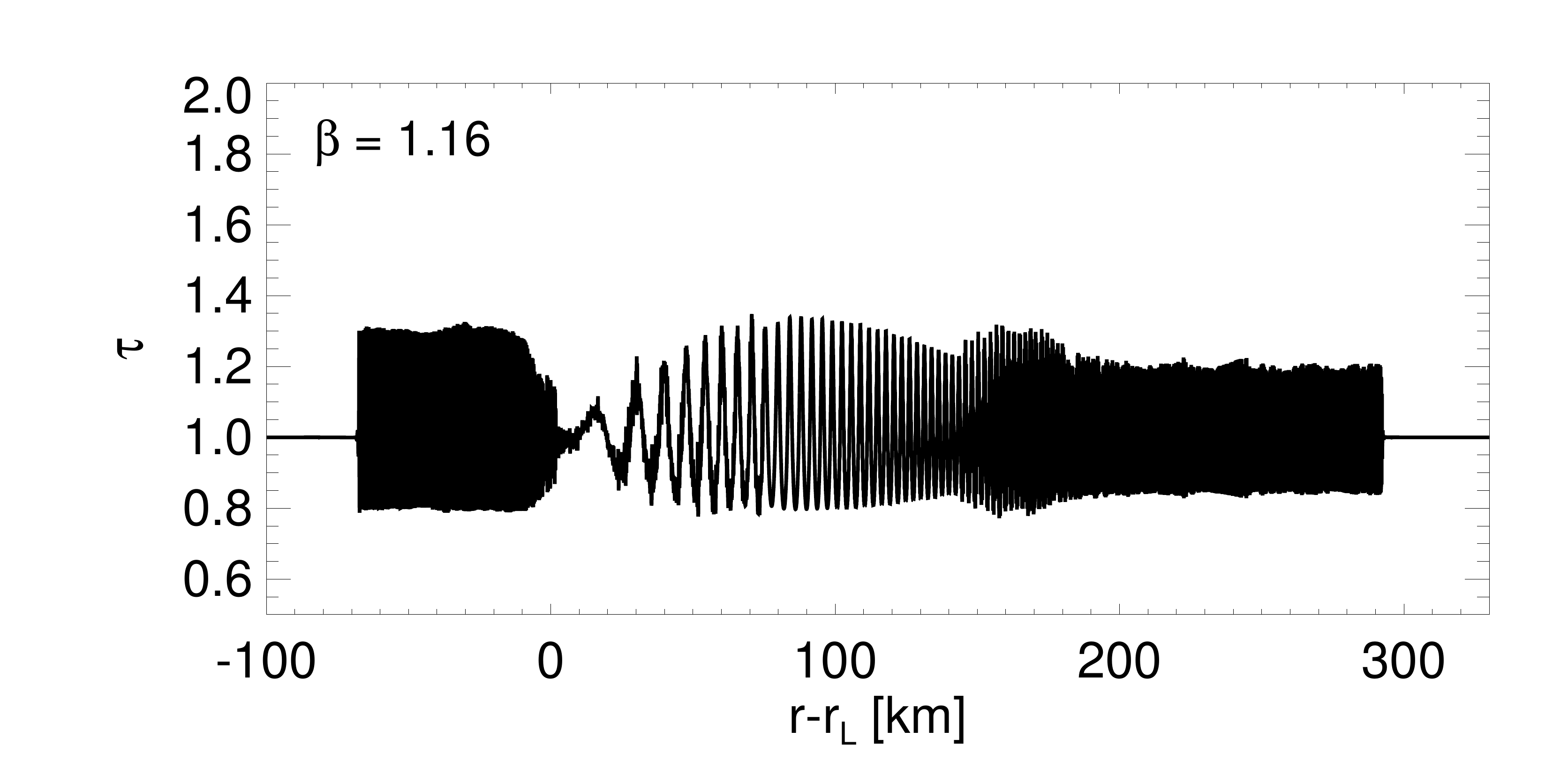}
\includegraphics[width = 0.44 \textwidth]{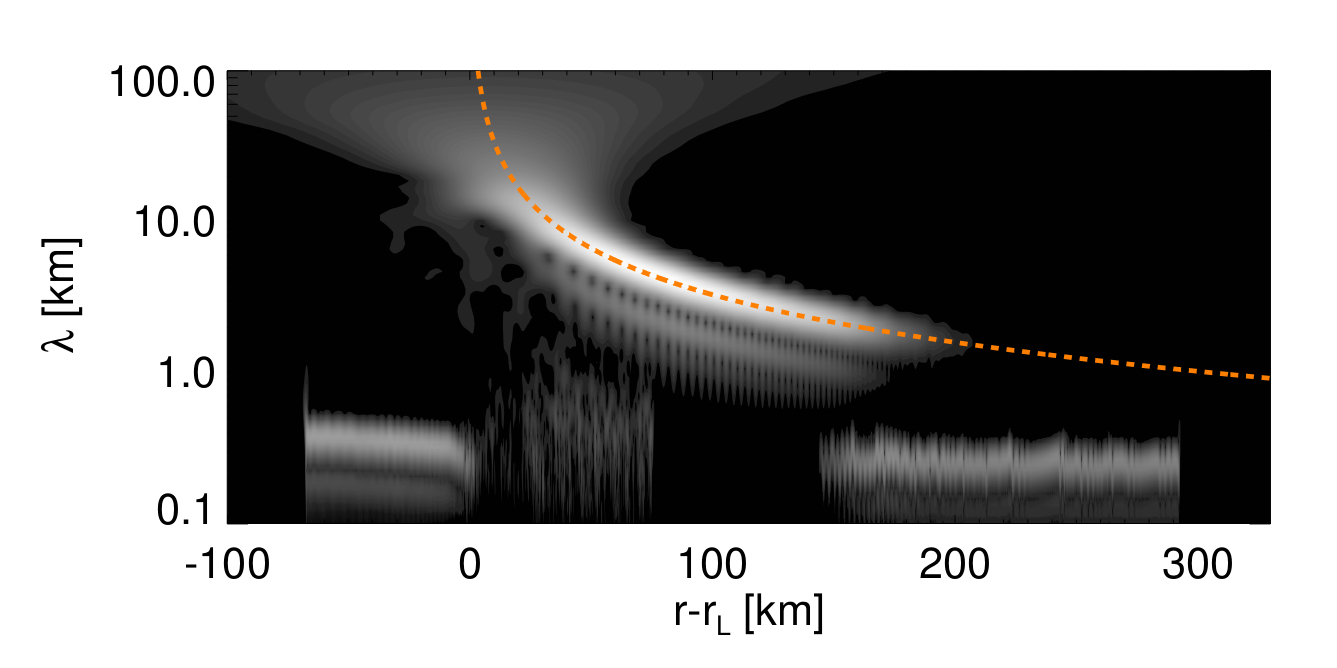}\\
\includegraphics[width = 0.44 \textwidth]{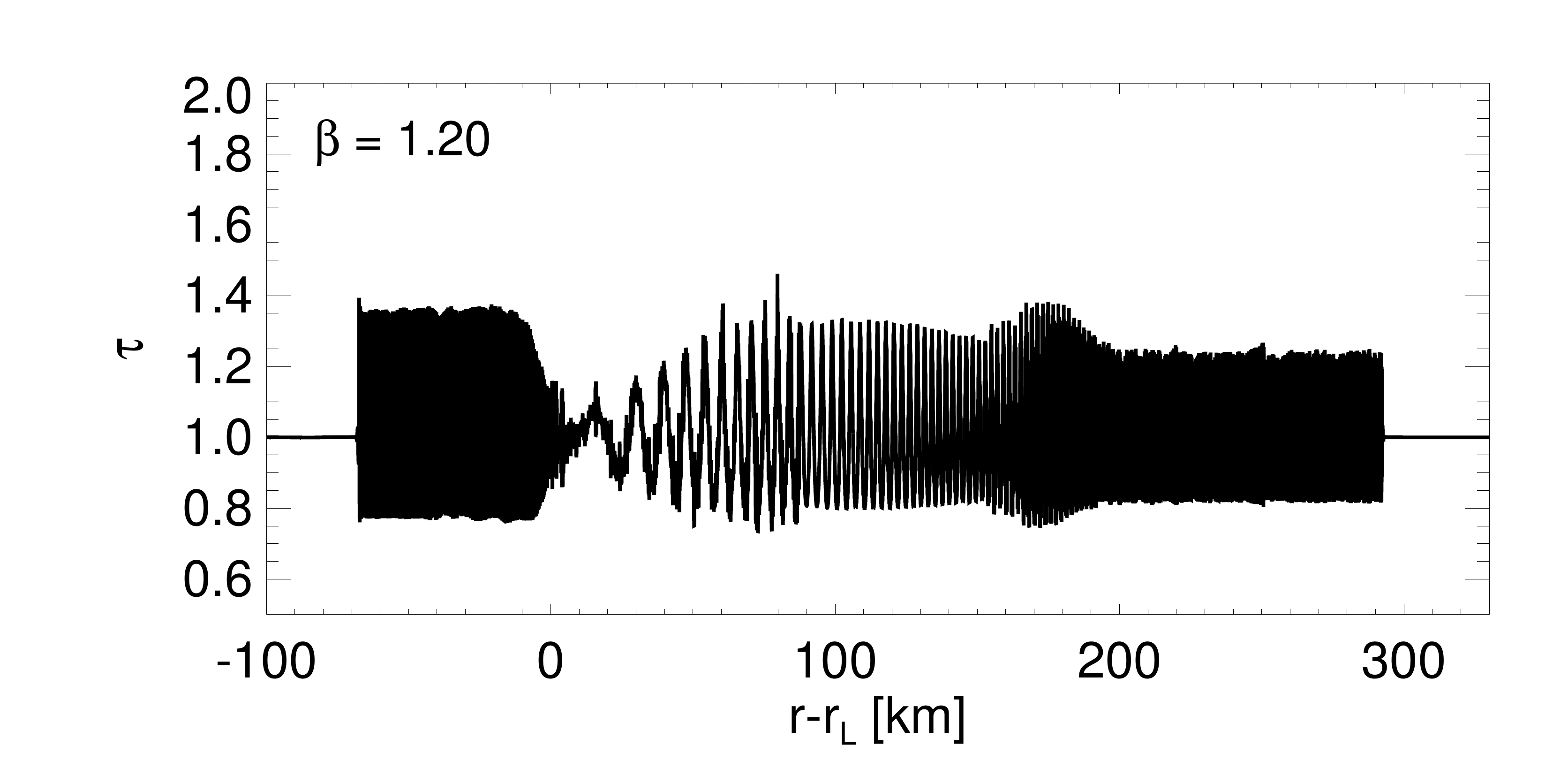}
\includegraphics[width = 0.44 \textwidth]{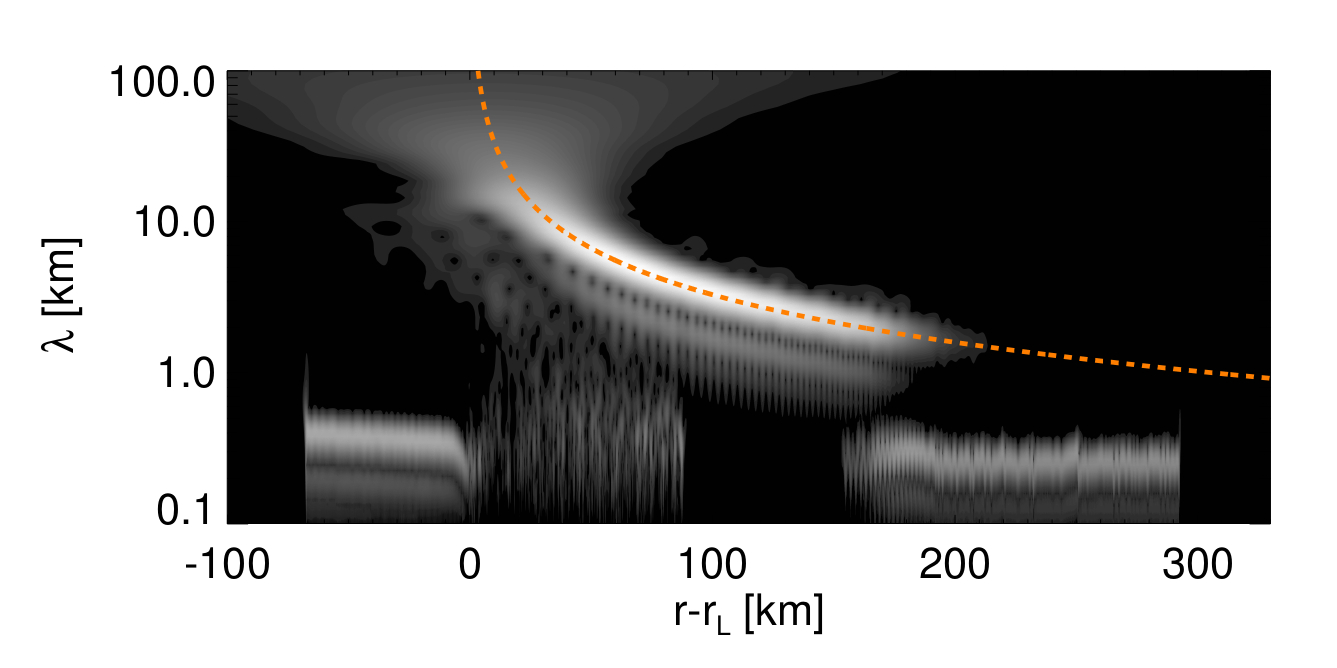}\\
\includegraphics[width = 0.44 \textwidth]{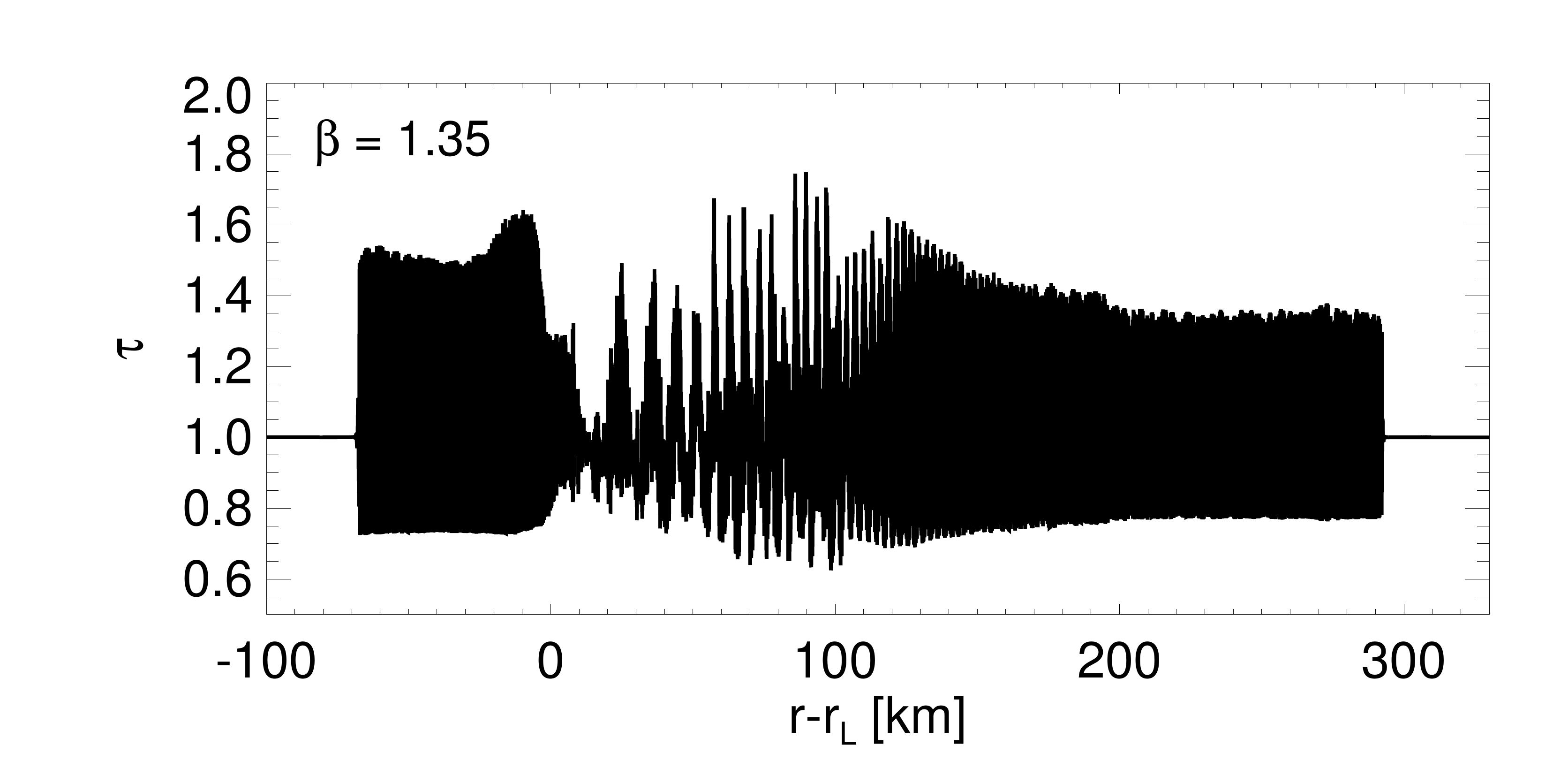}
\includegraphics[width = 0.44 \textwidth]{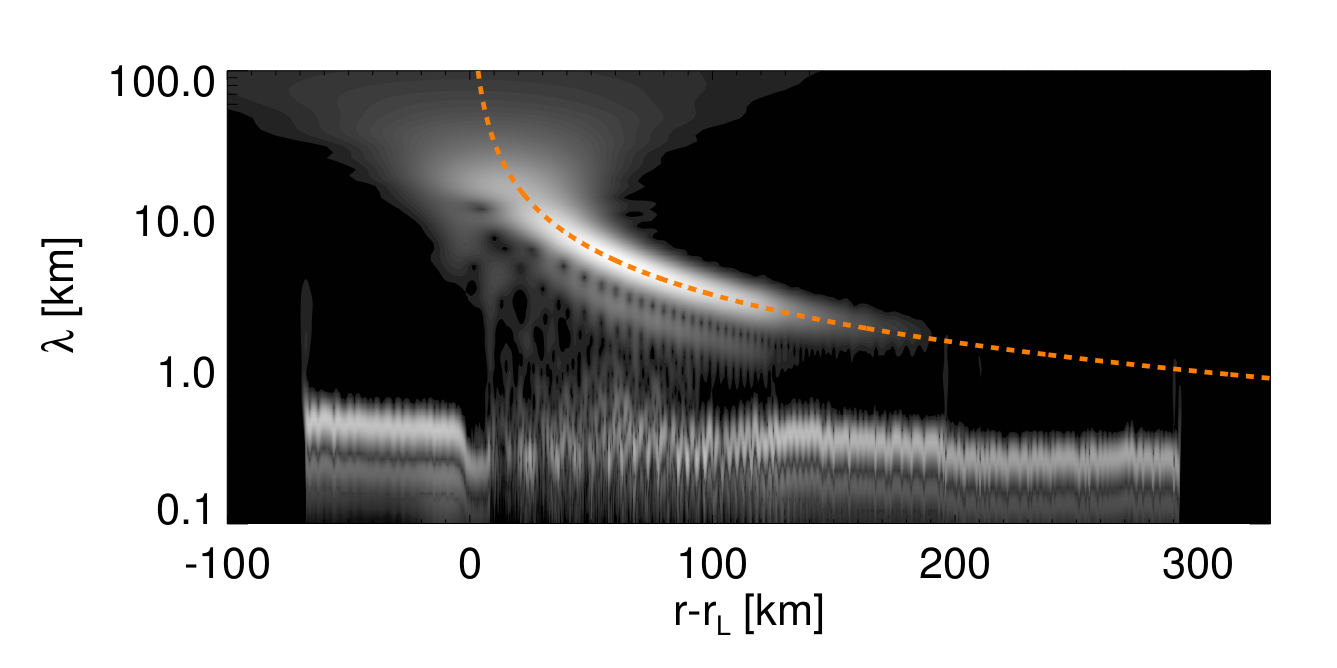}\\
\caption{Hydrodynamical integrations using the $Pr76$-parameters (Table \ref{tab:hydropar}) with increasing value of the viscosity parameter $\beta=0.85-1.35$ 
(see Figure \ref{fig:lambet}) 
from top to bottom. The surface density profiles (left) as well as their wavelet-powers (right) reveal co-existence of a resonantly forced density wave and 
short-scale viscous overstability for the cases $\beta =1.10-1.35$. All integrations use a scaled torque
$\tilde{T}^{s}=9 \cdot 10^{-2}$, a grid of size $L_{r}=450\,\text{km}$ and resolution 
$h=25\, \text{m}$. The plots correspond to times $t \gtrsim 
20,000\,\text{ORB}$. }
\label{fig:pr76waveletbetvarhr}
\end{figure}

\FloatBarrier

\begin{figure}[h!]
\centering
\includegraphics[width = 0.44 \textwidth]{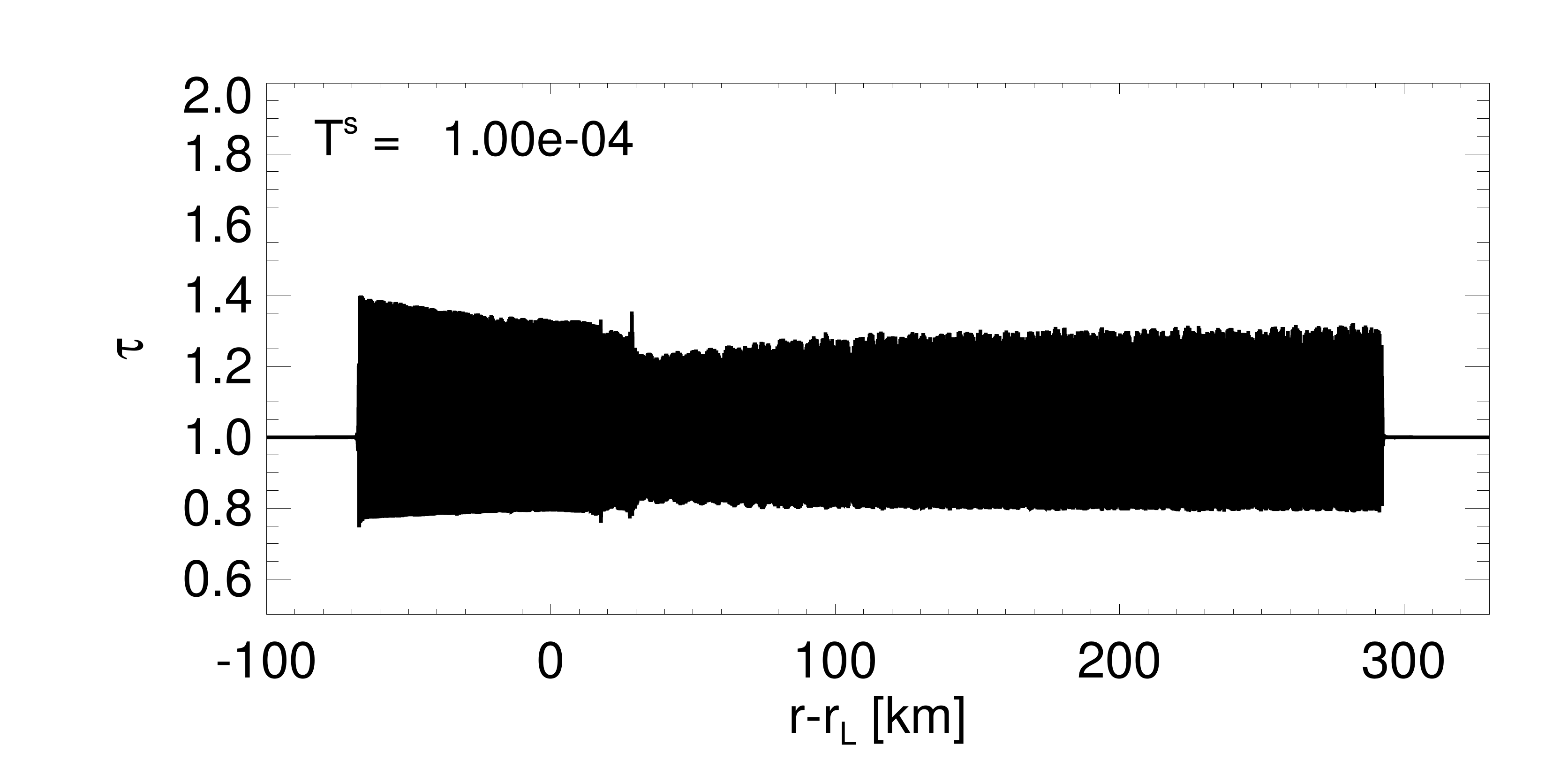}
\includegraphics[width = 0.44 \textwidth]{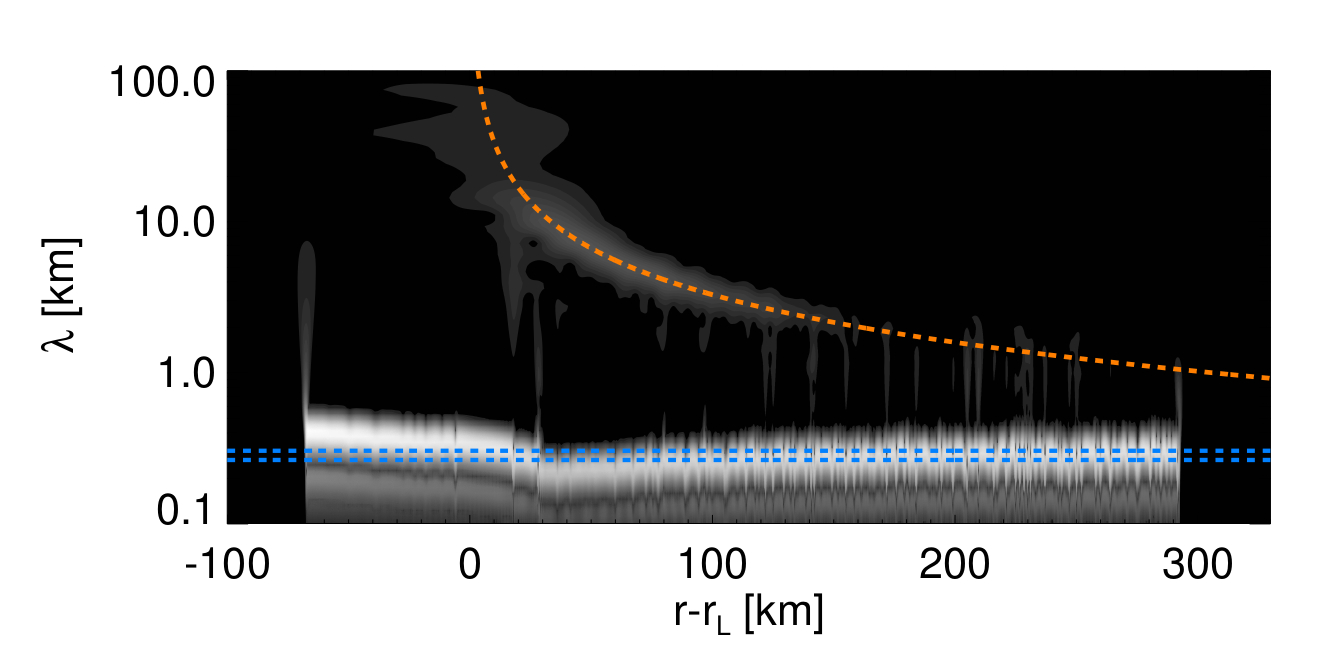}\\
\includegraphics[width = 0.44 \textwidth]{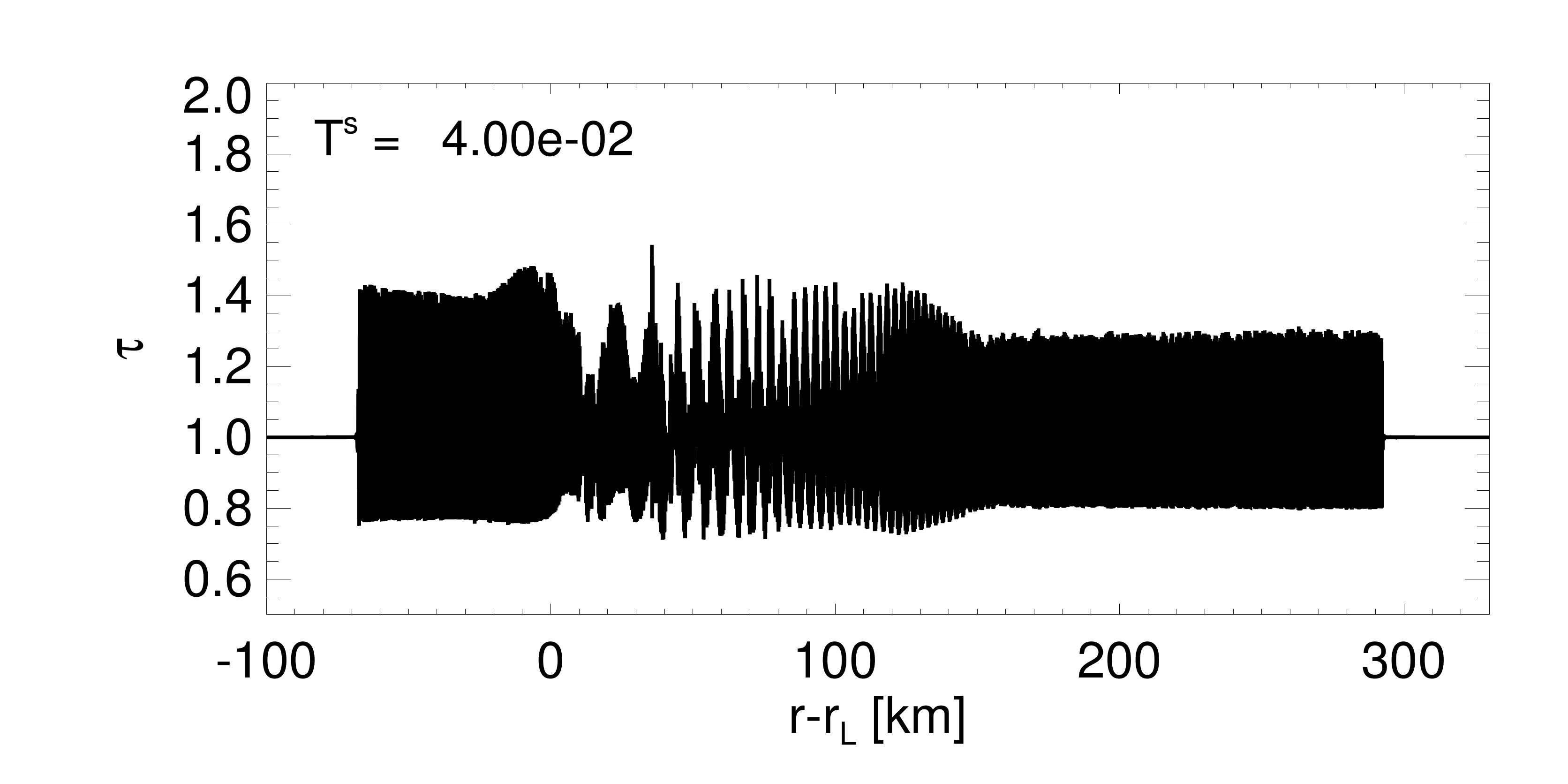}
\includegraphics[width = 0.44 \textwidth]{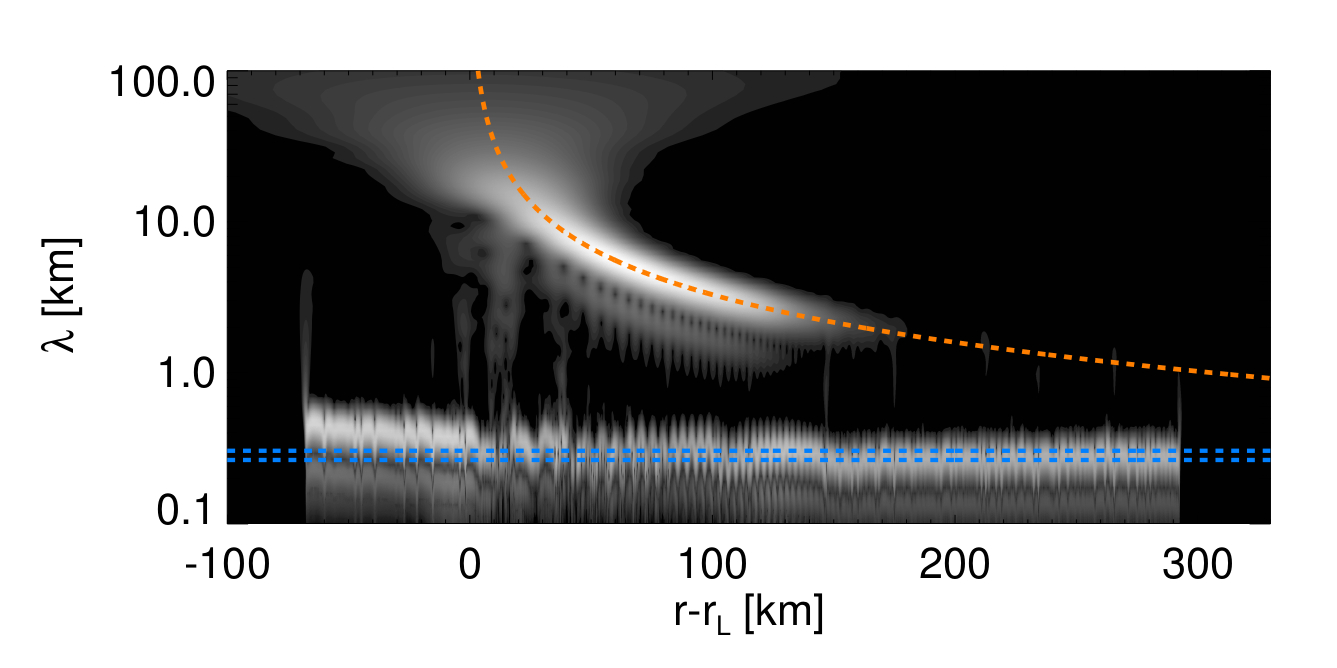}\\
\includegraphics[width = 0.44 \textwidth]{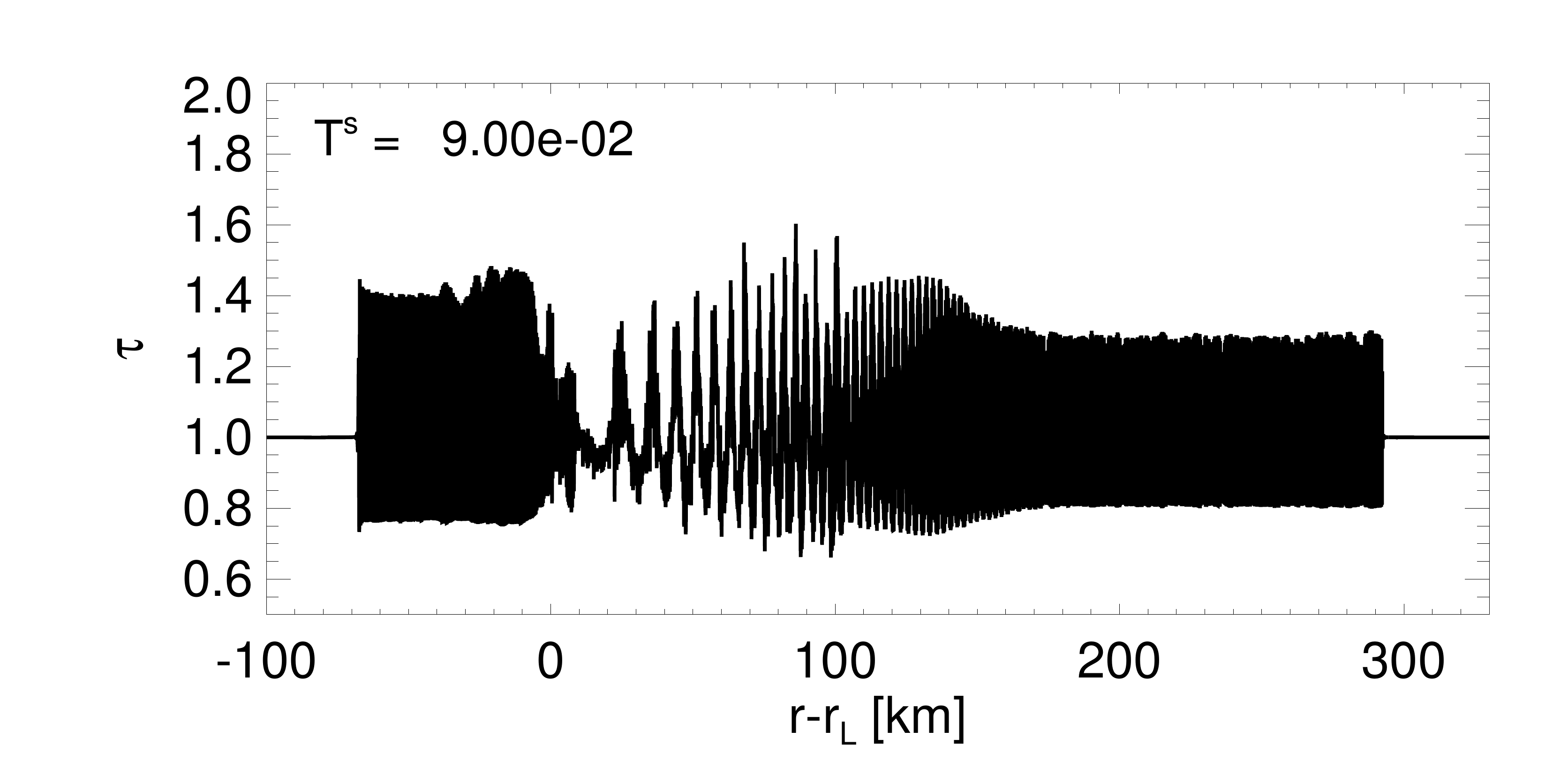}
\includegraphics[width = 0.44 \textwidth]{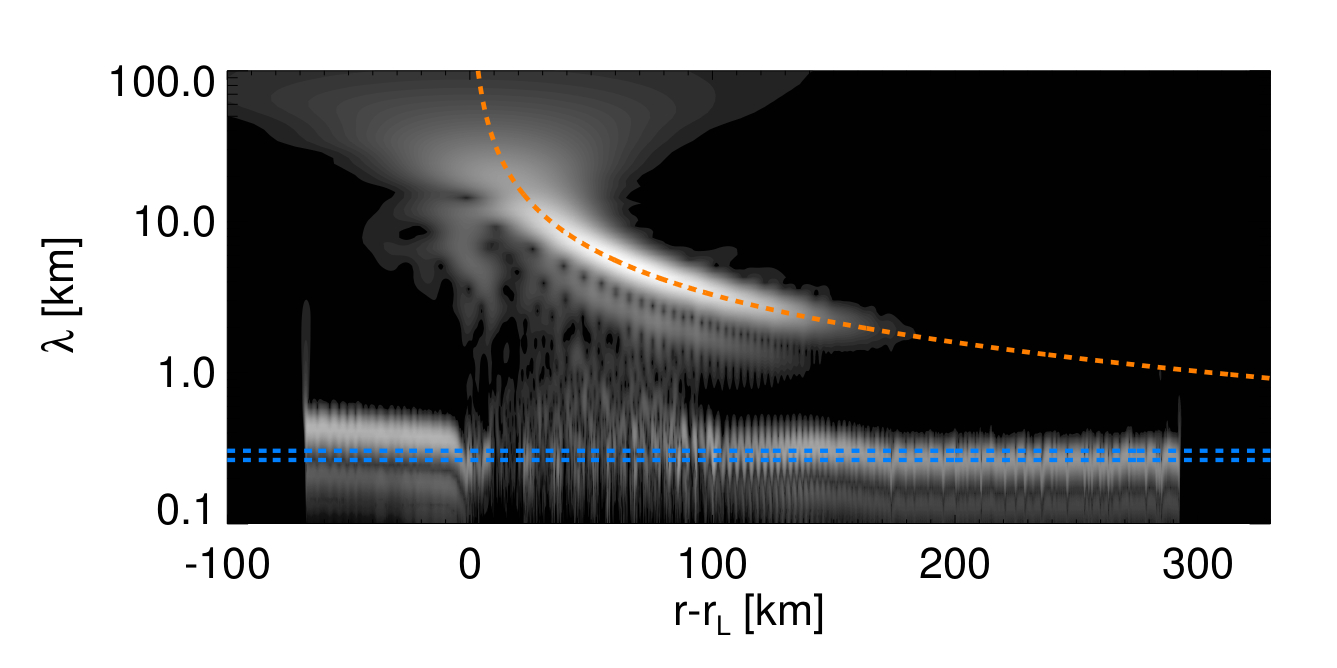}\\
\includegraphics[width = 0.44 \textwidth]{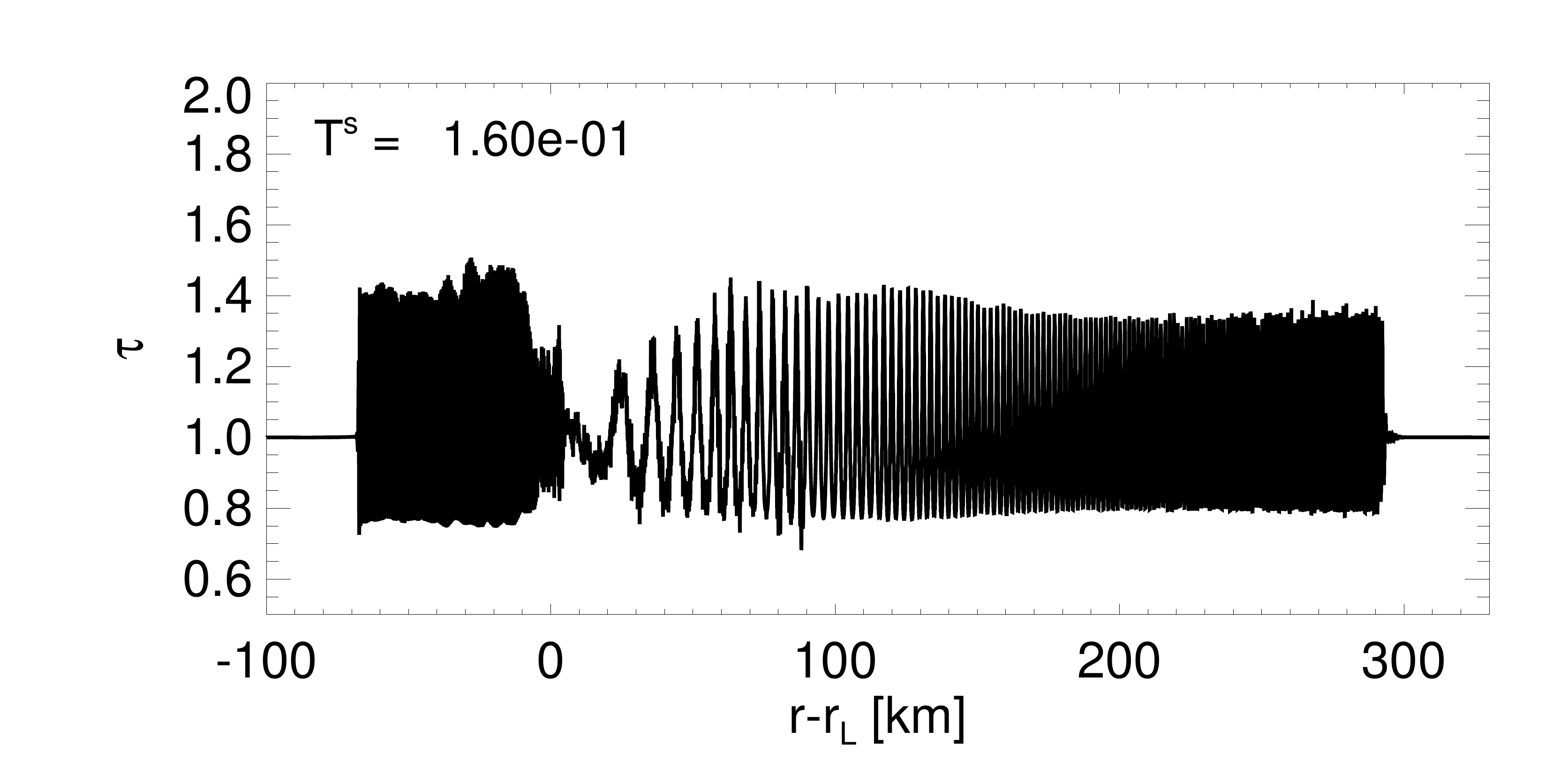}
\includegraphics[width = 0.44 \textwidth]{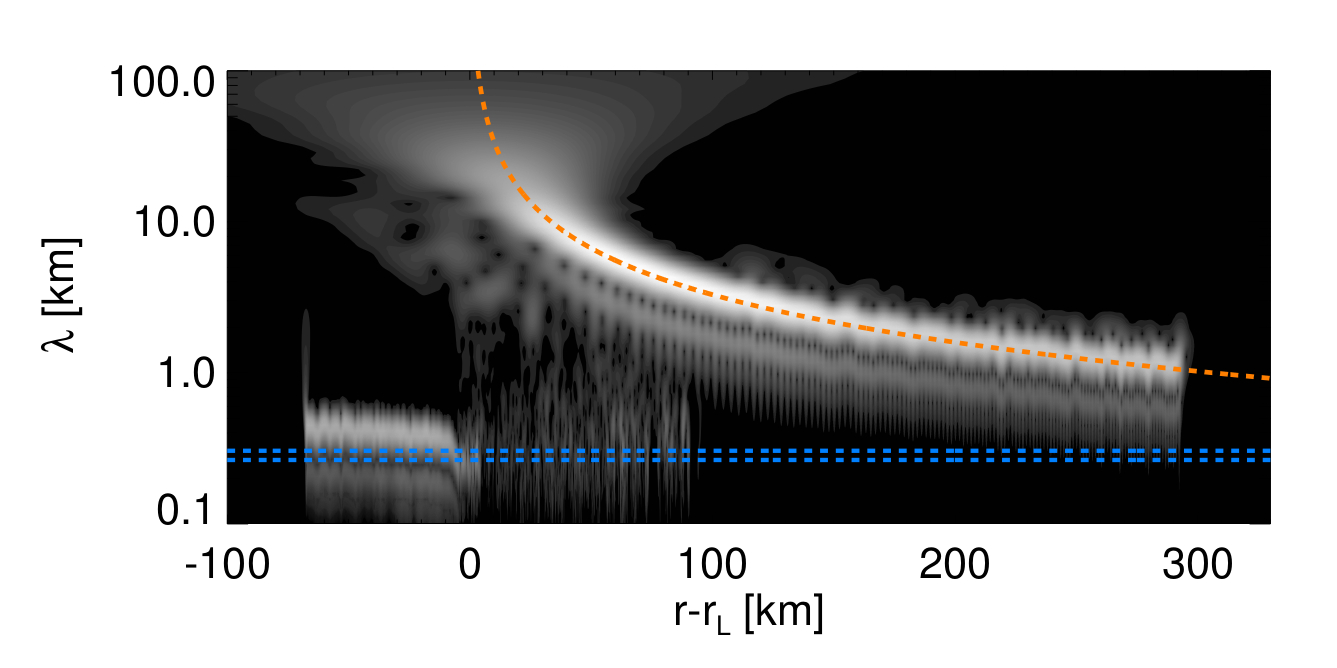}
\caption{Hydrodynamical integrations using the $Pr76$-parameters (Table \ref{tab:hydropar}) with increasing forcing strength from top to bottom 
($\tilde{T}^{s}=10^{-4}-0.16$). The surface density profiles (left) as well as their wavelet-powers (right) reveal co-existence of a resonantly forced density 
wave and short-scale viscous overstability. All integrations use 
$\beta=1.25$, a grid of size $L_{r}=450\,\text{km}$ and resolution 
$h=25\, \text{m}$. As in Figure \ref{fig:pr76osnofor} the blue dashed lines indicate the wavelength of vanishing nonlinear 
group velocity of overstable waves (by margins $\pm 20\,\text{m}$). The plots correspond to times $t \gtrsim 20,000\,\text{ORB}$.}
\label{fig:pr76waveletbet125hr}
\end{figure}

 \FloatBarrier
%
%
\clearpage

\begin{figure}[ht!]
\centering
\vspace{-0.5cm}
\includegraphics[width = 0.42 \textwidth]{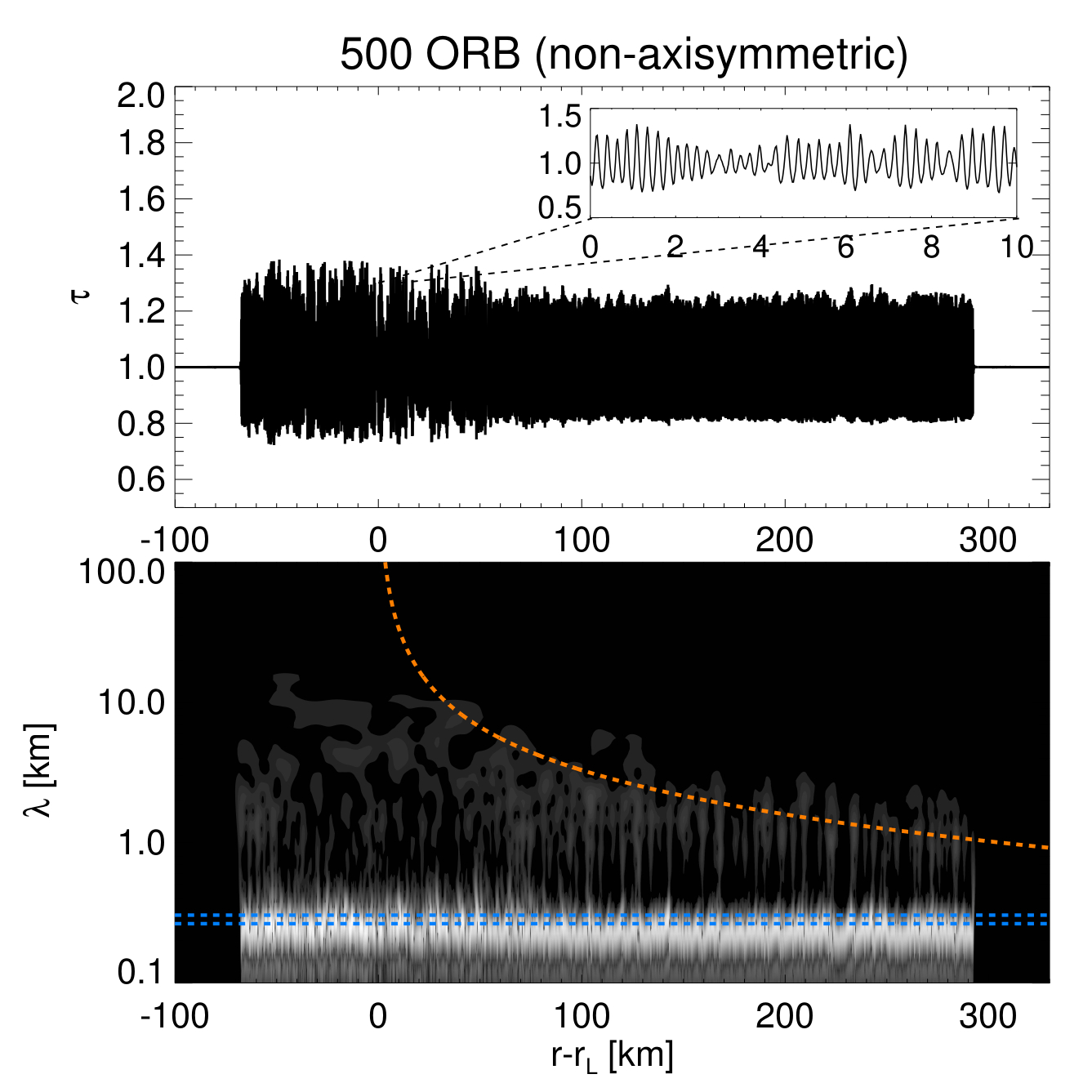}
\includegraphics[width = 0.42 \textwidth]{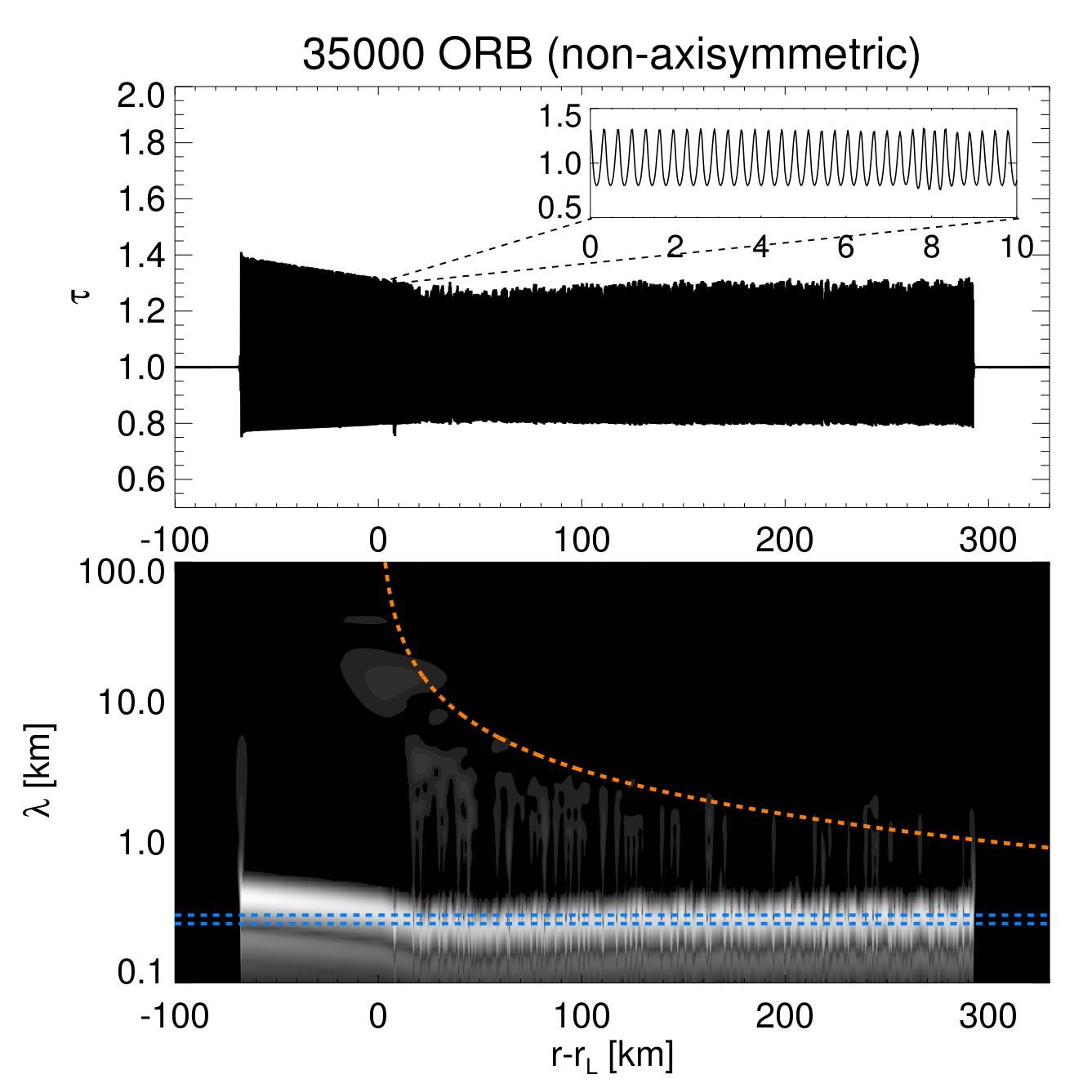}
\includegraphics[width = 0.42 \textwidth]{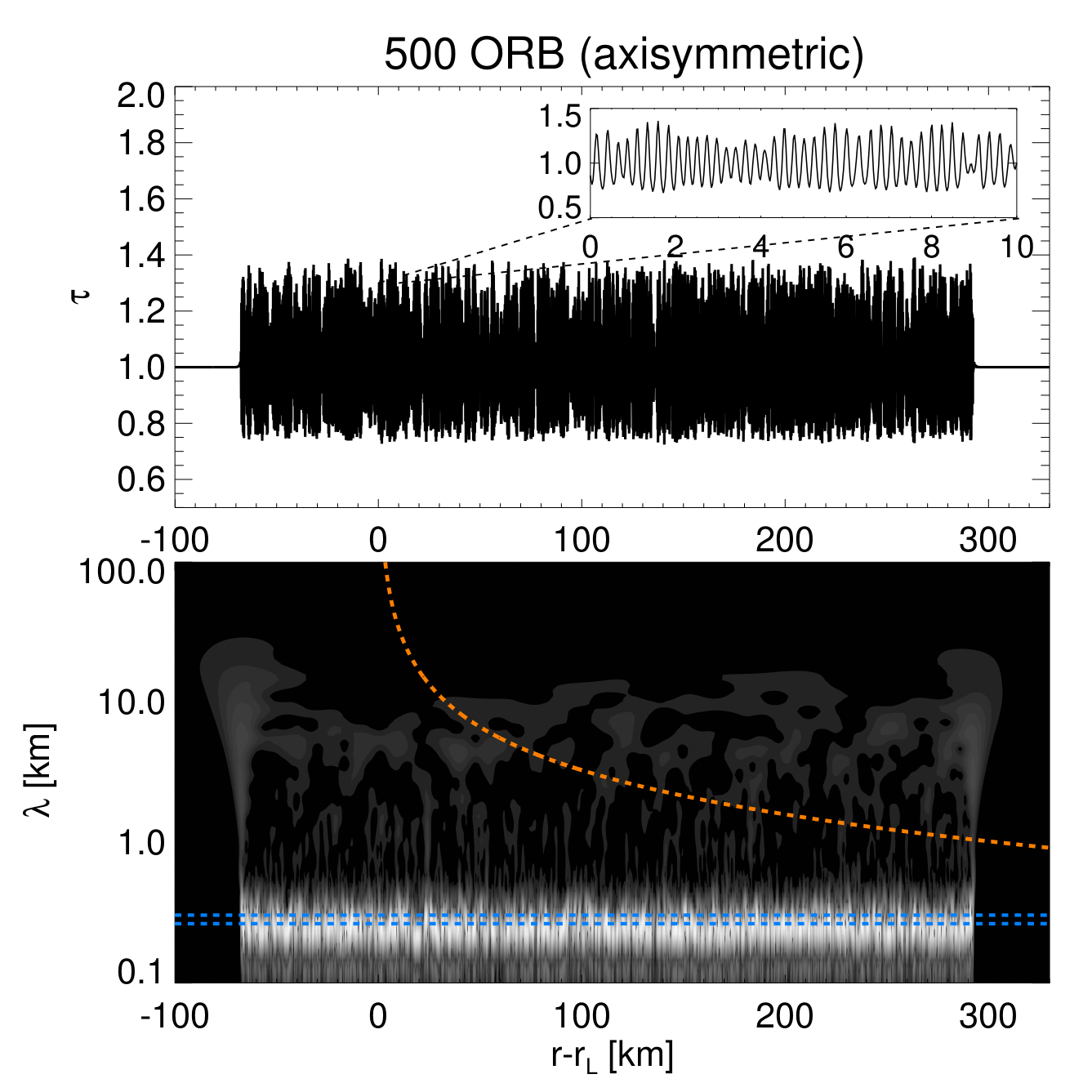}
\includegraphics[width = 0.42 \textwidth]{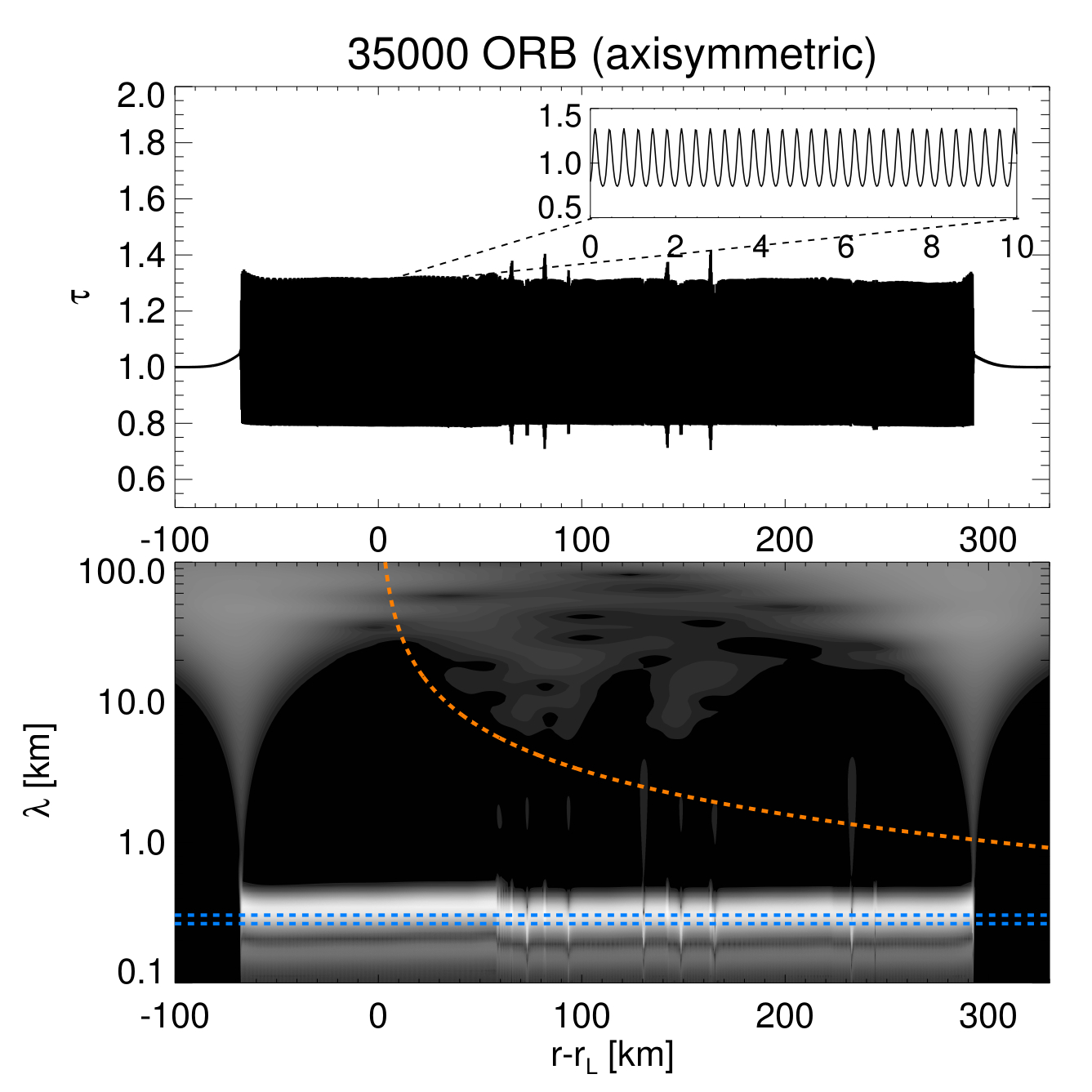}
\includegraphics[width = 0.33 \textwidth]{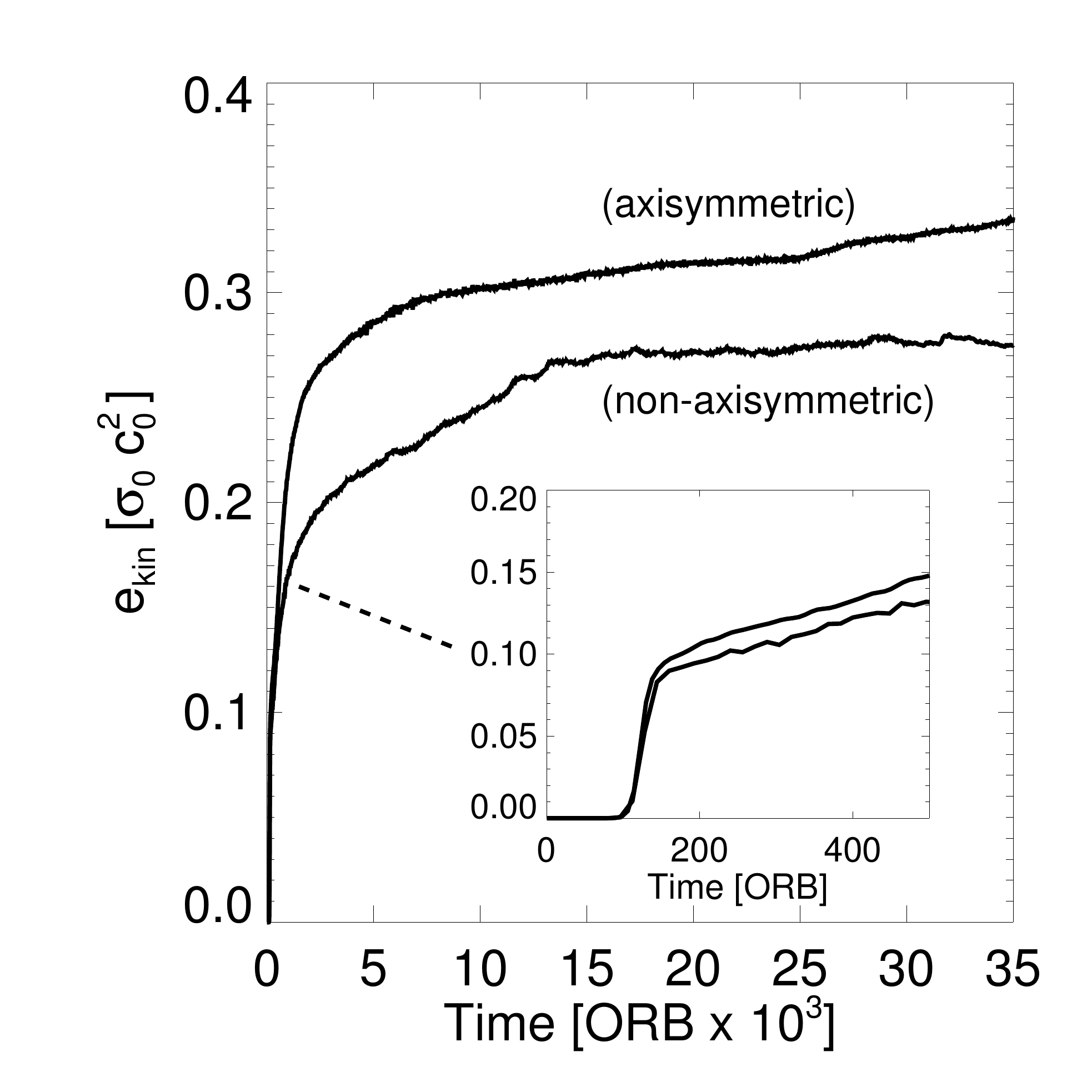}
\caption{Comparison of the nonlinear evolution of free ($T^{s}=0$) viscous overstability in hydrodynamical integrations with (non-axisymmetric, $m=7$) and 
without 
(axisymmetric, $m=0$) the azimuthal derivative terms (Section \ref{sec:azideriv}). Shown ae profiles of the surface density $\tau$ along with their 
wavelet powers for two different times ($t=500\,\text{ORB}$ and $t=35,000\,\text{ORB}$ ).  The blue dashed lines indicate the expected 
nonlinear saturation wavelength of axisymmetric ($m=0$) viscous overstability by margins $\pm 20\,\text{m}$ (See Section \ref{sec:osnofor}). The red dashed 
curves represent the linear density wave dispersion relation (\ref{eq:disprel}). Note that in the axisymmetric case this curve has no physical meaning.
The bottom frame displays the evolution of the kinetic energy density 
for both integrations. The insert plot indicates that the linear growth phases ($t\lesssim 200 \,\text{ORB}$) of non-axisymmetric and axisymmetric modes are  
practically identical, in agreement with our considerations in Section \ref{sec:theo}. The higher saturation energy of the axisymmetric integration is due to 
the slightly larger saturation wavelength (see Section \ref{sec:osnofor} for explanations).}
\label{fig:oscomp}
\end{figure}

 \FloatBarrier
%
%
\clearpage


\begin{figure}[ht!]
\vspace{-0.2 cm}
\centering
\includegraphics[width = 0.28 \textwidth]{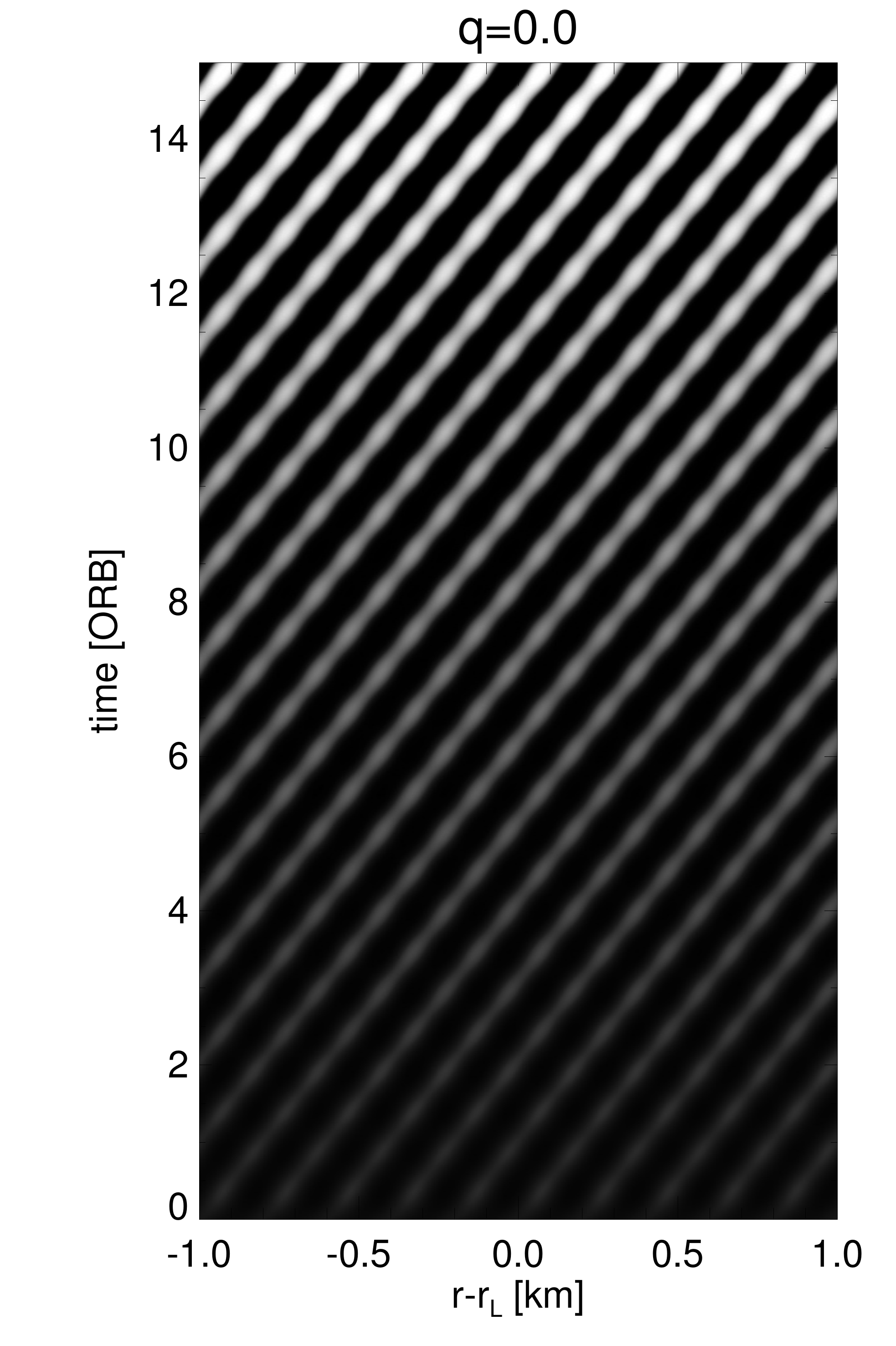}
\includegraphics[width = 0.28 \textwidth]{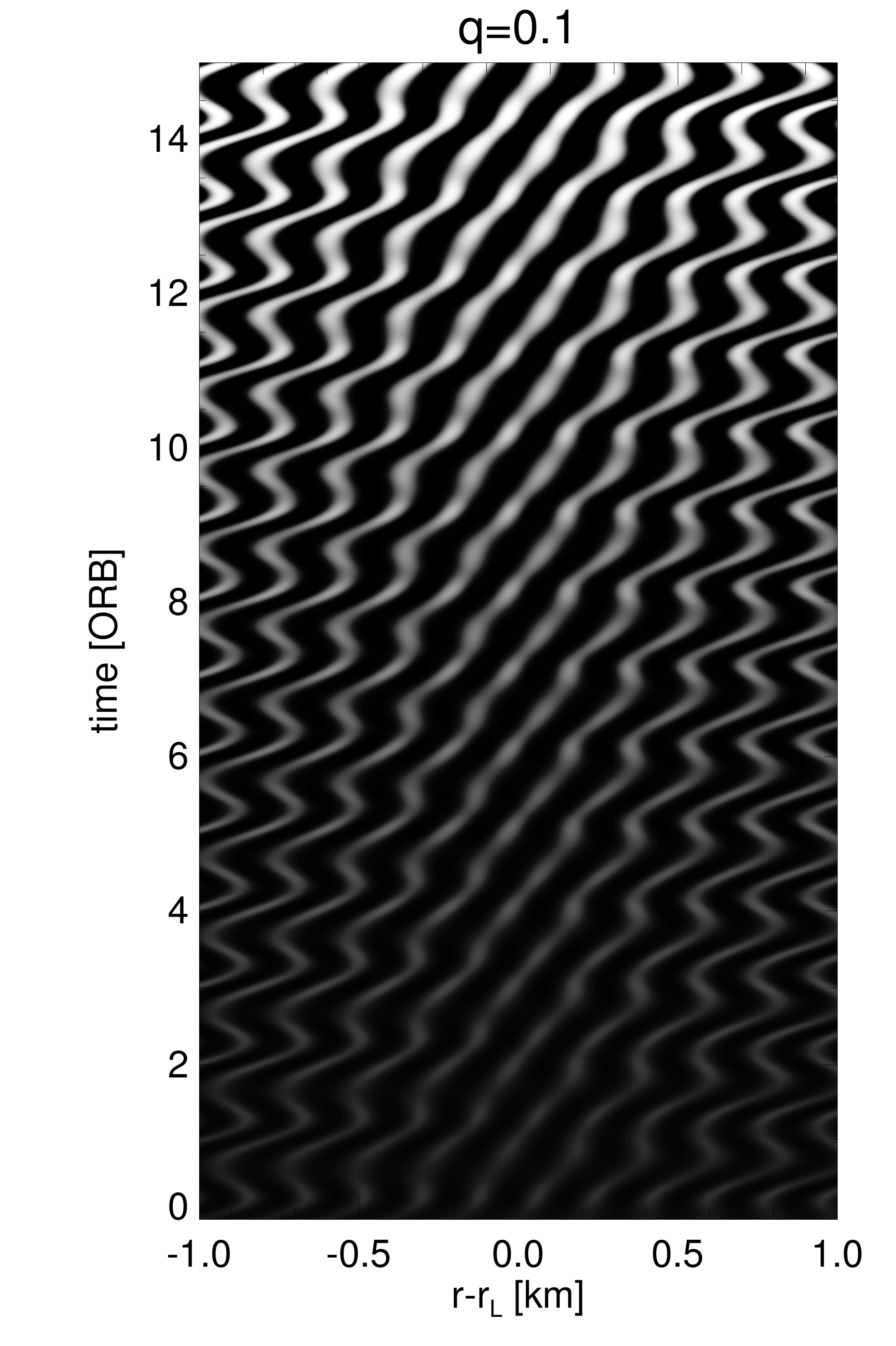}
\includegraphics[width = 0.28 \textwidth]{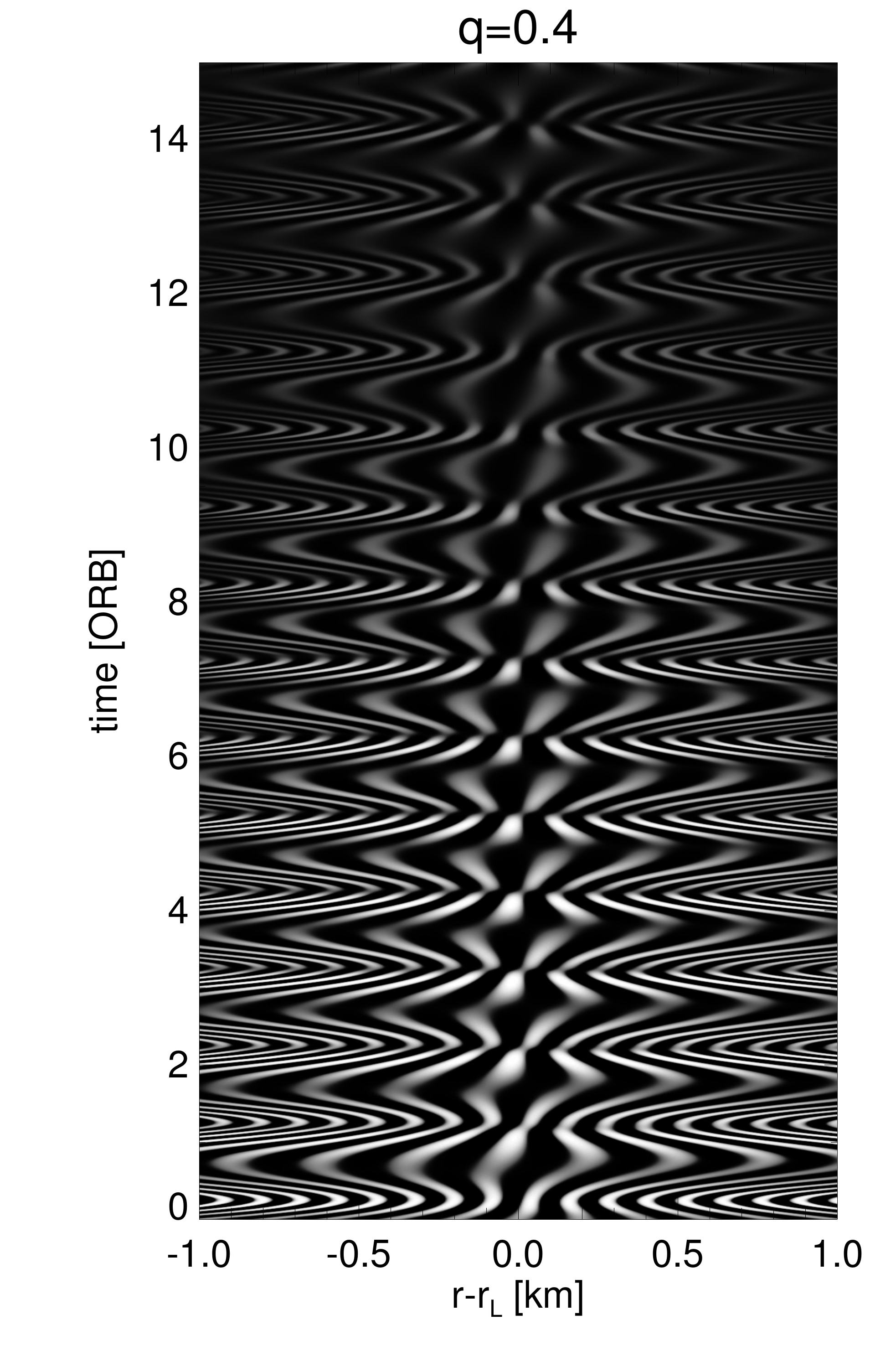}
\caption{Space-time diagrams showing the linear evolution of the radial velocity field $u'$ of an initially seeded traveling wave in a ring with $\beta=1.35$ 
perturbed by an ILR (at $r=r_{L}$). The perturbation, quantified by $q$, increases from left to right. At initial time $t=0$ the model ring is in the 
uncompressed state [Equation (\ref{eq:kpert})] and the initial wavelength $\lambda=200\,\text{m}$.}
\label{fig:sptdpert}
\end{figure}
\begin{figure}[hb!]
\vspace{-0.7 cm}
\centering
\includegraphics[width = 0.3 \textwidth]{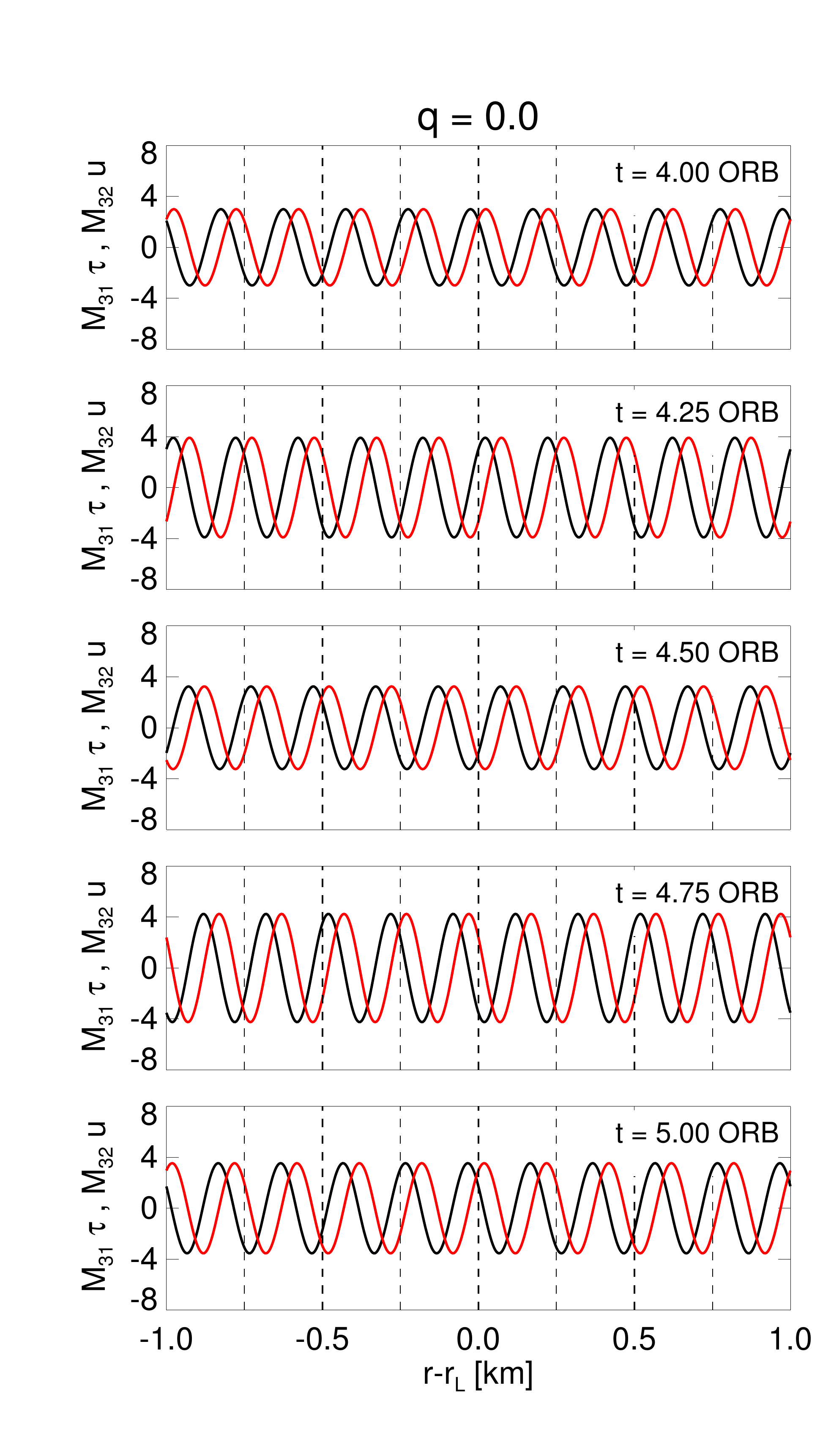}
\includegraphics[width = 0.3 \textwidth]{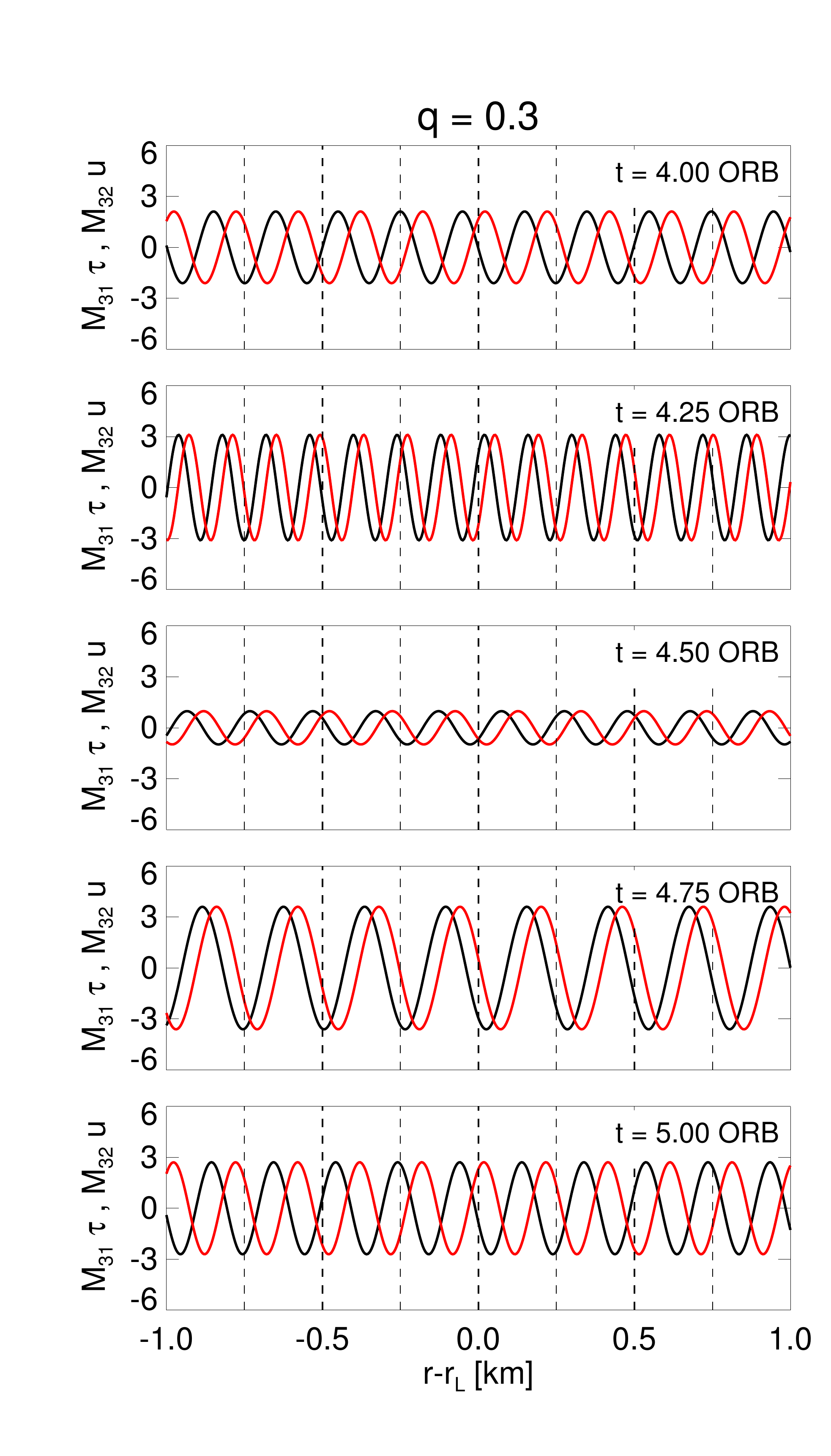}
\includegraphics[width = 0.3 \textwidth]{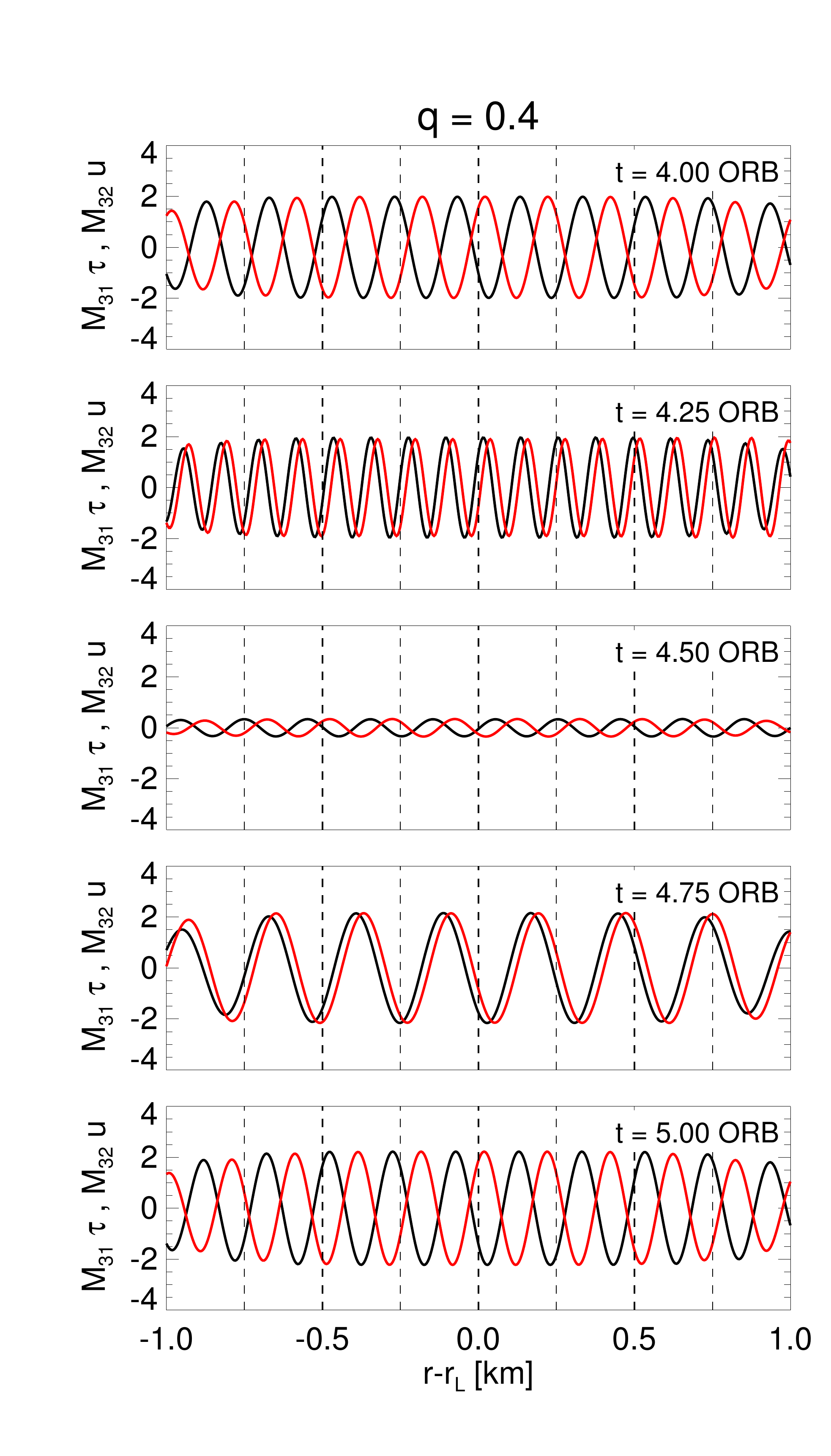}
\caption{Snapshots of the terms $M_{31}\tau{'}$ and $M_{32} u^{'}$ that appear in the equation for the azimuthal velocity perturbation and which must be 
sufficiently in phase for the viscous overstability mechanism to work. The snapshots are from integrations with $\lambda=200\,\text{m}$, $\beta=1.35$, and 
cover one orbital period in equal time-intervals. With increasing strength of the satellite perturbation (quantified through the nonlinearity parameter $q$) 
these terms become increasingly out of phase. For $q=0.4$ almost all possible phase differences in the range $0-2\pi$ occur, which explains the negative growth 
rates of overstable modes on all wavelengths (Figure \ref{fig:grqvar}, lower right panel). For clarity the two quantities have been rescaled so as to 
posses equal amplitudes in all plots.}
\label{fig:coup}
\end{figure}

 \clearpage
%

\section{Method B for the Azimuthal Derivatives}\label{sec:azimb}

\subsection{Linear Waves}
When restricting to linear density waves for the time being and adopting the notation (\ref{eq:vectorsim}), (\ref{eq:vecsim}), then the azimuthal derivative of 
the solution vector 
can be written as
\begin{equation}\label{eq:aziderivsplit}
\begin{split}
\partial_{\theta} \mathbf{\Psi}|_{\theta=0} & =\partial_{\theta}  \left[ \begin{pmatrix} \mathcal{A_{\tau}}\left(r,t\right)   \\ 
\mathcal{A}_{u}\left(r,t\right)   \\ \mathcal{A}_{v}\left(r,t\right)   \end{pmatrix} \exp \left\{i \left(m\theta - \omega^{s} t \right)\right\} 
\right]_{\theta=0} \\[0.4cm]
& = -m  \begin{pmatrix} \tau_{I}(r,t) \\ u_{I}(r,t) \\  v_{I}(r,t) \end{pmatrix} + i m   \begin{pmatrix} \tau_{R}(r,t) \\ u_{R}(r,t) \\  v_{R}(r,t) 
\end{pmatrix} , \\[0.4 cm]
\end{split}
\end{equation}
As a result the azimuthal derivatives (\ref{eq:aziderivsplit}) induce a coupling between the real and imaginary parts of (\ref{eq:nleq}).

\subsection{Nonlinear Waves}

For the description of nonlinear density waves (the amplitudes $\mathcal{A_{\tau}}$, $\mathcal{A}_{u}$, $\mathcal{A}_{v}$ are not small) for Equations 
(\ref{eq:nleq}) a splitting in real and imaginary part is not suitable. However, we can retain the description in terms of a coupled set of 
equations which can be seen as follows.
First we note that performing the azimuthal derivative of the vector of state (\ref{eq:vecsim}) is equal to a phase shift of $\pi/2$ and a multiplication by 
$m$, so that
\begin{equation}\label{eq:aziphase}
\begin{split}
\partial_{\theta} \mathbf{\Psi}(r,\theta,t) &  =  m \mathbf{\Psi}(r,\theta+ \pi/2,t)\\[0.4cm]
& =  m \left[\mathbf{\Psi}_{R}(r,\theta + \pi/2,t) + i\, 
\mathbf{\Psi}_{I}(r,\theta + \pi/2,t) +c.c. \right]
\end{split}
\end{equation}
This can also be expressed in terms of a time shift of $P/4$, i.e.\ 
\begin{equation*}
\partial_{\theta} \mathbf{\Psi}(r,\theta,t) =  m\mathbf{\Psi}(r,\theta,t+ P/4)
\end{equation*}
where $P=\frac{2 \pi}{\Omega_{L}}$.

Inspection of the forcing terms (\ref{eq:fsatrad}), (\ref{eq:fsatazi}) shows that the imaginary part of the forcing function equals the real part, but with a 
phase shift of $-\pi/2$.  This means that we can consider the real and imaginary parts of (\ref{eq:nleq}) to describe the same forced 
density wave, but with a phase shift of $-\pi/2$ so that
\begin{equation}\label{eq:realima}
 \begin{pmatrix} \tau_{R}(r,\theta,t) \\ u_{R}(r,\theta,t) \\  v_{R}(r,\theta,t) \end{pmatrix} =  \begin{pmatrix} \tau_{I}(r,\theta+\frac{\pi}{2},t) \\ 
u_{I}(r,\theta+\frac{\pi}{2},t) \\  v_{I}(r,\theta+\frac{\pi}{2},t) 
\end{pmatrix}.
\end{equation}
This observation is independent of whether the equations are linear or nonlinear.
The idea is now to define two sets of the same \emph{nonlinear} Equations (\ref{eq:nleq}), where one set is forced with the real parts of 
(\ref{eq:fsatrad}), (\ref{eq:fsatazi}) and is assumed to possess the solution vector
\begin{equation*}
\mathbf{\Psi}_{R}(r,\theta,t) = \begin{pmatrix} \tau_{R}(r,\theta,t) \\ u_{R}(r,\theta,t) \\  v_{R}(r,\theta,t) \end{pmatrix}.
\end{equation*}
The other set is forced with the imaginary parts of (\ref{eq:fsatrad}), (\ref{eq:fsatazi}) and is assumed to possess the solution vector
\begin{equation*}
\mathbf{\Psi}_{I}(r,\theta,t) = \begin{pmatrix} \tau_{I}(r,\theta,t) \\ u_{I}(r,\theta,t) \\  v_{I}(r,\theta,t) \end{pmatrix}.
\end{equation*}
The amplitudes $\mathcal{A_{\tau}}$, $\mathcal{A}_{u}$, $\mathcal{A}_{v}$ will now be affected by nonlinear terms in (\ref{eq:nleq}).

Combining Equations (\ref{eq:aziphase}) and (\ref{eq:realima}) yields
\begin{equation*}
 \partial_{\theta} \begin{pmatrix} \tau_{R}(r,\theta,t) \\ u_{R}(r,\theta,t) \\  v_{R}(r,\theta,t) \end{pmatrix}
 =m \begin{pmatrix} \tau_{I}(r,\theta,t+ \pi) \\ u_{I}(r,\theta,t+ \pi) \\  v_{I}(r,\theta,t+ \pi) \end{pmatrix} = -m\begin{pmatrix} \tau_{I}(r,\theta,t) \\ 
u_{I}(r,\theta,t) \\  v_{I}(r,\theta,t) \end{pmatrix}
\end{equation*}
and
\begin{equation*}
 \partial_{\theta} \begin{pmatrix} \tau_{I}(r,\theta,t) \\ u_{I}(r,\theta,t) \\  v_{I}(r,\theta,t) \end{pmatrix}
 = m\begin{pmatrix} \tau_{R}(r,\theta,t) \\ u_{R}(r,\theta,t) \\  v_{R}(r,\theta,t) \end{pmatrix} .
\end{equation*}
Note that although we retain the notation with subscripts $R$ and $I$, the interpretation of the expressions denoting real and imaginary parts is only valid in 
the linear regime. In the nonlinear case the two quantities $\mathbf{\Psi}_{R}$ and $\mathbf{\Psi}_{R}$ merely describe the same density wave up to a 
constant relative phase shift.

\vspace{0.5cm}

\section{WENO Reconstruction of the Flux-Vector}\label{sec:weno}

The computation of the flux derivative $\partial_{r}\mathbf{F}$ in (\ref{eq:nleqnum}) includes a splitting of $\mathbf{F}$ according to the method of 
\citet{liou1993} which was also used in LSS2017 and a WENO 
reconstruction of its individual components.
In short terms the reconstruction is as follows.
We have (\citet{shu1988})
\begin{equation}\label{eq:hcl}
 \partial_{r} \mathbf{F}(r_{j}) = \frac{\mathbf{f}_{j+1/2}-\mathbf{f}_{j-1/2}}{h}
\end{equation}
with the numerical flux $\mathbf{f}$, implicitly defined through
\begin{equation}\label{eq:numflux}
 \mathbf{F}(r_{j}) = \frac{1}{h} \int \limits_{r_{j-1/2}}^{r_{j+1/2}}  \mathbf{f}(\xi) \mathrm{d} \xi
\end{equation}
so that (\ref{eq:hcl}) is exactly fulfilled. In these expressions the subscripts $j \pm 1/2$ denote evaluations at radial locations $r_{j}\pm \frac{1}{2}h$.
Equation (\ref{eq:numflux}) can be used to obtain interpolating polynomials for $\mathbf{f}$ at a given location $r$, since the nodal values of the physical 
flux 
$\mathbf{F}(r_{j})$ are known for all $j$.
We denote the so obtained unique 5th-order accurate polynomial approximation for the numerical flux values at half nodes (see Section 4.1.1 of 
LSS2017) by $\hat{\mathbf{f}}_{j\pm 1/2}^{(5)} = \mathbf{f}_{j\pm 1/2} + \mathcal{O}\left( h^5\right)$ where  
$\hat{\mathbf{f}^{(5)}}_{j + 1/2}$ and $\hat{\mathbf{f}}^{(5)}_{j - 1/2}$ use the 5-point stencils $[r_{j-2},r_{j-1},\ldots,r_{j+2}]$ and 
$[r_{j-3},r_{j-2},\ldots,r_{j+1}]$, respectively.

The starting point of the WENO reconstruction is the replacement of $\hat{\mathbf{f}}^{(5)}_{j\pm 1/2}$ by
\begin{equation}\label{eq:weno}
 \hat{\mathbf{f}}_{j\pm 1/2} = \sum \limits_{k=0}^{2} w_{k} \hat{\mathbf{f}}^{k}_{j\pm 1/2} 
\end{equation}
where 
\begin{equation*}
 \hat{\mathbf{f}}^{k}_{j + 1/2}  = \sum\limits_{l=0}^{2} c_{kl} \mathbf{F}_{j-k+l} 
\end{equation*}
are the (unique) third-order accurate polynomial approximations for $\mathbf{f}_{j+ 1/2}$ using the three 3-point stencils $[r_{j},r_{j+1},r_{j+2} ]$, 
$[r_{j-1},r_{j},r_{j+1}]$ and $[r_{j-2},r_{j-1},r_{j} ]$, respectively (for $\mathbf{f}_{j- 1/2}$ these are shifted accordingly by $-1$).

For a particular choice of the ``weights'' $w_{k}$ Equation (\ref{eq:weno}) 
does yield $ \hat{\mathbf{f}}_{j \pm 1/2}  = \hat{\mathbf{f}}_{j\pm 1/2}^{(5)}$.
The key point of the decomposition (\ref{eq:weno}) is an adequate assignment of the weights $w_{k}$ so that these yield the 
standard 5th-order accurate Lagrange interpolation $\hat{\mathbf{f}}_{j\pm 1/2}^{(5)}$ wherever $\mathbf{F}$ behaves smoothly across the entire 5-point stencil.
If, however, in some region the solution vector 
contains a 
discontinuity in one of the three sub-stencils, the corresponding weight should diminish in order to avoid spurious oscillations of the solution vector. We use 
the WENO-Z weights introduced by \citet{borges2008} which yield improved accuracy near extrema, as compared with the original WENO weights 
(\citet{jiang1996}). This improved accuracy is important as we are modeling wave systems that exhibit a wide range of length scales, where the shortest 
length scales will span only several grid points, and where the state variables can contain sharp gradients.

Since we apply a splitting of the flux $\mathbf{F} \to \mathbf{F}^{+} + \mathbf{F}^{-}$ so that $\partial (\mathbf{F}^{+/-}) 
/ \partial \mathbf{U}$ possess only non-negative/non-positive eigenvalues, the reconstruction outlined above applies to $\mathbf{f}^{+}_{j\pm 1/2}$, whereas 
$\mathbf{f}^{-}_{j\pm 1/2}$ is reconstructed using stencils that are shifted by $+1$ so as to ensure a correct upwinding (cf.\ 
LSS2017).

\vspace{0.5cm}

\section{WKB-Approximation for Self-Gravity}\label{sec:wkbsg}

For integrations of linear density waves one can implement the self-gravity terms that arise from the solution for the self-gravity potential $\phi^{d}$ in the 
WKB-approximation [cf. Equation (\ref{eq:poissonazi})]
\begin{equation}\label{eq:wkbsg1}
\begin{split}
f^{d} & = f^{d}_{R} +i\, f^{d}_{I}\\[0.1cm]
 & = -\partial_{r} \, \phi^{d}(r,t) = i \,2 \pi G \sigma_{0} \, \tau(r,t)\\[0.1cm]
 & =  i \,2 \pi G \sigma_{0} \left[ \tau_{R}(r,t) + i \tau_{I}(r,t) \right]\\[0.1cm]
  & = -2 \pi G \sigma_{0}\, \tau_{I}(r,t) + i\, 2 \pi G \sigma_{0} \, \tau_{R}(r,t).
\end{split}
\end{equation}
In this approximation, the self-gravity force at a certain grid point is governed by the value of the surface mass density at this particular grid point only.
This implementation of the self-gravity force couples the real and imaginary parts of (\ref{eq:nleq}).

An alternative way to implement the WKB self-gravity which does not induce an additional coupling between the equations and which turns out to work also in the 
nonlinear regime is derived from Equations (35), (45), 
(52) and (53a) in LSS2016.
From these relations follows that the disk potential $\phi^{d}_{(l)}$ and radial velocity $u_{(l)}$ are related through
\begin{equation*}
 \phi_{(l)}^{d} = \frac{\mathcal{D}   \epsilon r_{L}  }{\Omega_{L}} u_{(l)}
\end{equation*}
where $l=1,2$ denote the first and second harmonics of these quantities. 
This relation holds to the lowest order in $\frac{r-r_{L}}{r_{L}}$.
The exact relation (in the rotating frame) is [cf.\ Equation (\ref{eq:azideriv})]
\begin{equation}\label{eq:wkbsg}
 \phi_{(l)}^{d} = -   \frac{\mathcal{D} \epsilon r_{L} }{ \omega^{s} - m\left[\Omega-\Omega_{L} \right]} u_{(l)}.
\end{equation}
If we assume this relation holds for all higher harmonics $l=3,4,\ldots$, we can write the self-gravity force as
\begin{equation}\label{eq:wkbforce}
\begin{split}
f^{d} & = -  \partial_{r} \phi^{d} = \frac{2 \pi G \sigma_{0}}{ \omega^{s} - m\left[ \Omega - \Omega_{L} \right] } \partial_{r} u.
\end{split}
\end{equation}
For sufficiently linear waves the implementations (\ref{eq:wkbsg1}) and (\ref{eq:wkbforce}) yield identical results.



\vspace{0.5cm}

\section{Derivation of the Perturbed Ground State}\label{sec:gspert}

We start with Equation (\ref{eq:streamline}) describing an $m$-lobed fluid streamline
and the expressions for the radial and azimuthal velocities (\citet{Borderies83})
\begin{align}
 u & = \Omega a e \sin m\left(\phi +\Delta \right),\\
 v & = r \Omega \left[ 1 + 2 e \cos m \left(\phi + \Delta \right) \right],
\end{align}
where the latter expressions are valid in an inertial frame $(r,\phi)$ with $\varphi = \phi + \Omega^{s} (t-t_{0})$.
Radial compression of the ring matter is described by
\begin{equation}\label{eq:jac}
 J = \partial_{a} r= 1 - q \cos \left(m \phi + m \Delta + \hat{\gamma}\right)    
\end{equation}
with
\begin{align}
 q \cos \hat{\gamma} & =  \partial_{a}(a e) ,\label{eq:q1}\\
 q \sin \hat{\gamma} & = m a e \partial_{a} \Delta\label{eq:q2},
\end{align}
where $q$ is the nonlinearity parameter [Equation (\ref{eq:qpar})] and $\partial_{a}$ denotes the derivative with respect to 
$a$.
This results in the scaled surface mass density
\begin{equation}\label{eq:taudens}
 \tau = \frac{1}{J}.
\end{equation}

The linearized velocity fields near $x=x_{0}$ are given by
\begin{align}
 u(x) & =  u(x_0) + x \left[\partial_{r}u\right]_{x=x_0} ,\\
 v(x) & =  v(x_0) + x \left[a \partial_{r} (v/r)\right]_{x=x_0} ,
\end{align}
where $\partial_{r} = (1/J) \,\partial_{a}$ by Equation (\ref{eq:jac}).
For the radial velocity $u$ we need to compute
\begin{equation}\label{eq:uderiv}
\begin{split}
 \partial_{a} u & = \sin m \left(\phi + \Delta \right) \left[ a e \partial_{a} \Omega  + \Omega e + \Omega a \partial_{a} e\right] 
 +\Omega a e \cos m \left(\phi + \Delta \right) m \partial_{a} \Delta\\
  \quad & = \sin m \left(\phi + \Delta \right) \left[ -\frac{3}{2} \Omega e + \Omega e + \Omega q \cos \hat{\gamma} \right] + \Omega q \sin \hat{\gamma} \cos m 
\left(\phi 
+ \Delta \right)\\
  \quad & = -\frac{\Omega e }{2} \sin m \left(\phi + \Delta \right) + \Omega q \left[\sin m \left(\phi + \Delta\right)\cos \hat{\gamma} + \cos m \left(\phi + 
\Delta 
\right) \sin \hat{\gamma} \right]\\
\quad & = -\frac{\Omega e }{2} \sin m \left(\phi + \Delta \right) + \Omega q \sin \left[ m \left(\phi + \Delta \right) + \hat{\gamma} \right].
 \end{split}
\end{equation}
Thus, we have
\begin{equation}\label{eq:ufin}
 \begin{split}
 u  = & \left[\Omega a e \sin m \left(\phi + \Delta\right) \right]_{x=x_0}\\
 \quad & +\frac{\Omega x}{J} \left[ -\frac{e}{2} \sin m \left(\phi + \Delta \right)  +  q \sin\left[m \left(\phi + \Delta\right) + 
\hat{\gamma} \right]\right]_{x=x_0} .
 \end{split}
\end{equation}

For the azimuthal velocity $v$ consider
\begin{equation}\label{eq:vderiv}
\begin{split}
 \partial_{a} (v/r) & = \partial_{a} \left[ \Omega \left(1 + 2 e \cos m \left(\phi + \Delta \right) \right) \right]\\
 \quad & = -\frac{3\Omega}{2 a} \left ( 1 +2 e \cos m \left(\phi + \Delta \right) \right) + 2 \Omega \left[ \partial_{a} e \cos m \left(\phi + \Delta \right) 
 -e \sin m \left(\phi + \Delta\right) m \partial_{a} \Delta \right]\\
 \quad & = -\frac{3\Omega}{2 a} \left ( 1 +2 e \cos m \left(\phi + \Delta \right) \right) +  \frac{2 \Omega }{a} \left[ q \cos \hat{\gamma} \cos m \left(\phi + 
\Delta \right) - q \sin \hat{\gamma} \sin m \left(\phi + \Delta \right) \right]\\
\quad & = \frac{\Omega}{a} \left[ - \frac{3}{2} -3 e \cos m \left(\phi + \Delta\right) + 2 q \cos  \left[m \left(\phi + \Delta \right) \right]\right].
  \end{split}
\end{equation}
This leads to
\begin{equation}\label{eq:vfin}
 \begin{split}
 v  = & \left[r \Omega \left(1+ 2 e \cos m \left(\phi + \Delta\right) \right) \right]_{x=x_0}\\
 \quad & +\frac{\Omega x}{J} \left[   - \frac{3}{2} -3 e \cos m \left(\phi + \Delta\right)   + 2 q \cos  \left[m 
\left(\phi + \Delta \right)+\hat{\gamma} \right] \right]_{x=x_0} .
\end{split}
\end{equation}
\citet{mosqueira1996} assumes that nonlinearity (i.e.\ $q >0$) arises solely from an eccentricity gradient, implying $\hat{\gamma}=0$ [c.f.\ 
(\ref{eq:q1}), (\ref{eq:q2})], and that the eccentricity $e$ 
vanishes at $x=x_{0}$. These assumptions are expected to be fulfilled in the evanescent region of the density wave, close to the Lindblad resonance, i.e.\ for 
$x_{0} \lesssim 0$.
In the frame rotating with $\Omega(x=x_{0}))$ we then have
\begin{align}
\tau & = \frac{1}{1 - q \cos m \left(\phi + \Delta \right) }\label{eq:taumosq},\\
 u & =   \Omega  q x \frac{\sin m \left(\phi + \Delta\right)  }{1 - q \cos m \left(\phi + \Delta \right)  }\label{eq:umosq},\\
v & = -\frac{3}{2} \Omega x \frac{1- \frac{4}{3} q \cos  m \left(\phi + \Delta \right)   }{1 - q \cos m \left(\phi + \Delta \right)}\label{eq:vmosq}.
\end{align}

If, on the other hand, we are in the density wave propagation region ($x>0$) we can assume that nonlinearity arises solely due to the variation of the phase 
angle $\Delta$, such that $\hat{\gamma} \sim \pi/2$.
Let us rewrite (\ref{eq:ufin}) and (\ref{eq:vfin}) using the first line of (\ref{eq:uderiv}) and the second line of 
(\ref{eq:vderiv}) such that
\begin{align}
\begin{split}
u & = \left[\Omega a e \sin m \left(\phi + \Delta\right) \right]_{x=x_0}\\
\quad & +  \Bigg[\sin m \left(\phi + \Delta \right) \left[ \partial_{a} \Omega a 
e + \Omega e + \Omega a \partial_{a} e\right]  +\Omega a e \cos m \left(\phi + \Delta \right) m \partial_{a} \Delta \Bigg] \frac{x}{J},\\
 v & = \left[r \Omega \left(1+ 2 e \cos m \left(\phi + \Delta\right) \right) \right]_{x=x_0}\\
 \quad & +  \left[ -\frac{3\Omega}{2 } \left ( 1 +2 e \cos m 
\left(\phi + \Delta \right) \right) + 2 \Omega a \left[ \partial_{a} e \cos m \left(\phi + \Delta \right) 
 -e \sin m \left(\phi + \Delta\right) m \partial_{a} \Delta \right]\right] \frac{x}{J}.
 \end{split}
\end{align}
Since we now have $m a e \partial_{a} \Delta \gg \partial_{a} (a e)$  as well as $m x \partial_{a} \Delta  \gg 1 $ (the WKB-approximation) the last term 
within the brackets in front the factor $\frac{x}{J}$ dominates for both velocities and we arrive (in the frame rotating with $\Omega(x=x_{0}))$) at
\begin{align}
\tau & = \frac{1}{1 + q \sin m \left(\phi + \Delta \right) },\\
 u & =   \Omega  q x \frac{\cos m \left(\phi + \Delta\right)  }{1 + q \sin m \left(\phi + \Delta \right)  }\label{eq:uwkb},\\
v & = -\frac{3}{2} \Omega x \frac{1 + \frac{4}{3} q \sin  m \left(\phi + \Delta \right)   }{1 + q \sin m \left(\phi + \Delta \right)}\label{eq:vwkb},
\end{align}
which is identical to (\ref{eq:taumosq})-(\ref{eq:vmosq}), up to an irrelevant constant phase shift of $\pi /2$.

%
%

 \end{document}